\documentclass[11pt]{article}
\usepackage{amsmath,amsfonts,epsf}
\usepackage{amssymb}
\usepackage{graphicx}
\usepackage{grffile}
\input epsf
\usepackage[nosort]{cite}
\usepackage{amssymb,amsmath,amsthm}
\usepackage{epstopdf}
\usepackage{enumerate}
\usepackage{tensor}
\usepackage{mathrsfs}
\usepackage{comment}
\usepackage{braket}
\usepackage{dcolumn}
\usepackage{bm}
\usepackage[hidelinks]{hyperref}
\usepackage{slashed}
\usepackage{mathtools}
\usepackage{appendix}
\usepackage{breqn}

\usepackage{comment}

\makeatletter\@addtoreset{equation}{section}\makeatother

\title{Bifundamental Multiscalar Fixed Points in $d=3-\epsilon$}
\author{Samarth Kapoor${}^a$ and Shiroman Prakash${}^b$}
\date{}

\newcommand{\tlambda}{{\lambda}}
\begin{document}

\maketitle

\begin{center}

 {\small{\it {$^a$\, Department of Physics and Astronomy, University of British Columbia, \\6224 Agricultural Road, Vancouver, B.C. V6T 1Z1, Canada\\E-mail: samarth4@student.ubc.ca}} \\
 {\it {\center $^b$\, Department of Physics and Computer Science, \\Dayalbagh Educational Institute,  Agra 282005, India\\ E-mail: sprakash@dei.ac.in}}}
\end{center}

\begin{abstract}We study fixed-points of scalar fields that transform in the bifundamental representation of $O(N)\times O(M)$ in $3-\epsilon$ dimensions, generalizing the classic tricritical sextic vector model. In the limit where $N$ is large but $M$ is finite, we determine the complete beta function to order $1/N$ for arbitrary $M$. We find a rich collection of large-$N$ fixed-points in $d=3$, as well as fixed-points in $d=3-\epsilon$, that can be studied to all orders in the parameter $\hat{\epsilon}=N\epsilon$.  With the goal of defining a large-$N$ nonsupersymmetric conformal field theory dominated by a web of planar diagrams, we also study fixed-points in the ``bifundamental'' large-$N$ limit, in which $M$ and $N$ are both large, but the ratio $M/N$ is held fixed. We find a unique infrared fixed-point in $d=3-\epsilon$, which we determine to order $\epsilon^2$. When $M/N \ll 1$, we also find an ultraviolet fixed-point in $d=3$ and $d=3-\epsilon$ that merges with the infrared fixed-point at $\epsilon \sim O(M/N)$. We expect at least one of two candidate fixed-points in integer dimensions -- the infrared fixed-point in $d=2$ and the ultraviolet fixed-point in $d=3$ -- to survive for finite values of $M/N$. 
\end{abstract}

\newpage

\tableofcontents 

\newpage
\section{Introduction}

In this paper, we study multiscalar fixed-points in $3-\epsilon$ dimensions with $O(N)\times O(M)/\mathbb Z_2$ symmetry. Our primary motivation for studying large-$N$ fixed-points with bifundamental matter originates from string theory and the AdS/CFT correspondence, which we describe below, although we believe such fixed-points could be of more general interest in other contexts as well. 

The $\mathcal N =6$ supersymmetric $U(N)_k\times U(M)_{-k}$ Chern-Simons theory with matter in the bifundamental representation known as ABJ theory \cite{ABJ}, plays an important and unique role in our understanding of the AdS/CFT correspondence, as first observed in \cite{ABJTriality}. When $N$ is large, but $M \ll N$, the theory is a large-$N$ vector model \cite{Moshe:2003xn}, and as per general expectations in \cite{Giombi:2011kc}, is a dual to a higher-spin gauge theory \cite{Fronsdal:1978rb, Fradkin:1987ks, Fradkin:1986qy, Vasiliev:1995dn, Vasiliev:1999ba, Sezgin:2002ru}, where the gauge fields are dressed with Chan-Paton factors transforming under $U(M)$. When $N=M$, the theory becomes the ABJM theory \cite{ABJM}, whose large-$N$ limit consists of a web of planar diagrams dual to type IIA string theory on $AdS_4\times CP_3$ when $\lambda=\frac{N}{k}$ is held fixed. The dual reduces to type IIA supergravity when $\lambda$ is taken to strong coupling. In the language of the higher-gauge theory, the parameter $M/N$ plays the role of a gravitational 't Hooft coupling. When $M/N$ is small, the theory contains an infinite tower of higher-spin gauge fields, and as $M/N$ increases, the tower of higher-spin gauge fields somehow coalesce to form strings. (See \cite{Hirano:2015yha, Honda:2015sxa} for more discussion.)

Does a similar phenomenon occur for other examples of conformal field theories with higher-spin duals, with less (see \cite{Honda:2017nku}) or no supersymmetry? For example $U(N)_k$ Chern-Simons theory coupled to fundamental fermions (or bosons) is believed to be dual to a one-parameter family of party-violating higher-spin gauge theories \cite{Giombi:2011kc, Aharony:2011jz}. One can generalize these theories to $U(N)_k\times U(M)_{-k}$ or $O(N)_k\times O(M)_{-k}$ Chern-Simons theories with bifundamental matter \cite{Gurucharan:2014cva, GuruCharan:2017ftx} -- when $M/N$ is small these theories would be virtually identical to their vector model counterparts, and therefore dual to a higher-spin gauge theory -- but when $M/N$ is order 1, the large-$N$ limit of such theories, like the ABJM theory, becomes a web of planar diagrams suggesting that they would be dual to some sort of string theory \cite{HOOFT1974461}. If the behaviour of these theories when $\alpha=1$ and the 't Hooft coupling $\lambda=\frac{N}{k}$ is large is similar to that of ABJM, it is natural to conjecture that such theories could conceivably be dual to simple non-supersymmetric theories of Einstein gravity in $AdS_4$.

In an extension of the weak gravity conjecture \cite{Arkani-Hamed:2006emk} as part of the swampland program, the authors of \cite{Ooguri:2016pdq} (see also \cite{Freivogel:2016qwc}) conjecture that  non-supersymmetric AdS vacuua do not exist in a consistent quantum theory of gravity. This translates into the statement that a strongly interacting\footnote{By ``strongly interacting,'' we essentially mean that the theory satisfies the gap condition -- i.e., anomalous dimensions of all but a finite number of primary operators diverge.} non-supersymmetric conformal field theory whose large-$N$ limit consists of a web of planar diagrams should not exist, or at least there should be some reason for it to not have a simple holographic dual. Therefore a simple way to test of the conjecture of \cite{Ooguri:2016pdq} is to construct examples of such large-$N$ CFT's and attempt to demonstrate or rule out their existence at strong coupling. 

If the conjecture of \cite{Ooguri:2016pdq} is correct, then it should be impossible for the bifundamental Chern-Simons theories in \cite{Gurucharan:2014cva, GuruCharan:2017ftx} to attain strong coupling. Unfortunately, it is hard to say anything about non-supersymmetric bifundamental Chern-Simons theories in the limit where $M/N=1$ and $\lambda$ is large. But there are other simpler examples of conformal field theories with higher-spin duals. Specifically, one can ask, is it possible to generalize the critical $O(N)$ vector model in $3$ dimensions \cite{Wilson:1972cf} to a critical $O(N)\times O(M)/\mathbb Z_2$ model, with $M$ and $N$ both large, and $\alpha \equiv \frac{M}{N}$ held fixed? As $\alpha \rightarrow 0$, this theory would reduce to the critical $O(N)$ model, whose dual is conjectured to be a higher-spin gauge theory \cite{KlebanovPolyakov, Giombi:2009wh, Giombi:2010vg, Giombi:2011ya, Giombi:2016ejx}, modified slightly, as per the discussion in \cite{ABJTriality}, to include Chan-Paton factors transforming in $O(M)$. In the dual description, $\alpha=M/N$ is again the 't Hooft coupling for the gravitational coupling $1/N$, (which evidently possesses a sort of S-duality as our field theory is dual under interchange of $M$ and $N$) and one can ask whether the higher-spin gauge fields coalesce into string like objects as $\alpha$ increased to $1$. 

In the context of phase transitions, $O(N)\times O(M)/\mathbb Z_2$ fixed-points of multiscalar $\phi^4$ theory in $4-\epsilon$ dimensions have been extensively studied in the literature using both the epsilon expansion and bootstrap \cite{Pisarski:1980ix,Kawamura:1988zz, Kawamura1990, Kawamura, Pelissetto:2001fi,Gracey:2002pm,Gracey:2002ze, Osborn:2017ucf, Rychkov:2018vya, Kompaniets:2019xez, Henriksson:2020fqi}, as reviewed in \cite{Osborn:2017ucf,Henriksson:2020fqi}. It turns out\footnote{SP thanks Simone Giombi, Igor Klebanov, Fedor Popov, and Gregory Tarnopolsky for discussions on this point a few years ago.} that a real bifundamental fixed-point, whose large-$N$ limit is dominated by all planar diagrams, only exists in the epsilon expansion \cite{Osborn:2017ucf} for 
\begin{equation}
    M/N < \alpha_{critical}=\left(5-2 \sqrt{6}\right)+\left(\frac{25}{2 \sqrt{6}}-5\right) \epsilon+O\left(\epsilon^2\right),
\end{equation} or, $M/N>1/\alpha_{critical}$. For $1/\alpha_{critical}>\alpha>\alpha_{critical}$, the fixed-point can still be studied formally, although it is no longer unitary and some anomalous dimensions become complex. The anomalous dimension of the elementary scalar field $\phi_{ij}$ remains real, and is given by \cite{Osborn:2017ucf} $$\gamma_\phi = \frac{\alpha ^2}{8 \left(\alpha ^2+1\right)^2}.$$ $\gamma_\phi$ attains its maximum at $\alpha=1$, confirming the expectation that $\alpha=1$ represents strong coupling for the theory. We therefore conclude that, for this theory, the fixed-point becomes complex before strong coupling is attained, in accordance with the conjecture of \cite{Ooguri:2016pdq}. A physical interpretation for complex fixed-points was put forth in \cite{Gorbenko2018, Gorbenko-2-2018}, so formally, one could attempt to define a holographic dual for the complex $\alpha=1$ fixed-point. This dual might contain a very small number of massless fields (possibly only the graviton) if a gap develops when $d=3$, but also a few fields with masses below the Breitenlohner-Freedman bound, corresponding to primary operators of complex dimension \cite{Klebanov:2016xxf}.  

Similar questions can also be asked about the existence of bifundamental multiscalar critical fixed-points starting from sextic interactions in $d=3-\epsilon$ and cubic interactions in $d=6-\epsilon$ dimensions. In this paper, we focus our attention on bifundamental $\phi^6$ fixed-points in $3-\epsilon$ dimensions. 

Is it possible to construct a strongly interacting conformal field theory whose large-$N$ limit is dominated by a web of planar diagrams, via a bifundamental multiscalar theory in $d=3-\epsilon$? We have discovered two candidate fixed-points in integer dimensions -- the infrared bifundamental fixed-point in $d=3-\epsilon$ for $\epsilon=1$, and the ultraviolet bifundamental fixed-point in $d=3$. Our calculations were perturbative --  we require $\epsilon \ll 1$ for the IR fixed-point, and in $\alpha \ll 1$ for the UV fixed-point -- so we are unable to conclusively demonstrate the existence of a strong-coupling limit. Nevertheless, we expect that, for any value of $\alpha$, at least one of these two candidate fixed-points exist. 

The IR fixed-point always exists for $d$ sufficiently close to $3$. For $\alpha \ll 1$, the UV fixed-point also exists in and near $d=3$, and merges with the IR fixed-point at a critical value of $d_*=3-\frac{9}{2\pi^2} \alpha +O(\alpha^2)$ from Eq. \eqref{Critical-epsilon}. Therefore, at small $\alpha$, the IR fixed-point does not extend to $d=2$, but the UV bifundamental fixed-point exists in $d=3$. Our arguments for the existence of the UV fixed-point require $\alpha \ll 1$ (or $\alpha \gg 1$) -- therefore the UV bifundamental fixed-point may or may not exist $d=3$ at any finite value of $\alpha$. However, if the UV fixed-point does not exist for some finite value of $\alpha$, then the IR fixed-point almost certainly survives until $d=2$, because there is no other candidate fixed-point for it to merge with at finite $\epsilon$. 

\subsection{A brief review of the sextic $O(N)$ vector model}

\begin{figure}
    \centering
    \includegraphics[scale=.8]{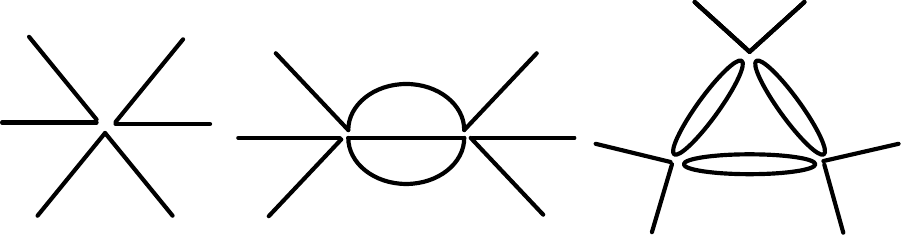}
    
    \includegraphics[scale=.8]{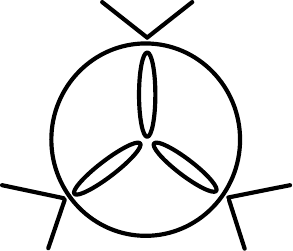}
    \caption{Some Feynman diagrams for corrections to the vertex are pictured above. Here solid lines denote $O(N)$ fundamental index contractions, and $O(M)$ index contractions are not pictured. The interaction vertices can be chosen so that the two-loop and four-loop diagrams in the top row are proportional to $\alpha$ in the bifundamental large-$N$ limit, or $1/N$ in the vector model large-$N$ limit. However, any diagram with six or more loops, such as the diagram in the second row, is suppressed by at least $\alpha^2$ in the bifundamental large-$N$ limit, or $1/N^2$ in the vector model large-$N$ limit.}
    \label{fig:beta-fn-diagrams}
\end{figure}

The large-$N$ limit of the sextic $O(N)$ vector model in $3-\epsilon$ dimensions is quite different from the the large-$N$ limit of the quartic $O(N)$ vector model in $4-\epsilon$ dimensions. Classic references include, e.g., \cite{Townsend:1975kh, Townsend:1976sy, Townsend:1976sz,Appelquist:1981sf, Pisarski1982,  Pisarski:1983gn,Bardeen:1983rv,Gudmundsdottir:1984rr,Gudmundsdottir:1984vyf, David:1985zz}. In particular, Pisarski \cite{Pisarski1982} showed that it is possible to demonstrate the existence of a large-$N$  interacting ultraviolet fixed-point for the sextic $O(N)$ vector model in $d=3$. 
The argument of \cite{Pisarski1982} leading to a fixed-point in $d=3$ is essentially as follows. In the large-$N$ limit, the beta function for the sextic triple-trace interaction vanishes to all orders in the 't Hooft coupling. Therefore, in order to determine large-$N$ fixed-points, one must consider the first $1/N$ correction to the beta function. A simple graph theoretic argument shows that the first $1/N$ correction to the beta function also vanishes beyond four-loops \cite{Pisarski1982}. Let us briefly review this argument. Consider a $O(N)$ symmetric scalar theory with sextic interaction vertex $\sim g(\phi_i \phi_i)^3$. Note that, unlike for quartic scalar theories, an $L$-loop correction to the interaction vertex is proportional to $g^n$ where $n=L/2+1$. Any such diagram contains exactly $e_{int}=3(n-1)$ internal edges (i.e., propagators), as seen in Fig. \ref{fig:beta-fn-diagrams}. The largest possible power of $N$ for such a diagram is the total number of disconnected cycles formed by the internal edges, following  $O(N)$ index-contractions. (For example, in the four-loop diagram in Fig. \ref{fig:beta-fn-diagrams}, there are three disconnected cycles.) Since our graph can contain no tadpoles (or it would vanish), each such disconnected cycle must contain at least 2 internal edges, so the diagram is proportional to, at most, $N^{\lfloor e_{int}/2 \rfloor}=N^{\lfloor \frac{3}{2}(n-1) \rfloor}$. Using the 't Hooft coupling $\lambda=gN^2$, we find each the $L$-loop correction to the vertex is proportional to $g \lambda^{L/2} N^{\lfloor{-L/4}\rfloor}$.
Therefore the only contributions to the beta function at order $1/N$ come from two-loop and four-loop diagrams. An analogous result applies to the bifundamental $O(N)\times O(M)$ theories we will study in this paper, in the limit $M \ll N$.

For sextic theories at finite $N$, the four-loop beta function is cubic in the coupling constants, and therefore only allows one to determine next-to-leading order corrections, (i.e., $O(\epsilon^2)$), in the epsilon expansion. However, in the vector model large-$N$ limit, the four-loop beta function is the complete beta function, to first nontrivial order in $1/N$. Any higher-order corrections to the beta function are proportional to $1/N^2$ and can be made arbitrarily small by making $N$ sufficiently large -- so any interacting fixed-point (with no marginal directions) of the $O(1/N)$ four-loop beta function will be reliable in the large-$N$ limit, even those that do not become free when $d=3$. Extending this analysis to $d=3-\epsilon$, it is easy to see that the four-loop beta function allows one to determine fixed-points valid to all orders in the ``rescaled'' epsilon parameter $\hat{\epsilon}=N \epsilon$. 

\cite{Pisarski1982} finds two fixed-points in $d=3-\epsilon$ -- an IR fixed-point which becomes free in $d=3$ and a UV fixed-point, which is not free when $d=3$. These two merge into a complex fixed-point at a critical value of $\hat{\epsilon}$, $\hat{\epsilon}_C={36 \pi^2} + O(1/N)$, and there is no real fixed-point for $\epsilon<\hat{\epsilon}_C/N$. Therefore, at least when $N$ is large, the sextic $O(N)$ fixed-point does not extend to a CFT in 2 dimensions.  The UV fixed-point, however, appears to be an example of an ultraviolet-stable large-$N$ fixed-point in $d=3$, that may survive for $N$ sufficiently large but finite.

An important aspect of sextic $O(N)$ model is the phenomena of spontaneous breaking of conformal invariance, \cite{Bardeen:1983rv, Amit:1984ri, David:1984we, David:1985zz} and this continues to receive attention in the literature: see, e.g., \cite{Omid:2016jve, Fleming:2020qqx, Litim:2017cnl}. We do not provide a review of this here. Let us also remark that more recent discussion about the asymptotic safety and flows between this UV fixed-point and the (quartic) Wilson Fisher fixed-point in $d=3$ appears in \cite{Litim:2018pxe}. Some operator-product-expansion coefficients and anomalous dimensions in the UV fixed-point were computed more recently, using the background field method, in \cite{Goykhman:2020ffn}. A similar fixed-point arising from a triple-trace deformation to ABJM was studied in \cite{Craps:2009qc}.  

When we study the bifundamental $O(N)\times O(M)/\mathbb Z_2$ model below, we will use similar arguments to \cite{Pisarski1982} to determine interacting fixed-points in $d=3$ when $N$ is large, and $M$ is finite. These arguments also apply when $M$ and $N$ are both large, but the ratio $M/N \ll 1$. However, if $M/N \sim O(1)$, higher loop contributions to the beta function are no longer suppressed by $1/N$ (or $M/N$), and it is not possible to determine interacting fixed-points in $d=3$ from the four-loop beta function.

\subsection{Summary of results}
In this paper, we generalize the sextic $O(N)$ vector model in $3-\epsilon$ dimensions, to a bi-fundamental $O(N)\times O(M)/Z_2$ model. We study this theory in two natural large-$N$ limits. The first is a bifundamental large-$N$ limit motivated by ABJ theory in which both $M$ and $N$ are large, and the ratio $\alpha=\frac{M}{N}$ is held fixed. The second is the conventional ``vector model'' large-$N$ limit, where $N$ is large and $M$ is held finite. We also briefly discuss fixed-points when $M=2$ and $N$ is finite.

In the bifundamental limit, we find a unique, nontrivial infrared fixed-point in $d=3-\epsilon$ for the appropriate 't Hooft couplings, in the $\epsilon$ expansion. We compute anomalous dimensions of all the spin-zero sextic, quartic and quadratic operators that are invariant under $O(N)\times O(M)$ symmetry group at this fixed-point. Unlike the quartic theory in $4-\epsilon$ dimensions, the fixed-point we study here exists as a real fixed-point for all finite values of $\alpha$, including $\alpha=1$. However, as $\alpha \rightarrow 0$, the large-$N$ beta function vanishes in $d=3$ (as expected from \cite{Pisarski1982}). 

To better understand the $\alpha \to 0$ limit, we study the bifundamental theory in the limit where $\alpha \ll 1$, to first order in $(M/N)$. Diagrammatically, this limit is very similar to the vector model large-$N$ limit. All our sextic couplings are exactly marginal when $\alpha=0$, and the four-loop beta function is the complete beta function for the theory to first order in $\alpha$ but all orders in the sextic 't Hooft couplings. In this limit, we are therefore able to determine an interacting ultraviolet fixed-point  in $d=3-\epsilon$, which remains non-interacting in $d=3$, in addition to the infrared fixed-point which becomes free when $d=3$. Both fixed-points can be determined to all orders in the parameter $\tilde{\epsilon}=\epsilon/\alpha$, For $d<3$ the ultraviolet fixed-point is ultraviolet-stable, and may be used to provide an asymptotically safe definition of the theory. For $d=3$, however, one of the couplings is marginal, so higher-order corrections are needed to determine its existence. Finite $M$ calculations indicate that the ultraviolet fixed-point does exist when $d=3$. It would be interesting to know whether or not the $d=3$ ultraviolet fixed-point survives at small but finite $\alpha$, or, when $\alpha=1$. If so, it might constitute an interacting on-supersymmetric large-$N$ CFT in $3$ dimensions, with a potentially simple gravitational dual. 

Does the infrared bifundamental fixed-point in $d=3-\epsilon$ extend to $d=2$? Using the results from studying the theory at small $\alpha$, we show that the infrared fixed-point merges with the ultraviolet fixed-point and becomes complex at $\epsilon_C=\frac{9}{2\pi^2}\alpha + O(\alpha^2)$. For small $\alpha$, we can therefore conclude that the fixed-point does not extend to $d=2$. But, unlike the $\phi^4$ theory, the range of $\epsilon$ for which the fixed-point is real increases with $\alpha$. It is therefore conceivable, but not-at-all certain, that the fixed-point could extend to $d=2$ for $\alpha=1$. It may be possible to calculate the order $\alpha^2$ term in $\epsilon_C$ with higher loop calculations, which could perhaps shed more light on this question. Of course, a merger is only possible if the ultraviolet fixed-point exists in $d=3$. Therefore we expect that, at any value of $\alpha$, either the UV fixed-point exists in $d=3$, or the IR fixed-point survives until $d=2$. Of course, our computations are perturbative in nature, and it remains to be seen whether either of these fixed-points are strongly interacting at $\alpha=1$. 

We then study the theory in the more familiar vector model large-$N$ limit, in which $N$ is large while $M$ is held finite. Again, as in \cite{Pisarski1982}, the large-$N$ beta function is exactly computable to first nontrivial order in $1/N$, and we are able to use it to determine large-$N$ interacting fixed-points in $d=3$ for any finite value of $M$. For sufficiently large values of $M$, we find a total of $6$ nontrivial interacting fixed-points in $d=3$, including two ultraviolet-stable fixed-points that may provide asymptotically safe definitions of the theory when $N$ is large. We can extend these fixed-points to $d=3-\epsilon$ to all orders in the parameter $\hat{\epsilon}=N\epsilon$, and thereby determine the range of $\epsilon$ for which the fixed-points exist in the large-$N$ limit. We  find a rich collection of infrared and ultraviolet fixed-points, generalizing the fixed-points of \cite{Pisarski1982}, that generically merge into complex fixed-points at various finite values of $\hat{\epsilon}$. However, unlike the case of $M=1$ in \cite{Pisarski1982}, we also find that there also exist two (four, in the case of $M=3$) real fixed-points that persist for arbitrarily large values of $\hat{\epsilon}$, and therefore could conceivably extend to $d=2$.  

While the primary focus of this paper is on large-$N$ fixed-points, we also study perturbative fixed-points for $M=2$ and $N$ finite in section \ref{sec:finiteN}. 

Let us mention some closely-related work. For the special case of $M=N$, the theory we study was recently also studied perturbatively in \cite{Jepsen:2020czw}, where the focus was on generating non-conventional fixed-points by studying the theory at non-integer values of $N$, and a closely related supersymmetric model was recently studied in \cite{Jepsen:2021rhs}. $O(N)\times O(M)$ fixed-points can also be considered in the larger context of tensor models, e.g., \cite{Klebanov:2016xxf, Giombi:2017dtl, Prakash:2017hwq, Delporte:2020rce, Benedetti:2020yvb, Benedetti:2020sye}, and in particular, fixed-points of $O(N)^3$ tensor models with sextic interactions were discussed in \cite{GKPPT, Popov:2019nja, Prakash:2019zia, Benedetti:2019rja, Harribey:2021xgh}. The theory we consider can also be coupled to a Chern-Simons gauge field in $d=3$ -- some two-loop computations in the resulting theory appear in \cite{Banerjee:2013nca}. 

Let us also remark that the analysis of \cite{Yabunaka:2017uox, Yabunaka:2018mju, Yabunaka:2021fow}, who study the sextic $O(N)$ theory in $3-\epsilon$ using functional renormalization group techniques, suggests certain subtleties in the $N\to \infty$ limit of the sextic $O(N)$ model. Such subtleties could also be present in the $N\to \infty$ limit of the $O(N)\times O(M)$ models when $M$ is finite, although the analysis of \cite{Yabunaka:2017uox, Yabunaka:2018mju, Yabunaka:2021fow} would need to be redone in this case.  

This work can also be understood as part of the larger program of determining fixed-points of general multiscalar theories, which has attracted considerable attention in recent years, e.g., \cite{ Osborn:2017ucf, Rychkov:2018vya, Kompaniets:2019xez,Henriksson:2020fqi,Eichhorn:2015woa, Kompaniets:2017yct,  Stergiou:2019dcv, Zinati:2019gct, Ryttov:2019dtg, Osborn:2020cnf,  Codello:2020lta, Chai:2020hnu}, and in particular, our paper heavily relies on the very useful general results for multiscalar theories presented in \cite{Osborn:2017ucf}.

\section{$O(N)\times O(M)/\mathbb Z_2$ invariant multiscalar theory}
We study a theory that contains $MN$ real scalar fields, denoted as $\phi_{ab}$ where $a=1,~2, \ldots, N$ and $b=1,~2, \ldots, M$, that is invariant under the symmetry group $O(N)\times O(M)/\mathbb Z_2$. We can write the action for a scalar theory in $3-\epsilon$ dimensions with this symmetry group as:
\begin{equation}
    \int d^{3-\epsilon}x~\left(  \frac{1}{2}(\partial \phi_{ab})(\partial \phi_{ab})+V(\phi) \right). 
\end{equation}
Near $d=3$, the most general renormalizable potential can be written as a sum of sextic, quadratic and quartic terms: $V(\phi)=V^{(6)}(\phi)+V^{(4)}(\phi)+V^{(2)}(\phi)$. The most general classically-marginal sextic interaction $V^{(6)}(\phi)$ that is invariant under $O(N)\times O(M)/\mathbb Z_2$ is a sum of three sextic operators\footnote{In Eq. \eqref{sextic-operators}, each coupling constant is divided by $6!$, which is a convenient choice for our purposes. An alternative convention would have been to instead divide each coupling by the size of the automorphism symmetry group of each operator, i.e., $g_1$ by $|D_6|=12$, $g_2$ by $|D_2||D_4|=2\times8=16$, and $g_3$ by $|S_3||D_2|^3=3!2^3$. Here $S_3$ is the group of permutations and $D_n$ is the dihedral group on $n$ elements.},
\begin{equation}
    V^{(6)} = \frac{g_1}{6!} \mathcal O_1^{(6)} +\frac{g_2}{6!} \mathcal O_2^{(6)} + \frac{g_3}{6!} \mathcal O_3^{(6)},  \label{sextic-operators}
\end{equation}
a single-trace operator, 
\begin{equation}
    \mathcal O_1^{(6)} = ( \phi_{a_1b_1}\phi_{a_1b_2}\phi_{a_2b_2}\phi_{a_2b_3}\phi_{a_3b_3}\phi_{a_3b_1}),
\end{equation}
a double-trace operator,
\begin{equation}
    \mathcal O_2^{(6)} = (\phi_{a_1b_1}\phi_{a_1b_1})(\phi_{a_2b_2}\phi_{a_2b_3}\phi_{a_3b_3}\phi_{a_3b_2}),
\end{equation}
and a triple-trace operator,
\begin{equation}
   \mathcal O_3^{(6)} = (\phi_{ab}\phi_{ab})^3.
\end{equation}

The most general quartic interaction invariant under $O(N)\times O(M)/\mathbb Z_2$, is the sum of two quartic operators, \begin{equation}
    V^{(4)}=\nu_1 \mathcal O_1^{(4)}+ \nu_2 \mathcal O_2^{(4)},\label{quartic}
\end{equation}  
a single-trace quartic operator $\mathcal O_1^{(4)}= (\phi_{a_1b_1}\phi_{a_1b_2}\phi_{a_2 b_2}\phi_{a_2 b_1})$ and a double-trace quartic quartic operator, $\mathcal O_2^{(4)}= (\phi_{a_1b_1}\phi_{a_1b_1})^2$.

The only quadratic operator invariant under our symmetry group is the usual, single-trace, mass term:
\begin{equation}
   V^{(2)}= m^2\phi_{ab}\phi_{ab}.
\end{equation}
At any fixed-point, the two relevant quartic couplings and the mass term must be tuned to zero, in addition to any sextic couplings which turn out to be relevant when quantum corrections are taken into account.

Using the results for a general sextic multiscalar model in \cite{Osborn:2017ucf}, we can obtain the four-loop beta function for this theory for finite $N$ and $M$. These expressions, which are rather long, are presented in Appendix \ref{beta-function}.  The results of \cite{Osborn:2017ucf} also allow us to obtain expressions for the anomalous dimensions for the field $\phi_{ab}$, $\phi^2$, $\mathcal O^{(4)}_1$ and $\mathcal O^{(4)}_2$. These expressions are presented in Appendix \ref{anomalousApp}. 

In this paper, we are primarily interested in determining fixed-point solutions to these beta functions in the limit when $N$ is large. We consider two natural large-$N$ limits: a bifundamental large-$N$ limit, in which both $M$ and $N$ are large, and a vector-model large-$N$ limit, in which $N$ is large but $M$ is finite. Results for the bifundamental large-$N$ limit are presented in section \ref{sec:bifundamental}. Results for the vector-model large-$N$ limit are presented in section \ref{sec:vector-model}. For completeness, we also study the perturbative fixed-points of our model for $M=2$ and $N$ finite in section \ref{sec:finiteN}.

Note that, for a sextic theory, a two-loop beta function allows one to determine fixed-points up to order $\epsilon$, and a four-loop beta function allows one to determine fixed-points up to order $\epsilon^2$. However, as we will see below, in the limit where $N$ is large and $M$ is finite, or when $M/N \ll 1$, the four loop beta function is the complete beta function to first nontrivial order in $1/N$ and $M/N$.

\section{Fixed points in the bifundamental large-$N$ limit}
\label{sec:bifundamental}
\subsection{Unique fixed-point for $\frac{M}{N}$ finite}
\label{sec:unique-infrared-fixed-point}
In analogy with \cite{ABJTriality}, we define the bifundamental large-$N$ limit of the theory as follows: We take the limit $N\rightarrow \infty$, keeping the ratio $\alpha=M/N$ fixed. We will always assume that $N>M$.

In this large-$N$ limit, the following 't Hooft couplings for the single-trace, double-trace and triple-trace interactions are held fixed:
\begin{equation}
    \tlambda_1 \equiv \frac{g_1}{(8\pi)^2} N^2  ,~\tlambda_2 \equiv \frac{g_2}{(8\pi)^2}  N^{2}M,~\tlambda_3\equiv \frac{g_3}{(8\pi)^2}  N^2 M^2. \label{bifundamental-thooft}
\end{equation}
These definitions follow from a standard graph-theoretical analysis of the leading contribution to free energy graphs, as in, e.g., \cite{KPP, Prakash:2017hwq}. Alternatively, the definitions in \eqref{bifundamental-thooft} can be obtained by imposing the requirement that two-loop corrections to the scalar propagator remain finite in our large-$N$ limit. The factors of $(8\pi)^2$ are chosen for convenience. 

This choice of 't Hooft couplings does not preserve the symmetry between $N$ and $M$. An alternative choice of couplings which preserves the symmetry between $N$ and $M$ is given by $\bar{\tlambda}_1=\alpha \tlambda_1$, $\bar{\tlambda}_2=\sqrt{\alpha} \tlambda_2$, and $\bar{\tlambda}_3=\tlambda_3$. In the regime where $\alpha \sim O(1)$, the difference between $\bar{\tlambda}_i$ and $\tlambda_i$ is simply a matter of notation. However, in the regime where $\alpha \ll 1$, these two choices of 't Hooft couplings may define different large-$N$ limits. The large-$N$ limit defined by holding $\tlambda_i$ fixed is clearly non-singular for all $\alpha<1$ and reduces to the familiar vector model large-$N$ limit when $\alpha$ is very small, so it appears most suitable for our present discussion.

To two loops, the beta functions in this large-$N$ limit are:
\begin{eqnarray}
\beta^{2\text{ loop}}_{\tlambda_1} & = & -2 \tlambda _1 \epsilon + \frac{\alpha  \tlambda _1^2}{10}  \label{beta1}\\
\beta^{2\text{ loop}}_{\tlambda_2} & = & -2 \tlambda _2 \epsilon+\frac{\alpha}{30}  \tlambda _1 \left(9 (\alpha +1) \tlambda _1+4 \tlambda
   _2\right)  \label{beta2}\\
 \beta^{2\text{ loop}}_{\tlambda_3} & = & -2 \tlambda _3 \epsilon+\frac{\alpha}{90}   \left(21 \alpha  \tlambda _1^2+12 (\alpha +1) \tlambda _2
   \tlambda _1+4 \tlambda _2^2\right) \label{beta3}
\end{eqnarray}

The system of equations $\beta_{\tlambda_i}=0$ possess a unique nonzero solution, which is:
\begin{equation}
\tlambda _1\to \frac{20}{\alpha }\epsilon,~\tlambda _2\to -\frac{180 (\alpha 
    +1 )}{\alpha }\epsilon,~\tlambda _3\to \frac{20 \left(72 \alpha ^2
    +151 \alpha   +72  \right)}{3 \alpha } \epsilon. \label{bifundamental-fixed-point}
\end{equation}

We have also computed four-loop and $1/N$ corrections to this fixed-point in Appendix \ref{corrections}. In particular, we will make use of the four-loop corrections presented in  Appendix \ref{four-loop} in the next subsection, when presenting anomalous dimensions at this fixed-point.

\subsubsection*{Stability and anomalous dimensions}
In the large-$N$ limit, at the four-loop level, the anomalous dimension for the field and the mass operator at the fixed-point are independent of \(\alpha\):
\begin{equation}
    \gamma_{\phi}=\frac{\alpha ^2 {\tlambda_1}^2}{10800}=\frac{\epsilon^2}{27}, \quad \gamma_{\phi^2}=\frac{2 \alpha ^2 {\tlambda_1}^2}{675}=\frac{32\epsilon^2}{27}
\end{equation}
The quartic single-trace and double-trace operators given in \eqref{quartic} do not mix at this order, and have the  anomalous dimensions
\begin{equation}
    \begin{split}
     &\gamma_{\nu_1}=\frac{2 \alpha  \tlambda_1}{15}-\frac{\left(3 \pi ^2 \alpha ^3+\left(106+3 \pi ^2\right) \alpha ^2+3 \pi ^2 \alpha \right) \tlambda_ 1^2}{5400}=\frac{8 \epsilon}{3}-\frac{2(\pi^2+38\alpha+3\pi^2 \alpha+\pi^2\alpha^2)\epsilon^2}{27 \alpha}+O(\epsilon^3)\\
    &\gamma_{\nu_2}=\frac{4 \alpha ^2 \tlambda_1^2}{675}=\frac{64\epsilon^2}{27}+O(\epsilon^3)
    \end{split}
\end{equation}
The anomalous dimension of the double-trace quartic operator, is just twice that of $(\phi_{ab})^2$, as expected in the large-$N$ limit, while the single-trace quartic operator acquires a nontrivial anomalous dimension.

The stability matrix for the sextic couplings at the fixed-point is given by
\begin{equation}
\label{bifundamental-stabilitymatrix}
    M_{ab}\equiv\frac{\partial \beta_a(\lambda)}{\partial \lambda_b}\biggr|_{\vec{\lambda}_{\star}}.
\end{equation}
To four-loops, the eigenvectors of the matrix in \eqref{bifundamental-stabilitymatrix} are:
\begin{equation}
    \left(
\begin{array}{c}
 3+\frac{\left(-306 \pi ^2 \alpha ^4-1003 \pi ^2 \alpha ^3+252 \alpha ^3-1293 \pi ^2 \alpha ^2+492 \alpha ^2-1003 \pi
   ^2 \alpha +252 \alpha -306 \pi ^2\right) \epsilon }{4 \alpha  \left(72 \alpha ^2+151 \alpha +72\right)} \\
 -27 (\alpha +1)+ \frac{3 \left(774 \pi ^2 \alpha ^5+3409 \pi ^2 \alpha ^4+108 \alpha ^4+6075 \pi ^2 \alpha ^3+444 \alpha ^3+6075 \pi
   ^2 \alpha ^2+444 \alpha ^2+3409 \pi ^2 \alpha +108 \alpha +774 \pi ^2\right) \epsilon }{4 \alpha  \left(72 \alpha
   ^2+151 \alpha +72\right)}\\
 72 \alpha ^2+151 \alpha +72 \\
\end{array}
\right),
\end{equation}
\begin{equation}
\label{bifundamental-eigen-vectors}
    \left(
\begin{array}{c}
 0 \\
 -1+\frac{\left(20 \pi ^2 \alpha ^2+27 \pi ^2 \alpha -28 \alpha +20 \pi ^2\right) \epsilon }{60 \alpha } \\
 5 (\alpha +1) \\
\end{array}
\right)\quad \text{ and}\quad\left(
\begin{array}{c}
 0 \\
 0 \\
 1 \\
\end{array}
\right);
\end{equation}
with the eigenvalues \begin{equation}
   2\epsilon-\frac{\left(\pi ^2 \alpha ^2+34 \alpha +\pi ^2\right) \epsilon ^2}{9 \alpha },~ \frac{2 \epsilon }{3}-\frac{2 \left(\pi ^2 \alpha ^2+3 \pi ^2 \alpha +22 \alpha +\pi ^2\right) \epsilon ^2}{27 \alpha
   }~\text{and } -2\epsilon+\frac{32 \epsilon ^{2}}{9},
\end{equation} 
respectively. So, the fixed-point \eqref{bifundamental-fixed-point} is stable in two directions, and unstable in one direction.

The scaling dimension, $\Delta_{(\phi^6)_{(i)}}$ of a sextic operator $(\phi^6)_{(i)}$ is related to the corresponding eigenvalue of the stability matrix $(\partial_\lambda \beta_\lambda)_{(i)}$, via
\begin{equation}
    \Delta_{(\phi^6)_{(i)}}=d+(\partial_\lambda \beta)_{(i)}.
\end{equation}
The scaling dimensions of quartic and quadratic operators $\phi^n_i$ are related to their anomalous dimensions via
\begin{equation}
    \Delta_{(\phi^n)_{(i)}} = n(d-2)/2 + \gamma_{(\phi^n)_{(i)}}.
\end{equation}
We thus see that the scaling dimensions of the double-trace and triple-trace operators satisfy $\Delta_{(\lambda_2)}=\Delta_{\nu_1}+\Delta_{\phi^2}$, and $\Delta_{(\lambda_3)}=3 \Delta_{\phi^2}$, in the large-$N$ limit, as expected. We have calculated $O(1/N)$ corrections to the stability matrix and anomalous dimensions in Appendix \ref{corrections}. 

\begin{figure}
    \centering
    \includegraphics[width=3in]{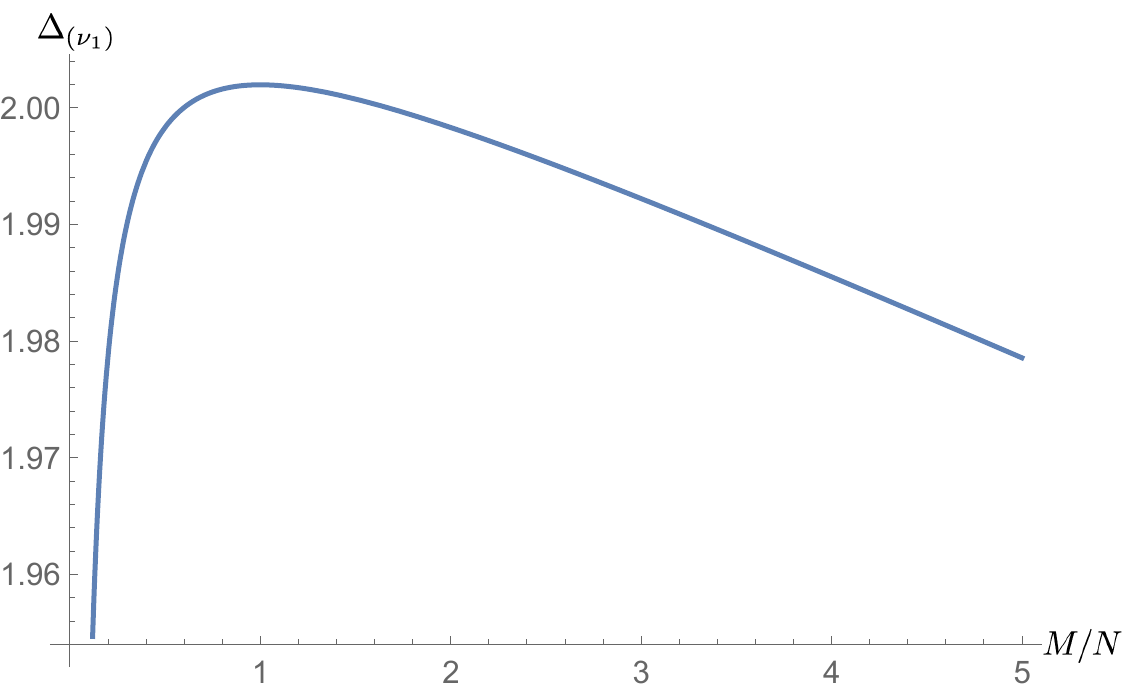}
    \includegraphics[width=3in]{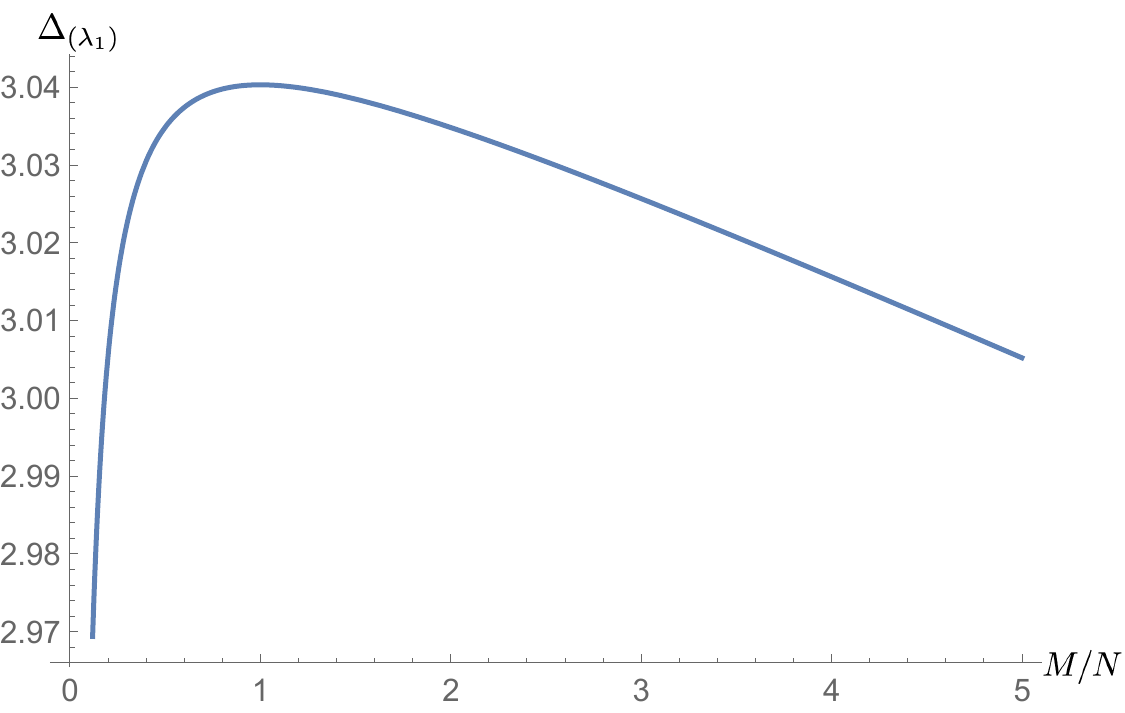}
    \caption{Plots of the scaling dimensions $\Delta_{\nu_1}$ and $\Delta_{\lambda_1}$ of the single-trace quartic and sextic operators, for $\epsilon=0.1$, as a function of $\alpha=M/N$. The maximum is attained at $M/N=1$.}
    \label{fig:quartic-and-sextic-anomalous-dimensions}
\end{figure}

Let us briefly discuss the anomalous dimensions of the unprotected single-trace operators. Although $\gamma_\phi$ and $\gamma_{\phi^2}$ are independent of $\alpha$ to the order we were able to compute, we expect that subsequent higher-order corrections would depend on $\alpha$ in a nontrivial way. Anomalous dimensions of the quartic and sextic single-trace operators do show explicit dependence on $\alpha$, which are plotted in Fig. \ref{fig:quartic-and-sextic-anomalous-dimensions}. As expected, both are symmetric under interchange of $\alpha$ and $\alpha^{-1}$, and the anomalous dimensions attain their maximum value at $\alpha=1$. This result is consistent with the expectation that $\alpha$ acts as a measure of the strength of interactions, and that anomalous dimensions of generic unprotected operators could diverge when $\alpha \to 1$ and $\epsilon \to 1$, as required for a simple gravitational dual description. 

We also observe that the $\epsilon^2$ terms in the anomalous dimensions of $\mathcal O^{(4)}_1$ and $\mathcal O^{(6)}_1$ are negative. If we fix $\epsilon$, and allow $\alpha=M/N$ to become arbitrarily small (or large), these anomalous dimensions can become negative, and arbitrarily large in magnitude, leading to a violation of unitarity. This suggests that, at least in the limit $\alpha \ll 1$ (or $\alpha \gg 1$), our fixed-point only exists up to some critical value of $\epsilon$, which we denote as $\epsilon_C(\alpha)$,  that vanishes as $\alpha \to 0$. This is consistent with our expectations -- when $\alpha \to 0$ our model reduces to the classic tricritical vector model studied in \cite{Pisarski1982}. In the large-$N$ limit, \cite{Pisarski1982} shows that the tricritical vector model fixed-point only exists for $\epsilon<\frac{36\pi^2}{N}$, which tends to $0$ as $N \to \infty$. 
 
\subsection{The $\frac{M}{N} \rightarrow 0$ limit}
\label{small-alpha-section}

In this subsection, we study our theory in the limit of $N$ large and $\alpha \ll 1$ in more detail, and compute the critical value of $\epsilon_C(\alpha)$ to first order in $\alpha$. Of course, the results of this section also determine $\epsilon_C$ to leading order in $\alpha^{-1}$, since our theory has a symmetry $\alpha \leftrightarrow \frac{1}{\alpha}$.  

Observe that the large-$N$ beta-functions in \eqref{beta1}-\eqref{beta3} vanish when $\alpha \rightarrow 0$, as expected from \cite{Pisarski1982, Osborn:2017ucf}. Let us now understand this limit better by keeping only those terms linear in $\alpha$ in the four-loop large-$N$ beta function:
\begin{eqnarray}
\beta_{\lambda_1} & = &-2 \lambda _1 \epsilon +\alpha  \left(\frac{\lambda _1^2}{10}-\frac{\pi ^2 \lambda
   _1^3}{3600}\right) + O(\alpha^2) \label{alpha-beta1}\\
   \beta_{\lambda_2} & = & -2 \lambda _2 \epsilon +\alpha  \left(\frac{3 \lambda
   _1^2}{10}+\frac{2 \lambda _1 \lambda _2}{15}-\frac{\pi ^2 \lambda
   _1^3}{1200}-\frac{\pi ^2  \lambda _1^2\lambda _2}{1800}\right)
   + O(\alpha^2) \label{alpha-beta2}\\
   \beta_
  {\lambda_3} & = & -2 \lambda _3  \epsilon +
   \alpha  \left(\frac{2 \lambda _1 \lambda _2}{15}+\frac{2 \lambda _2^2}{45}-\frac{\pi ^2 \lambda _1^2 \lambda
   _2}{1800}-\frac{\pi ^2 \lambda _1
   \lambda _2^2}{2700}-\frac{\pi ^2 \lambda _2^3}{24300}\right)
  + O(\alpha^2) \label{alpha-beta3}
\end{eqnarray}
The complete beta function of our theory can be expanded as a power series in $\alpha$ and $\lambda_i$. We now argue that the 4-loop beta function given in equations \eqref{alpha-beta1}-\eqref{alpha-beta3} is actually the \textit{complete} beta function, valid to all orders in $\lambda_i$, but only to first order in $\alpha$.

Recall that an expansion in powers of $\alpha=M/N$, near $\alpha=0$ for a bifundamental theory is closely related to an expansion in powers of $1/N$ for the vector model obtained by setting $M=1$.
In \cite{Pisarski1982}, it was argued that the only contributions to the $1/N$ terms in the beta function for the vector model come from diagrams with 4 loops or fewer. Applying the discussion in \cite{Pisarski1982} regarding the $1/N$ terms in the sextic vector model to our bifundamental model, we observe that the diagrams which contribute to the beta function at order $\alpha^p$ in the bifundamental model are a proper subset of the diagrams that contribute to a vector model at order $1/N^p$, i.e., those diagrams which are also proportional to $M^p$ and therefore planar in the usual graph-theoretic sense.  It therefore follows that the only terms in the beta function proportional to $\alpha$ originate from two-loop and four-loop diagrams. So, the 4-loop beta function given in equations \eqref{alpha-beta1}-\eqref{alpha-beta3} is actually the \textit{complete} beta function, valid to all orders in $\lambda_i$ but only to first order in $\alpha$.

Because this is beta function contains all terms linear in $\alpha$, we can use it to compute  fixed-points that are valid to all orders in the parameter $\tilde{\epsilon} \equiv \epsilon/\alpha$. Moreover, we can also use the beta function to determine interacting fixed-points in $d=3$, since any higher-loop corrections to the beta function can be made arbitrarily small by reducing $\alpha$. Higher-loop contributions to the beta function will therefore only give rise to corrections to these fixed-points that are suppressed by powers of $\alpha$; unless a fixed-point has a marginal direction, in which case it may or may not survive higher-order corrections. Higher-loop corrections could also lead to new fixed-points, at sufficiently strong coupling.

\begin{figure}
    \centering
    \includegraphics{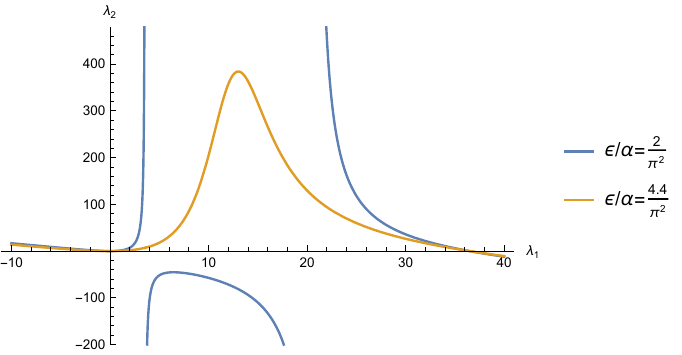}
    \caption{A plot of $\lambda_2^*(\lambda_1)$ determined using \eqref{critical-curve}. For $\epsilon<\frac{4}{\pi^2} \alpha$, there are two poles, and the critical curve determined by \eqref{critical-curve} consists of three disconnected components. For $\epsilon>\frac{4}{\pi^2} \alpha$, there are no poles, and the critical curve consists of a single component. }
    \label{fig:lambda2}
\end{figure}

Let us now look for zeros of the beta function. For any fixed-point, $\lambda_2$ and $\lambda_3$ are uniquely determined from the value of $\lambda_1$ via equations \eqref{alpha-beta2} and \eqref{alpha-beta3}, which depend linearly on $\lambda_2$ and $\lambda_3$ respectively: 
\begin{equation}
    \begin{split}
        \lambda_2^* & =  -\frac{3 \lambda _1^2 \left(\pi ^2 \lambda _1-360\right)}{2 \left(\pi ^2 \lambda _1^2-240 \lambda _1+3600 \tilde{\epsilon} \right)} \\
        \lambda_3^* & = -\frac{ \left(2 \pi ^2 \lambda _2^3+18 \pi ^2 \lambda _1 \lambda _2^2-2160 \lambda _2^2+27 \pi ^2 \lambda _1^2 \lambda _2-6480 \lambda _1 \lambda _2\right)}{97200 \tilde{\epsilon} } \label{critical-curve}
    \end{split}
\end{equation}Any fixed-point lies on the one dimensional ``critical curve'', determined by the solutions to these two equations. For $\tilde{\epsilon}<4/\pi^2$, there are two poles at $ \lambda_1^\pm=\frac{60 \left(2 \pm \sqrt{4-\pi ^2 \tilde{\epsilon} }\right)}{\pi ^2}$, so this critical curve consists of three disconnected components, while for $\tilde{\epsilon}>4/\pi^2$ it consists of a single connected component, as can be seen from the plot of $\lambda_2^*$ in Fig. \ref{fig:lambda2}. 

As in \cite{Pisarski1982}, we find two solutions. The first is a generalization of the conventional IR fixed-point of \cite{Pisarski1982}, given by
\begin{eqnarray}
    \tlambda_1^{IR} & = & \frac{60 \left(3  -\sqrt{9-2 \pi ^2 \tilde{\epsilon} }\right)}{\pi ^2 } + O(\alpha) \approx 20 \tilde{\epsilon}+\frac{10 \pi ^2 }{9}\tilde{\epsilon}^2 + O(\tilde{\epsilon}^3)+O(\alpha), \label{smallalpha1}
\end{eqnarray}
using which we can determine $\lambda_2$ and $\lambda_3$ from Eq. \eqref{critical-curve} to be:
\begin{eqnarray}
    \tlambda_2^{IR} & =&-\frac{360 \tilde{\epsilon} }{-1+\sqrt{9-2 \pi ^2 \tilde{\epsilon} }} +O(\alpha)\approx  -180 \tilde{\epsilon} -{30 \pi ^2 \tilde{\epsilon} ^2} + O(\tilde{\epsilon}^3)+O(\alpha), \label{smallalpha2} \\
    ~\tlambda_3^{IR} & \approx & {480 \tilde{\epsilon} }+\frac{620 \pi ^2 }{3}\tilde{\epsilon}^2 + O(\tilde{\epsilon}^3)+O(\alpha). \label{smallalpha3}
\end{eqnarray}
If we truncate this solution to order $\epsilon^2$, we recover the small $\alpha$ limit of the fixed-point in the previous subsection; but, because we know the complete beta function in this limit, the solutions are valid to all orders in $\tilde{\epsilon}$.

The second solution is a generalization of the UV fixed-point of \cite{Pisarski1982}, and is given by
\begin{eqnarray}
\tlambda_1^{UV} & = &  \frac{60 \left(3  +\sqrt{9-2 \pi ^2 \tilde{\epsilon} }\right)}{\pi ^2 } + O(\alpha) \approx  \frac{360}{\pi ^2} -20 \tilde{\epsilon} - \frac{10 \pi ^2 \tilde{\epsilon}^2}{9} + O(\tilde{\epsilon}^3)+O(\alpha),
\end{eqnarray} using which we can determine $\lambda_2$ and $\lambda_3$ from Eq. \eqref{critical-curve} to be:
\begin{eqnarray}
\tlambda_2^{UV} & =&\frac{360 \tilde{\epsilon} }{1+\sqrt{9-2 \pi ^2 \tilde{\epsilon} }} +O(\alpha)\approx  {90 \tilde{\epsilon} }+ \frac{15 \pi ^2}{2}\tilde{\epsilon} ^2 + O(\tilde{\epsilon}^3)+O(\alpha) \\
\tlambda_3^{UV} & \approx & -\frac{1080}{\pi ^2}-{210 \tilde{\epsilon} } -\frac{205 \pi^2 }{6}\tilde{\epsilon}^2 + O(\tilde{\epsilon}^3)+O(\alpha).
\end{eqnarray}
Ordinarily, one would regard the ``UV'' fixed-point as an artifact of perturbation theory that would not survive higher-loop corrections -- but, because we know the complete beta function to first order in $\alpha$, this solution is physically meaningful.  Notice that this solution is only valid when $\tilde{\epsilon} \neq 0$, because we had to divide by $\tilde{\epsilon}$ in Eq. \eqref{critical-curve}. We discuss the case of $d=3$ in subsection \ref{d=3smallalpha}, in which case $\lambda_3$ is undetermined and marginal.

\begin{figure}
    \centering
    \includegraphics[scale=.6]{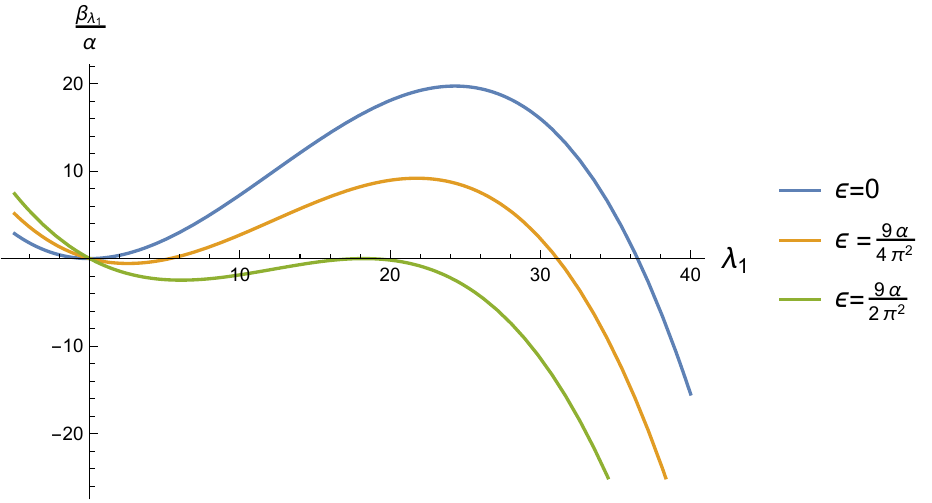}
    \caption{The beta function for $\lambda_1$, given in Eq. \eqref{alpha-beta1}, is plotted for various values of $\epsilon/\alpha$. When $0<\epsilon<\frac{9}{2\pi^2} \alpha$, the UV fixed-point flows to the IR fixed-point. When $\epsilon>\frac{9}{2\pi^2}$ the fixed-points disappear.}
    \label{fig:beta1}
\end{figure}
The beta function for $\lambda_1$ is shown in Fig. \ref{fig:beta1} for various values of $\epsilon/\alpha$. Fig. \ref{fig:beta1} illustrates that, as in \cite{Pisarski1982}, we find both the IR and UV fixed-points merge and become complex (i.e., cease to exist) for $\epsilon>\epsilon_C$, where 
\begin{equation}
    \epsilon_C=\frac{9}{2\pi^2} \alpha +O(\alpha^2). \label{Critical-epsilon}
\end{equation}
This implies that, for small $\alpha$, the bifundamental fixed-point does not extend to $d=2$, much like the vector model.

If we possessed higher-order corrections in $\alpha$ to the beta functions, we would be able to determine higher-order corrections to the UV and IR fixed-points above, and the corresponding corrections to Eq. \eqref{Critical-epsilon}. For $\alpha \sim O(1)$, we therefore expect the bifundamental fixed-point to remain real up to some unknown function of $\alpha$, $\epsilon_C(\alpha) \sim O(\alpha)$. 

\begin{figure}
    \centering
    \includegraphics[scale=.5]{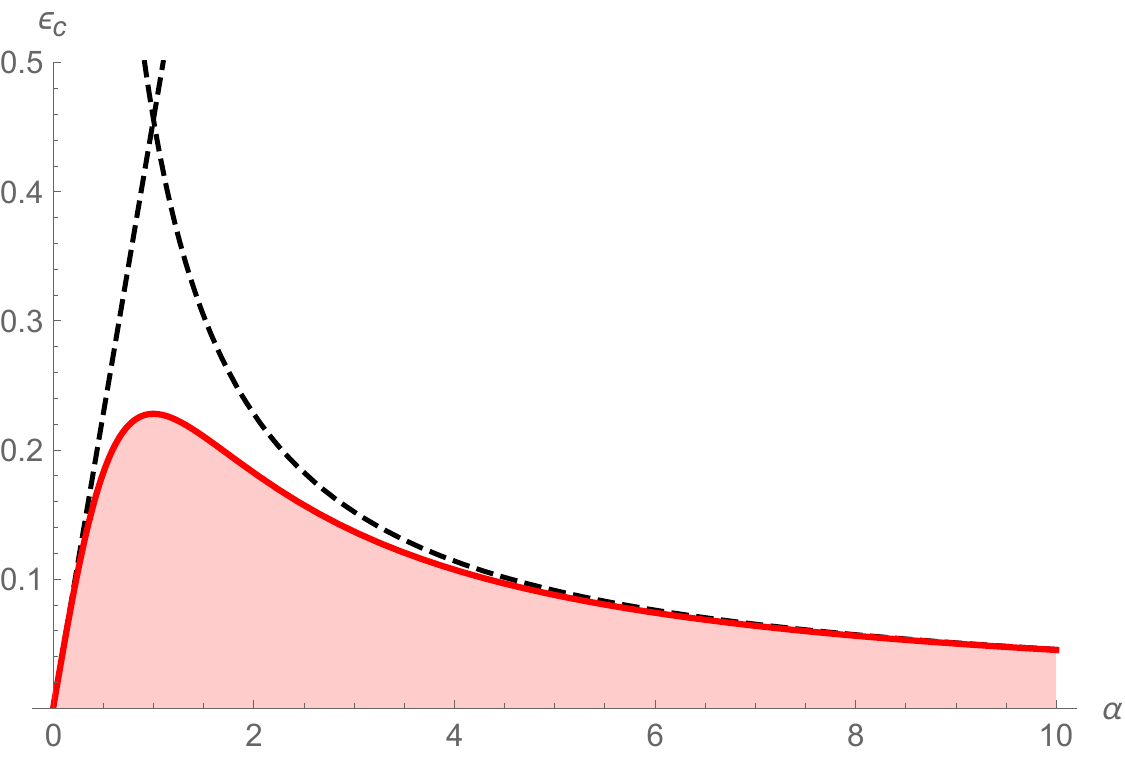}
    \caption{The bifundamental fixed-point exists for $\epsilon<\epsilon_C(\alpha)$, which satisfies $\epsilon_C(\alpha)=\epsilon_C(\alpha^{-1})$. Our results determine $\epsilon_C(\alpha)$ to first order in $\alpha$ and $1/\alpha$, shown as dashed black lines. With an $\alpha^2$ calculation, one could determine a Pad\'{e} approximation that interpolates between both of these curves, schematically shown in red.}
    \label{fig:critical-epsilon}
\end{figure}

If we use Eq. \eqref{Critical-epsilon} (which is only valid to first order in $\alpha$) to obtain a crude estimate for the range of $\epsilon$ for which the fixed-point is real, we find that $\epsilon_C \approx 0.45$, when $\alpha=1$. However, because $\epsilon_C(\alpha)$ must be symmetric in $\alpha \leftrightarrow \alpha^{-1}$, we can attempt to improve this estimate using a two-sided Pad\'{e} approximation of the form $\epsilon_C = \frac{A_0 + A_2 \alpha +A_0 \alpha^2}{B_0 + B_1 \alpha +B_0 \alpha^2}$. Substituting the order $\alpha$ and order $\alpha^2$ results, we find:
\begin{equation}
    \epsilon_C(\alpha) \approx \frac{9 \alpha }{2 \pi ^2 \left(1+B_1\alpha + \alpha ^2\right)}.
\end{equation} To determine $B_1$ we would need to know $\epsilon_C(\alpha)$ to order $\alpha^2$. This requires, in principle, an eight-loop calculation. If we simply set $B_1=0$, we find $\epsilon_C \approx .2$, as shown in Fig. \ref{fig:critical-epsilon}. This seems to suggest that the fixed-point may not extend to $d=2$, even for $\alpha=1$. 

Let us now discuss anomalous dimensions at this fixed-point. At the leading order in $\alpha$ and four-loop level, the anomalous dimensions are:
\begin{equation}
    \begin{split}
       & \gamma_{\phi}=\frac{\alpha ^2 {\tlambda_1}^2}{10800}, \quad \gamma_{\phi^2}=\frac{2 \alpha ^2 {\tlambda_1}^2}{675}, \\
     &\gamma_{\nu_1}=\frac{2 \alpha  \tlambda_1}{15}-\frac{3 \pi ^2 \alpha  \tlambda_ 1^2}{5400}, \text{ and} \quad \gamma_{\nu_2}=\frac{4 \alpha ^2 \tlambda_1^2}{675}.
    \end{split}
\end{equation}
Diagrammatic arguments, similar to those given earlier for the corrections to the vertex, imply that higher-loop corrections to these anomalous dimensions are suppressed by additional powers of $\alpha$ \cite{Pisarski1982}. Hence these expressions are valid to all orders in the 't Hooft couplings, but to leading nontrivial order in $\alpha$. 

The anomalous dimensions for $\phi$ and $\phi^2$ at the IR fixed-point are given by
\begin{eqnarray}
    \gamma_{\phi} &= & \frac{\alpha ^2 (\tlambda_1^{IR})^2}{10800} \approx \frac{\alpha ^2 \tilde{\epsilon} ^2}{27}+\frac{1}{243} \pi ^2 \alpha ^2 \tilde{\epsilon} ^3+O\left(\tilde{\epsilon} ^4\right)+O(\alpha^3), \\
    \gamma_{\phi^2} &= &32 \gamma_{\phi} 
\end{eqnarray}
The quartic single-trace and double-trace operators decouple, with the  anomalous dimensions 
\begin{equation}
    \gamma_{\nu_1}=\frac{4 \alpha  \left(-3+\pi ^2 \tilde{\epsilon} +\sqrt{9-2 \pi ^2 \tilde{\epsilon} }\right)}{\pi ^2}+O(\alpha^2)\approx \alpha\left(\frac{8   \tilde{\epsilon} }{3}-\frac{2\pi ^2 }{27} \tilde{\epsilon} ^2+O\left(\tilde{\epsilon}\right) ^3\right)+O(\alpha^2),
\end{equation}
and
\begin{equation}
    \gamma_{\nu_2}=2\gamma_{\phi^2}=\frac{4\alpha^2(\tlambda_1^{IR})^2}{675}.
\end{equation}
For the UV fixed-point, the anomalous dimensions are:
\begin{eqnarray}
    \gamma_{\phi} &= & \frac{\alpha ^2 (\tlambda_1^{UV})^2}{10800} \approx \frac{12 \alpha ^2}{\pi ^4}-\frac{4 \alpha ^2 \tilde{\epsilon} }{3 \pi ^2}-\frac{\alpha ^2 \tilde{\epsilon} ^2}{27}-\frac{1}{243}
   \left(\pi ^2 \alpha ^2\right) \tilde{\epsilon} ^3+O\left(\tilde{\epsilon} ^4\right)+O(\alpha^3) \label{uvphi}\\
\gamma_{\phi^2} &= & 32 \gamma_{\phi}, \label{uvphi2}\\
\gamma_{\nu_1}&=&\frac{4 \alpha  \left(-3+\pi ^2 \tilde{\epsilon} -\sqrt{9-2 \pi ^2 \tilde{\epsilon} }\right)}{\pi ^2}+O(\alpha^2) \\ & \approx & \alpha\left(-\frac{24}{\pi ^2}+\frac{16 \tilde{\epsilon} }{3}+\frac{2 \pi ^2 \tilde{\epsilon} ^2}{27}+O\left(\tilde{\epsilon}^3\right) \right)+O(\alpha^2), \label{uvphi4}
\end{eqnarray}
and $\gamma_{\nu_2}=2\gamma_{\phi^2}$.

Let us now discuss stability of the fixed-points. The eigenvalues of the stability matrix \eqref{bifundamental-stabilitymatrix}, which is lower triangular, at the IR fixed-point are:
\begin{equation}
    \frac{2 \alpha \sqrt{9-2 \pi ^2 \tilde{\epsilon} } \left(3-\sqrt{9-2 \pi ^2 \tilde{\epsilon} }\right)}{\pi^2},\frac{\alpha\left(3-\sqrt{9-2 \pi ^2 \tilde{\epsilon} }\right) \left(-1+\sqrt{9-2 \pi ^2 \tilde{\epsilon} }\right)}{\pi ^2} \text{, and } -2 \alpha\tilde{\epsilon}, \label{uvphi6}
\end{equation}
corresponding to $\lambda_1$, $\lambda_2$ and $\lambda_3$ respectively.
For $\frac{\epsilon}{\alpha}<\frac{4}{\pi^2}$, the first two of these are positive. Hence, the IR fixed-point has two stable and one unstable directions. When $\frac{\epsilon}{\alpha}>\frac{4}{\pi^2}$, the second eigenvalue becomes negative, and there are two unstable directions. Some of the matrix elements become singular at $\frac{\epsilon}{\alpha}=\frac{4}{\pi^2}$ because $\tlambda_2$ becomes singular, but eigenvalues of the stability matrix remain finite. 

The eigenvalues of the stability matrix \eqref{bifundamental-stabilitymatrix}, at the UV fixed-point are:
\begin{equation}
   -\frac{2\alpha \sqrt{9-2 \pi ^2 \tilde{\epsilon} } \left(3+\sqrt{9-2 \pi ^2 \tilde{\epsilon} }\right)}{\pi
   ^2},-\frac{\alpha\left(3+\sqrt{9-2 \pi ^2 \tilde{\epsilon} }\right)\left(1+\sqrt{9-2 \pi ^2 \tilde{\epsilon} }\right) }{\pi ^2}\text{, and }-2 \alpha \tilde{\epsilon}.
\end{equation}
All eigenvalues are negative when $\epsilon>0$, which seems to suggest that the theory is asymptotically safe in this limit. However, in exactly three dimensions, the triple-trace coupling is marginal -- higher-order corrections would be needed to determine whether the operator is stable, unstable, or if the fixed-point ceases to exist, as discussed in subsection \ref{d=3smallalpha}. 

Let us now discuss flows between these fixed-points. In Fig. \ref{fig:beta1}, there is an apparent flow from the ``UV'' fixed-point to the IR fixed-point along the critical curve defined by \eqref{critical-curve},  which is stable with respect to deformations in $\lambda_1$ if $\epsilon$ is nonzero. For $\frac{9}{2\pi^2}>\tilde{\epsilon}>\frac{4}{\pi^2}$ the critical curve consists of a single connected component, and the flow from the UV fixed-point to the IR fixed-point implied by Fig. \ref{fig:beta1} does exist. However, for $\tilde{\epsilon}<\frac{4}{\pi^2}$ the critical curve actually consists of three disconnected components, with the two fixed-points on different components, and the ``apparent flow'' along the critical curve implied by Fig. \ref{fig:beta1}, does not actually exist. In this range of $\epsilon$, which includes $d=3$, there do, however, exist more complicated flows between the UV and IR fixed-point, which exit the critical curve of Eq. \eqref{critical-curve}. 
\begin{figure}
    \centering
    \includegraphics[width=.46\textwidth]{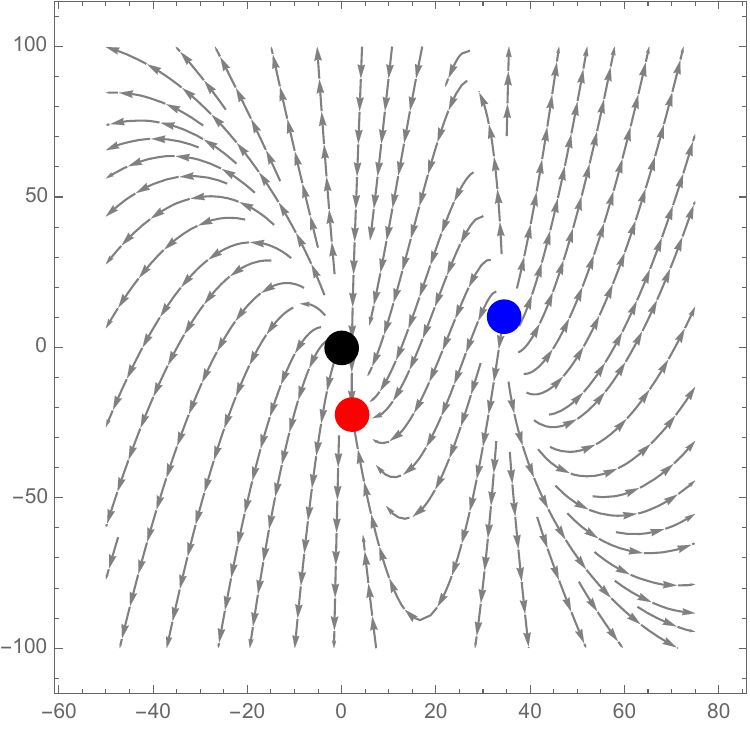}
    \caption{Flows in the $\lambda_1$-$\lambda_2$ plane from the UV fixed-point (blue) to the IR fixed-point (red) in $d=3-\frac{\alpha}{\pi^2}$. The free fixed-point (black) also flows to the IR fixed-point.  $\lambda_3$ is tuned so that $\beta_{\lambda_3}$ vanishes along these flows.}
    \label{fig:streams}
\end{figure}

\begin{figure}
    \centering
    \includegraphics[width=.5\textwidth]{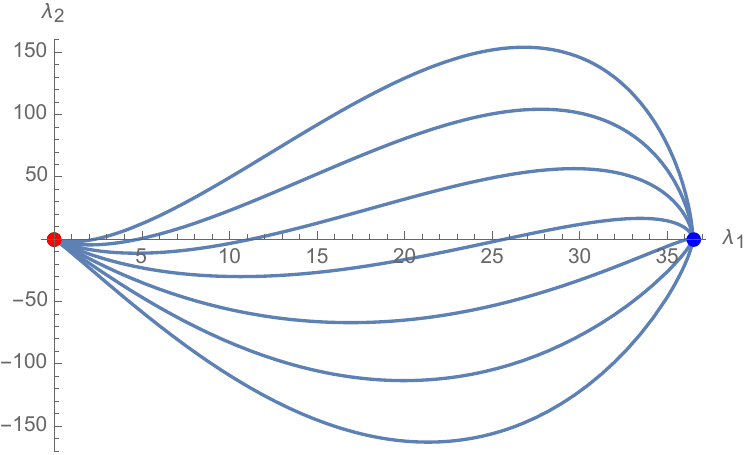}
    \caption{Flows from the UV fixed-point to the IR fixed-point in $d=3$.}
    \label{fig:estream}
\end{figure}

It is actually possible to solve for flows in the $\lambda_1$-$\lambda_2$ plane exactly, even when $\tilde{\epsilon}\neq 0$. We have:
\begin{equation}
\begin{split}
     \lambda_2(\lambda_1) & = \frac{\lambda_1 \left(\pi ^2 (\lambda_1+30 \tilde{\epsilon} )-360\right)}{10 \left(4-\pi ^2 \tilde{\epsilon}\right)} + C \lambda_1 \sqrt{\lambda_1 \left(360-\pi ^2 \lambda_1\right)-7200 \tilde{\epsilon} } \exp \left(\frac{\tan ^{-1}\left(\frac{\pi ^2 \lambda_1-180}{60 \sqrt{2 \pi ^2 \tilde{\epsilon} -9}}\right)}{\sqrt{2 \pi ^2 \tilde{\epsilon} -9}}\right) 
    \end{split}.
\end{equation}
When $\epsilon=0$, this simplifies to
\begin{equation}
    \lambda_2(\lambda_1) =  -9 \lambda_1 +\frac{\pi ^2 \lambda_1^2}{40} - \frac{C}{\pi^{1/3}} \lambda_1^{4/3} \left(360-\pi ^2 \lambda_1\right)^{2/3}. \label{exact-flows}
\end{equation}
Here, $C$ an integration constant. For flows between the UV fixed-point and the IR fixed-point, $C$ should be taken to be real, and the branch cut in the cube root in Eq. \eqref{exact-flows} chosen appropriately. We assume $\lambda_3$ is tuned to solve $\beta_{\lambda_3}=0$. Some flows between the UV and IR fixed-points for small $\tilde{\epsilon}$ and for $d=3$ are presented in Fig. \ref{fig:streams} and \ref{fig:estream}.

In the small $\alpha$ limit, flows from the UV fixed-point to the IR fixed-point may be affected by the spontaneous breaking of conformal invariance, as discussed for the sextic vector model in \cite{Bardeen:1983rv, Amit:1984ri}. Investigation of this effect is left for future work.

\subsubsection{Ultraviolet fixed-point in $d=3$}
\label{d=3smallalpha}
Here we discuss the fixed-points to the beta functions \eqref{smallalpha1}-\eqref{smallalpha3} when $d=3$ exactly.

When $d=3$, solving $\beta_{\lambda_1}=0$, we have 
\begin{equation}
    \lambda_1^{UV}=\frac{360}{\pi^2}. \label{type1}
\end{equation}
Then $\beta_{\lambda_2}=0$ implies
\begin{equation}
    \lambda_2^{UV}=0.
\end{equation}
This then implies that $\beta_{\lambda_3}$ vanishes identically for all $\lambda_3$. Because $\lambda_3$ is marginal, we cannot be certain of the existence of such a fixed-point until higher-order corrections in $\alpha$ or $1/N$ are included in the beta function. Following the discussion in \cite{Pisarski1982}, an order $\alpha^2$ calculation would require an 8-loop calculation. Notice that, when $\epsilon>0$ it is possible to determine $\lambda_3^{UV}$ and, using this result, we obtain a finite value for $\lambda_3^{UV}=-\frac{1080}{\pi^2}$ in the limit $\epsilon \to 0$; however, as far as we can tell, this result is not reliable in $d=3$ and would be modified when $\alpha^2$ or $1/N^2$ corrections to the beta function are included. 

To leading order in the $1/N$ expansion, anomalous dimensions do not depend on $\lambda_3$, so they can be determined for this putative fixed-point. Their values when $d=3$ can be read off from \eqref{uvphi}, \eqref{uvphi2}, \eqref{uvphi4} and \eqref{uvphi6}. In the large-$N$ limit, we know that $\Delta_{\lambda_3}=3 \Delta_{\phi^2}$. Because $\Delta_{\phi^2}>0$ we expect that in $d=3$, deformations to the UV fixed-point corresponding to $\lambda_3$ become irrelevant when higher-order corrections in $\alpha$ are taken into account. 

Another nontrivial solution in exactly $d=3$ can be obtained by
\begin{equation}
    \lambda_1 = 0, ~ \lambda_2 = \frac{1080}{\pi^2}, \label{type2}
\end{equation}
with $\lambda_3$ again marginal and undetermined.

Note also that, we can also obtain a third class of potentially nontrivial solutions by setting 
\begin{equation}
    \lambda_1=\lambda_2=0, \label{type3}
\end{equation} which leaves $\lambda_3$ marginal and undetermined to this order in $\alpha$. 

In the next section we study fixed-points of the theory when $N$ is large and $M$ is finite.  We are able to determine several fixed-points without marginal directions that approach one of the possible forms above when $M \to \infty$. In particular, the $d=3$ fixed-point $[A^+]$ defined in the next subsection approaches a fixed-point of the form given in Eq. \eqref{type3}, and the $d=3$ fixed-points $[D^+]$, $[D^-]$, $[B^+]$. $[E^+]$ and $[E^-]$ approach fixed-points of the form given in Eq. \eqref{type1}.   The solution $[E^+]$ appears to correspond to the ultraviolet fixed-point at finite $\epsilon$, but it develops a discontinuity in the solution for $\lambda_3$ near $\epsilon=0$ as $M \to \infty$. We did not find any solutions of the form given in Eq. \eqref{type2}. These results correspond to the inclusion of a particular combination of $\alpha$ and $1/N$ corrections to the beta functions in equations \eqref{smallalpha1}-\eqref{smallalpha3}, such that the four-loop beta function remains valid.

The finite $M$ results indicate that an ultraviolet bifundamental fixed-point in $d=3$ exists, but it would be nice to confirm this by a higher-order calculation in $\alpha$. Assuming the fixed-point does exist, we can then speculate about the possibility of it extending to finite $\alpha$. There would be no in-principle obstacle to computing arbitrarily high order corrections in $\alpha$, to the fixed-point. The power series in $\alpha$ obtained in this way approximates a function that interpolates between two large-$N$ saddle points -- hence it is rather different from the $1/N$ expansion, and one might hope that it has a finite radius of convergence.  Because the theory has a symmetry in $\alpha \leftrightarrow \alpha^{-1}$, we could use a few terms of a power series in $\alpha$ to construct a two-sided Pad\'{e} approximation as in Fig. \ref{fig:critical-epsilon}, which could be used to estimate the radius of convergence, or to estimate various observables in the theory near $\alpha=1$.

\section{Fixed points for $N$ large and $M$ finite}
\label{sec:vector-model}
In this section, we describe fixed-points of the $O(N)\times O(M)/\mathbb Z_2$ theory in the limit where $N$ is large and $M$ is finite. This large-$N$ limit is a conventional vector model large-$N$ limit, as the three interactions $\mathcal O_1$, $\mathcal O_2$ and $\mathcal O_3$ are effectively triple-trace interactions with respect to $O(N)$ index contractions. Our results and analysis are therefore an extension classic results for the tricritical vector model presented in \cite{Pisarski1982}, which is a special case of our model corresponding to $M=1$.  

The natural 't Hooft couplings in this limit remain those given in the previous section. To zeroth order in the $1/N$ expansion the beta function vanishes, as in \cite{Pisarski1982}, essentially due to the fact that our interactions are all effectively triple-trace in this limit. Keeping terms up to first order $1/N$, the beta functions in this limit are: 
\begin{eqnarray}
\beta_{\lambda_1} & = & -2 \lambda _1 \epsilon+ \frac{96 \lambda _2 \lambda _1+8 \lambda _2^2+9 \lambda _1^2 M (M+4)}{90 M N} \nonumber -\frac{\pi ^2 \lambda _1 \left(48 \lambda _2 \lambda _1+8 \lambda _2^2+3 \lambda _1^2 M (M+4)\right)}{10800 M N}  \nonumber\\
\beta_{\lambda_2} & = &-2 \lambda _2 \epsilon+ \frac{8 \lambda _2 \left(7 \lambda _2+6 \lambda _3\right)+27 \lambda _1^2 M^2+12 \lambda _1 \left(12 \lambda _3+\lambda _2 M (M+2)\right)}{90 M N} \nonumber \\ && -\frac{\pi ^2 \left(24 \lambda_2 \lambda _1 \left(7 \lambda _2+6 \lambda _3\right) +8 \lambda _2^2 \left(4 \lambda _2+3 \lambda _3\right)+27 \lambda _1^3 M^2+18 \lambda _1^2 \left(12 \lambda _3+\lambda _2 M (M+2)\right)\right)}{32400 M N} \nonumber\\
\beta_{\lambda_3} & = & -2 \lambda _3 \epsilon + \frac{36 \lambda _3^2+48 \lambda _2 \lambda _3+6 \lambda _1 \lambda _2 M^2+2 \lambda _2^2 (M+1) M}{45 M N} \nonumber \\ && -\frac{\pi ^2}{48600 M N} \Big(27 \lambda _2 \lambda _1^2 M^2+2 \left(144 \lambda _3 \lambda _2^2+162 \lambda _3^2 \lambda _2+54 \lambda _3^3+\lambda _2^3 \left(M^2+M+28\right)\right) \nonumber \\ && +18 \lambda _1 \left(24 \lambda _3 \lambda _2+18 \lambda _3^2+\lambda _2^2 M (M+1)\right)\Big) \label{finiteMbetafunction}
\end{eqnarray}
Again, as argued in \cite{Pisarski1982} and reviewed in the introduction, this four-loop beta function is actually the full beta function to order $1/N$. All higher loop corrections to this beta function are suppressed by $1/N^2$. Therefore, we can use this beta function to determine interacting fixed-points in $d=3$ as well as fixed-points that are valid to all orders in the parameter $\hat{\epsilon}=N\epsilon$.

These beta-functions can be written as a gradient of a potential $\beta_a=T_{ab}\frac{\partial U}{\partial{\lambda_b}}$, given in Eq. \eqref{potentialVector} in Appendix \ref{potential}, with the inverse metric:
\begin{equation}
    (T^{-1})_{ab}=\left(\begin{array}{ccc}\frac{1}{120} \left(M^3+3 M^2+4 M\right) & \frac{M+1}{30} & \frac{1}{15 M} \\\frac{M+1}{30} & \frac{M^2+M+4}{90 M} & \frac{1}{15 M} \\\frac{1}{15 M} & \frac{1}{15 M} & \frac{1}{15 M} \\
    \end{array}\right). \label{inverse-metric}
\end{equation}

While we are mainly interested in integer values of $M$, theories with $O(M)$ symmetry for non-integer $M$ may be of interest in some contexts, as described in \cite{Binder:2019zqc}. For $M\leq 2$, this metric is not positive definite. In particular, theories with $O(M)$ symmetry group for non-integer $M$ may not be unitary, and could display unusual RG behaviour, such as limit cycles\cite{Jepsen:2020czw, Jepsen:2021rhs}. Because the beta functions are the gradient of a potential, it appears that limit cycles should be impossible even for non-integer values of $M$. However, the metric in Eq. \eqref{inverse-metric} is only positive definite for $M>2$. when $M=2$ the metric possesses a zero eigenvalue, due to the fact that the three interaction terms $\mathcal O_1$, $\mathcal O_2$ and $\mathcal O_3$ are not all independent. When $-3<M<2$, the metric contains positive and negative eigenvalues, and therefore gives rise to non-unitary flows, as in \cite{Jepsen:2020czw}. When $M<-3$ the metric is again negative definite, so no unusual RG behaviour is possible. We briefly study fixed-points in our model for non-integer $-3<M<2$, in Appendix \ref{spooky-appendix}, with a view to find unconventional fixed-points similar to those in \cite{Jepsen:2020czw, Jepsen:2021rhs} -- however, we find that our model does not give rise to limit cycles.

When solving for the zeros of the beta function, we find fixed-points which become free in the limit $\epsilon \to 0$. Following \cite{Pisarski1982} we refer to these as IR fixed-points. There are also several fixed-point solutions of the beta function  that do not vanish as $\hat{\epsilon} \to 0$. Following \cite{Pisarski1982} we refer to these as ``UV'' fixed-points. These correspond to interacting fixed-points in $d=3$. 

We are able to study the existence of fixed-points for all values of $M \sim O(1)$ and $\tilde{\epsilon} \sim O(1)$, i.e., study fixed-points in the $M$-$\tilde{\epsilon}$ plane. This analysis is reminiscent of the analysis of \cite{Yabunaka:2017uox, Yabunaka:2018mju, Yabunaka:2021fow} which studies fixed-points of the sextic $O(N)$ model in the $N-\epsilon$ plane, via functional renormalization.  
However, the $O(N)$ model is not solvable when $N$ and $\epsilon$ are both finite, any conclusions about the dynamics of the theory in that regime necessarily involve uncontrolled approximations (which may still be self-consistent and physically justified). In contrast, our analysis of fixed-points in the  $M$-$\tilde{\epsilon}$ plane is controlled by the small parameter $1/N$, which can be made arbitrarily small. Interestingly, in our results for the theory when $N$ is large and $M$ is finite, we also find that the $\tilde{\epsilon} \to 0$ limit and the $M\to \infty$ limit do not commute.

At finite $\hat{\epsilon}$, some of these fixed-points merge -- for example, when $M=1$, there are exactly two nontrivial fixed-points which merge at $\hat{\epsilon}_*=\frac{36}{\pi^2}+O(1/N)$. This implies that, the merger occurs at $d_*=3-\frac{36}{\pi^2 N} + O(1/N^2)$, which is extremely close to $d=3$ for $N$ large. To determine the order $1/N^2$ correction to $d_*$ would require an 8-loop calculation.\footnote{Note that interacting fixed-points and mergers were also seen in other theories, e.g., \cite{Roscher:2018ucp}. However, we emphasize that, in our theory, we are able to obtain the complete beta function to leading order in $1/N$. Any changes in the location of mergers due to higher loop contibutions to the beta function must be suppressed by additional powers of $1/N$, and can only contribute to the $1/N^2$ term in $d^*$. This follows from the large-$N$ arguments given in \ref{small-alpha-section}.} Of course, the location of the merger at finite $N$ remains unknown. An interesting feature of the new fixed-points we determine below for $M \geq 2$ not present in the $M=1$ theory, is that some fixed-points persist for all values of $\hat{\epsilon}$ -- these fixed-points could conceivably survive to $d=2$ although more work would be needed to demonstrate this conclusively.  

In subsection \ref{vector-fixed-point} we present two simple fixed-points that are valid for all $M$. These fixed-points are a straightforward generalization of the fixed-points of \cite{Pisarski1982}. In subsection \ref{all-m-solutions-2} we present two additional solutions which are valid for all $M>3$, but we are only able to determine their analytic form near $d=3$.
 
 The remaining zeros of the beta functions in Eq. \eqref{finiteMbetafunction} are rather complicated, and depend on $M$ in a nontrivial way. In addition,  so this case must be treated separately. In subsection \ref{m=2}, we discuss the fixed-points of the theory for $M=2$, which, as mentioned previously, must be treated separately as the three operators $\mathcal O_1$, $\mathcal O_2$ and $\mathcal O_3$ are not independent. In subsection \ref{m=3} we discuss the fixed-points of the theory for $M=3$, and in subsection \ref{m>3} we discuss the fixed-points for $M>3$.

\subsection{Some fixed-points for arbitrary $M$}
While we are forced to resort to numerics to determine the fixed-point solutions of the beta functions \eqref{finiteMbetafunction}, we are able to determine some solutions analytically for arbitrary $M$. These are presented in this section.

\subsubsection{$[A^{+}]$ and $[A^{-}]$}
\label{vector-fixed-point}
 There are two  fixed-points -- a ``UV'' fixed-point and an IR fixed-point -- that are valid for arbitrary $M$ and arbitrary $\hat{\epsilon}$. These are fixed-points for which $\lambda_1=\lambda_2=0$, and $\lambda_3$ is given by one of,
\begin{eqnarray}
~[A^{-}]:\lambda_3^{IR}  &= & \frac{30 \left(6- \sqrt{36-\pi ^2 M \hat{\epsilon} }\right)}{\pi ^2} + O\left(\frac{1}{N^2}\right) \approx \frac{5 M \hat\epsilon }{2}+ \frac{5}{288} \pi ^2 M^2 \hat{\epsilon} ^2+O\left(\hat{\epsilon}^3\right) \label{all-m-solution-IR} \\
~[A^{+}]:\lambda_3^{UV} & = & \frac{30 \left(6+ \sqrt{36-\pi ^2 M \hat{\epsilon} }\right)}{\pi ^2} + O\left(\frac{1}{N}\right) = \frac{360}{\pi ^2}-\lambda_3^{IR}
. \label{all-m-solution-uv}
\end{eqnarray}

For $M=1$, these solutions are those given in \cite{Pisarski1982,Osborn:2017ucf}. When $M=1$, there is only one independent coupling, so $g_1$ and $g_2$ should be set to zero, and these are the only fixed-points, as discussed in \cite{Pisarski1982}. Note that, when $M$ is large the IR solution does not approach the bifundamental fixed-point \eqref{bifundamental-fixed-point};  it corresponds to a new fixed-point that emerges when $1/N$ corrections are included in the large-$N$ bifundamental beta function, given in equation \ref{new-sol-1}. 

\begin{figure}
    \centering
    \includegraphics[scale=.8]{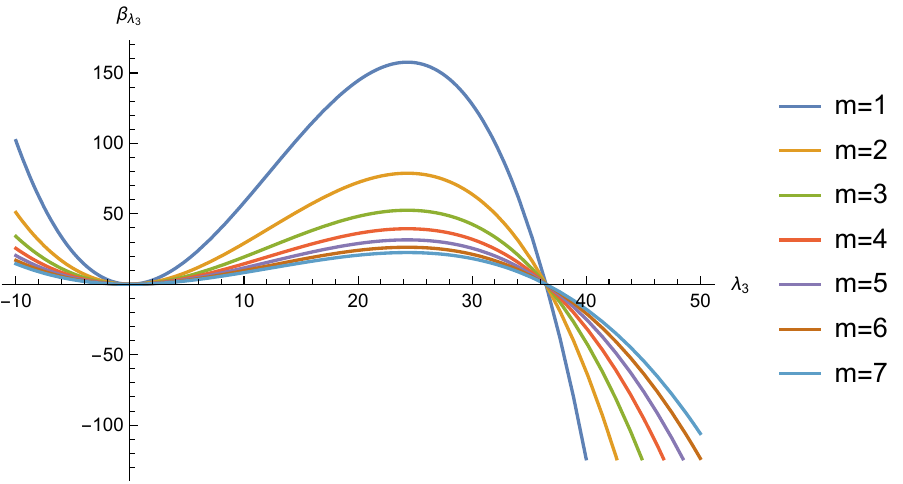}
    \caption{A plot of $\beta_{\lambda_3}$ in $d=3$ obtained from Eq. \eqref{finiteMbetafunction} for various values of $M$, when $\lambda_1=\lambda_2=0$. The zero at $\frac{360}{\pi ^2}$ is a UV-stable fixed-point when $M=1$. For $M>1$, there are two other directions not shown in this plot. }
    \label{fig:betafn3}
\end{figure}

The beta function for $\lambda_3$ in \eqref{finiteMbetafunction} is plotted in Fig. \ref{fig:betafn3}, for various values of $M$, when $\lambda_1=\lambda_2=0$. From Fig. \ref{fig:betafn3}, it is easy to see that there is a flow  from the UV fixed-point to the IR fixed-point for which $\lambda_1=\lambda_2=0$. (This flow, however, could be affected by spontaneous breaking of conformal invariance \cite{Bardeen:1983rv}.) The UV and IR fixed-points cease to exist when $\epsilon>\frac{36}{\pi^2 M N}$.

For the IR fixed-point, the anomalous dimension of $\phi$ is:
\begin{equation}
    \gamma_{\phi}=\frac{(\lambda_3^{IR})^2}{1350 M^2 N^2}\approx \frac{1}{N^2}\left(\frac{\hat{\epsilon} ^2}{216}+\frac{\pi ^2 M \hat{\epsilon} ^3}{15552}\right)+O\left(\hat{\epsilon} ^4\right)+O\left(\frac{1}{N^3}\right).
\end{equation}
For the UV fixed-point,
\begin{equation}
    \gamma_{\phi}=\frac{(\lambda_3^{UV})^2}{1350 M^2 N^2}\approx\frac{1}{N^2}\left(\frac{96}{\pi ^4 M^2}-\frac{4 \hat{\epsilon} }{3 \pi ^2 M}-\frac{\hat{\epsilon} ^2}{216}-\frac{\pi ^2 M \hat{\epsilon} ^3}{15552}\right)+O\left(\hat{\epsilon} ^4\right)+O\left(\frac{1}{N^3}\right)
\end{equation}

The quartic anomalous dimension matrix in this limit is:
\begin{equation}
    [\gamma_{\nu_1 \nu_2}]=\frac{240 \lambda_3-\pi ^2 \lambda_3^2}{450 M N}\left(
\begin{array}{cc}
 0 & 0 \\
 \frac{1}{M} & 1 \\
\end{array}
\right)
\end{equation}
Thus we obtain the following anomalous dimensions:
\begin{equation}
    \gamma_{\hat{\nu}_1}^{IR}=\frac{2 \left(6-\sqrt{36-\pi ^2 M \hat{\epsilon} }\right) \left(2+\sqrt{36-\pi ^2 M \hat{\epsilon} }\right)}{\pi ^2 M N}\approx \frac{1}{N}\left(\frac{4 \hat{\epsilon} }{3}-\frac{1}{216} \pi ^2 M \hat{\epsilon }^2\right)+O(\hat{\epsilon }^3) , \quad \gamma_{\hat{\nu}_2}^{IR}=0
\end{equation}
\begin{equation}
    \gamma_{\hat{\nu}_1}^{UV}\approx \frac{1}{N}\left(-\frac{96}{\pi ^2 M}+\frac{8 \hat{\epsilon} }{3}+\frac{1}{216} \pi ^2 M \hat{\epsilon} ^2\right)+O(\hat{\epsilon }^3) , \quad \gamma_{\hat{\nu}_2}^{UV}=0
\end{equation}
For small $\hat{\epsilon}$ the stability matrix $\partial_{\lambda_b} \beta_{a}$ at the IR fixed-point is:
\begin{equation}
    \left(
\begin{array}{ccc}
 -\frac{2 \hat{\epsilon} }{N} & 0 & 0 \\
 \frac{4 \hat{\epsilon} }{N} & -\frac{2 \hat{\epsilon} }{3 N} & 0 \\
 0 & \frac{8 \hat{\epsilon} }{3 N} & \frac{2 \hat{\epsilon} }{N} \\
\end{array}
\right)+O(\hat{\epsilon}^2)
\end{equation}
This is stable in one direction and unstable in two directions for small $\hat{\epsilon}$. For arbitrary $\hat{\epsilon}$, the eigenvalues of the stability matrix are:
\begin{equation}
    -\frac{2 \hat{\epsilon} }{N},-\frac{2 \left(\sqrt{36-\pi ^2 M \hat{\epsilon} }-2\right)
   \left(6-\sqrt{36-\pi ^2 M \hat{\epsilon} }\right)}{\pi ^2 M N} \text{ }\text{ and, } \frac{4 \sqrt{36-\pi ^2 M \hat{\epsilon} } \left(6-\sqrt{36-\pi ^2 M \hat{\epsilon} }\right)}{\pi ^2 M N}.
\end{equation}
We observe that at  $\hat{\epsilon}=\frac{32}{\pi^2 M}$, the second eigenvalue changes sign, so the IR fixed-point becomes stable in two directions and unstable in one direction.

For small $\hat{\epsilon}$, the stability matrix at the UV fixed-point is:
\begin{equation}
    \left(
\begin{array}{ccc}
 -\frac{2 \hat{\epsilon} }{N} & 0 & 0 \\
 \frac{576}{\pi ^2 M N}-\frac{4 \hat{\epsilon} }{N} & \frac{192}{\pi ^2 M N}-\frac{10 \hat{\epsilon} }{3 N} & 0 \\
 -\frac{864}{\pi ^2 M N}+\frac{12 \hat{\epsilon} }{N} & -\frac{480}{\pi ^2 M N}+\frac{28 \hat{\epsilon} }{3 N} & -\frac{288}{\pi ^2 M N}+\frac{6 \hat{\epsilon} }{N} \\
\end{array}
\right)+O(\hat{\epsilon}^2)
\end{equation}
Thus, the UV fixed-point is unstable in two  directions and stable in one direction. (When $\hat{\epsilon}=0$ one of the unstable directions becomes marginal direction, and higher-order corrections in $1/N$ will be required to determine stability.) For arbitrary $\hat{\epsilon}$, the eigenvalues of the stability matrix are:
\begin{equation}
    -\frac{2 \hat{\epsilon} }{N},\frac{2 \left(2+\sqrt{36-\pi ^2 M \hat{\epsilon} }\right)
   \left(6+\sqrt{36-\pi ^2 M \hat{\epsilon} }\right)}{\pi ^2 M N}\text{, and } -\frac{4 \sqrt{36-\pi ^2 M \hat{\epsilon} } \left(6+\sqrt{36-\pi ^2 M \hat{\epsilon} }\right)}{\pi ^2 M N}.
\end{equation}
These do not change sign as $\hat{\epsilon}$ is varied from $0$ to $\frac{36}{\pi^2 M}$.

\subsubsection{$[A_2^+]$ and $[B^+] $} \label{all-m-solutions-2}
Here we present two UV fixed-points, denoted $[A_2^+]$ and $[B^+] $, which exist for all $M \geq 3$. We present these solutions as power series in $\hat{\epsilon}$.

The fixed-point $[A_2^+]$ is given by
\begin{eqnarray}
   ~[A_2^+]:~ {\lambda} _1 &\to& \frac{20 M \hat{\epsilon} }{M^2+4 M-24}+O\left(\hat{\epsilon}^2\right)\\
   {\lambda}_2&\to&-\frac{60 M \hat{\epsilon} }{M^2+4 M-24}+O\left(\hat{\epsilon}^2\right)\\
   {\lambda}_3&\to&\frac{360}{\pi ^2}-\frac{5 M \left(M^2+4 M-40\right) \hat{\epsilon} }{2 \left(M^2+4 M-24\right)}+O\left(\hat{\epsilon}^2\right),
\end{eqnarray}
The field anomalous dimension for $[A_2^+]$ is:
\begin{equation}
    \gamma_{\phi}=\frac{1}{N^2}\left({\frac{96}{\pi ^4 M^2}-\frac{4 \hat{\epsilon} }{3 \pi ^2 M}}+\frac{\left(7 M^4+16 M^3-32 M^2+96 M-448\right) \hat{\epsilon} ^2}{216 \left(M^2+4 M-24\right)^2}\right)+O\left(\hat{\epsilon}^3\right)
\end{equation}
The quartic anomalous dimension matrix is:
\begin{equation}
    \frac{1}{N}\left(
\begin{array}{cc}
 \frac{8 \left(M^2+2 M-8\right) \hat{\epsilon} }{3 \left(M^2+4 M-24\right) } & 0 \\
 {-\frac{96}{\pi ^2 M^2}+\frac{16 (M-8) \hat{\epsilon} }{3 M \left(M^2+4 M-24\right)}} & {-\frac{96}{\pi ^2 M}+\frac{8 \hat{\epsilon}
   }{3}}\\
\end{array}
\right)+O(\hat{\epsilon}^2)
\end{equation}
with the eigenvalues:
\begin{equation}
    \gamma_{\nu_1}=\frac{8 \left(M^2+2 M-8\right) \hat{\epsilon} }{3 N \left(M^2+4 M-24\right) }+O(\hat{\epsilon}^2), \quad \gamma_{\nu_2}=-\frac{96}{\pi ^2 M N}+\frac{8 \hat{\epsilon}
   }{3 N}+O(\hat{\epsilon}^2)
\end{equation}
The stability matrix is:
\begin{equation}
    \frac{1}{N}\left(
\begin{array}{ccc}
 \frac{2 \left(M^2+4 M-8\right) \hat{\epsilon} }{M^2+4 M-24} & \frac{32 \hat{\epsilon}}{3 \left(M^2+4 M-24\right)} & 0 \\
 \frac{576}{\pi ^2 M}-\frac{32 (M-5) \hat{\epsilon} }{M^2+4 M-24} & \frac{192}{\pi ^2 M}+\left(\frac{4 \left(M^2-16\right)}{3 \left(M^2+4
   M-24\right)}-2\right) \hat{\epsilon}  & 0 \\
 -\frac{864}{\pi ^2 M}+\frac{4 \left(M^2+12 M-72\right) \hat{\epsilon} }{M^2+4 M-24} & -\frac{480}{\pi ^2 M}+\frac{4 \left(5 M^2+24 M-136\right)
  \hat{\epsilon} }{3 \left(M^2+4 M-24\right)} & -\frac{288}{\pi ^2 M}+6 \hat{\epsilon}  \\
\end{array}
\right)+O(\hat{\epsilon}^2)
\end{equation}
with the eigenvalues:
\begin{equation}
    \frac{2 \hat{\epsilon} }{N},\frac{192}{\pi ^2 M N}-\frac{2 \left(M^2+12 M-88\right) \hat{\epsilon} }{3 \left(M^2+4 M-24\right) N},-\frac{288}{\pi ^2 M N}+\frac{6 \hat{\epsilon} }{N}
\end{equation}
For small $\hat{\epsilon}$, we see $[A_2^+]$ is unstable in one direction and stable in the other two directions. 

The fixed-point $[B^+]$ is given by
\begin{eqnarray}
   ~[B^+]:~ {\lambda} _1 &\to& \frac{360}{\pi ^2}-\frac{20 \hat{\epsilon} }{M+4}+\frac{40\pi^2 \left(5  M+2 \right) \hat{\epsilon} ^2}{9 (M+4)^3}+O\left(\hat{\epsilon}^3\right)\\
   {\lambda}_2&\to&-\frac{5\pi ^2 \left(7  M^2+4  M\right) \hat{\epsilon} ^2}{8 (M+4)^2}+O\left(\hat{\epsilon}^3\right)\\
    {\lambda}_3&\to&\frac{15 M^2 \hat{\epsilon} }{2 (M+4)}+\frac{5 \pi ^2 \left(7 M^5+46 M^4-64 M^3+32 M^2\right) \hat{\epsilon} ^2}{96 (M+4)^3}+O\left(\hat{\epsilon}^3\right),
\end{eqnarray}
The field anomalous dimension is:
\begin{equation}
    \gamma_{\phi}=\frac{1}{N^2}\left(\frac{12 \left(M^2+3 M+4\right)}{\pi
   ^4}-\frac{4 \left(M^2+3 M+1\right) \hat{\epsilon} }{3 \pi ^2 (M+4)}+\frac{\left(127 M^3+52 M^2-368 M+64\right) \hat{\epsilon} ^2}{216 (M+4)^3}\right)+O\left(\hat{\epsilon}^3\right)
\end{equation}
The quartic anomalous dimension matrix is:
\begin{equation}
    \frac{1}{N}\left(
\begin{array}{cc}
 {-\frac{24 (M+2)}{\pi ^2}+\frac{16 (M+2) \hat{\epsilon} }{3 (M+4)}} & {-\frac{96}{\pi ^2}+\frac{64 \hat{\epsilon} }{3
   (M+4)}} \\
 {-\frac{24}{\pi ^2}-\frac{8 \hat{\epsilon} }{3 (M+4)}} & -\frac{8 M \hat{\epsilon} }{(M+4) } \\
\end{array}
\right)+O\left(\hat{\epsilon}^2\right)
\end{equation}
with the eigenvalues:
\begin{equation}
    -\frac{12 \left(M+2+\sqrt{ M^2+4 M+20}\right)}{ \pi ^2 N}+O\left(\hat{\epsilon}\right),\frac{12 \left(\sqrt{ M^2+4 M+20}-(M+2)\right)}{\pi ^2 N}+O\left(\hat{\epsilon}\right)
\end{equation}
The stability matrix is:
\begin{equation}
    \frac{1}{N}\left(
\begin{array}{ccc}
 -\frac{36 (M+4)}{\pi ^2}+6 \hat{\epsilon} &-\frac{192}{\pi ^2 M}+ \frac{128\hat{\epsilon}}{3 M (M+4)} & 0 \\
 -\frac{108 M}{\pi ^2} &  -\frac{24 (M+2)}{\pi ^2}-\left(\frac{8 (M-4)}{3 (M+4)}+2\right) \hat{\epsilon} & -\frac{288}{\pi ^2 M}+\frac{64
   \hat{\epsilon} }{M (M+4)} \\
 0 & -\frac{24 M}{\pi ^2}-\frac{32 M \hat{\epsilon} }{3 (M+4)} & \left(-\frac{24 M}{M+4}-2\right) \hat{\epsilon}  \\
\end{array}
\right)+O\left(\hat{\epsilon}^2\right)
\end{equation}
For small $\hat{\epsilon}$, one can show that $[B^+]$ is unstable in two directions and stable in the other direction. 

We had to resort to numerics to determine the behaviour of these solutions when $\hat{\epsilon}$ is finite, as discussed below for various values of $M$. 

\subsection{$M=2$}
\label{m=2}
In this section, we present all the fixed-points of the theory when $M=2$. We must study this case separately because, when $M=2$, the single-trace, double-trace and triple trace operators are not independent, and obey the relation
\begin{equation}
    3(\phi_{a_1b_1}\phi_{a_1b_1})(\phi_{a_2b_2}\phi_{a_2b_3}\phi_{a_3b_3}\phi_{a_3b_2})=2(\phi_{a_1b_1}\phi_{a_1b_2}\phi_{a_2b_2}\phi_{a_2b_3}\phi_{a_3b_3}\phi_{a_3b_1}) + (\phi_{ab}\phi_{ab})^3. 
\end{equation}

We choose the following two independent 't Hooft couplings:
\begin{equation}
    \tilde{\lambda}_1=\lambda_1+\frac{\lambda_2}{3}, \quad \tilde{\lambda}_3=\lambda_3+\frac{2\lambda_2}{3}
\end{equation}
The beta-functions for these redefined couplings reduce to
\begin{eqnarray}
\beta_{\tilde{\lambda}_1} & = & -2 \tilde{\lambda}_1 \epsilon+\frac{1}{N}\left(\frac{4}{15} \tilde{\lambda}_1 (3 \tilde{\lambda}_1+\tilde{\lambda}_3)-\frac{1}{900}\tilde{\lambda}_1 \left(2 \pi ^2\tilde{\lambda}_1^2+\pi ^2\tilde{\lambda}_1\tilde{\lambda}_3\right)\right)+O\Big(\frac{1}{N^2}\Big) \nonumber \\
\beta_{\tilde{\lambda}_3} & = & -2 \tilde{\lambda}_3 \epsilon + \frac{1}{900 N} \left(-\pi ^2 \tilde{\lambda}_1^3-2 \pi ^2 \tilde{\lambda}_1^2 \tilde{\lambda}_3+360 \tilde{\lambda}_1^2-3 \pi ^2 \tilde{\lambda}_1 \tilde{\lambda}_3^2+480 \tilde{\lambda}_1 \tilde{\lambda}_3-\pi ^2
   \tilde{\lambda}_3^3+360 \tilde{\lambda}_3^2\right) \nonumber\\ && + O\Big(\frac{1}{N^2}\Big) \label{finiteMbetafunction_m=2}
\end{eqnarray}
These beta functions can be written as the gradient of the following potential,
\begin{equation}
    \begin{split}
    U=&-\frac{\epsilon}{30} \left(7 \tilde{\lambda} _1^2+2 \tilde{\lambda} _3 \tilde{\lambda} _1+\tilde{\lambda} _3^2\right) +\frac{15 \tilde{\lambda} _1^3+9 \tilde{\lambda} _3 \tilde{\lambda} _1^2+3 \tilde{\lambda} _3^2 \tilde{\lambda} _1+\tilde{\lambda} _3^3}{225 N} \\ & -\frac{\pi ^2 \left(15
   \tilde{\lambda} _1^4+12 \tilde{\lambda} _3 \tilde{\lambda}_1^3+6 \tilde{\lambda} _3^2 \tilde{\lambda} _1^2+4 \tilde{\lambda} _3^3 \tilde{\lambda} _1+\tilde{\lambda}
   _3^4\right)}{108000 N}.
   \end{split}
\end{equation}
with the inverse metric
\begin{equation}
    (T^{-1})_{ab}=\left(
\begin{array}{cc}
 \frac{7}{30} & \frac{1}{30} \\
 \frac{1}{30} & \frac{1}{30} \\
\end{array}
\right). 
\end{equation}

In terms of the redefined 't Hooft couplings, the anomalous dimension of $\phi$ is given by
\begin{equation}
\label{anomalous_dim_field_m=2}
    \gamma_{\phi}=\frac{7 \tilde{\lambda}_1^2+2 \tilde{\lambda}_1 \tilde{\lambda}_3+\tilde{\lambda}_3^2}{5400 N^2}+O\left(\frac{1}{N^3}\right).
\end{equation}
In the leading order, $\gamma_{\phi^2}=32 \gamma_{\phi}$ for all the fixed-points. The quartic anomalous dimension matrix is
\begin{equation}
\label{anomalous_dim_quartic_m=2}
    [\gamma_{\nu_1\nu_2}]=\frac{1}{1800 N}\left(
\begin{array}{cc}
 {{960 \tilde{\lambda}_1}-4{\pi ^2 \tilde{\lambda}_1^2}} & {{960 \tilde{\lambda}_1}-4{\pi ^2 \tilde{\lambda}_1^2}} \\
 {-\pi ^2 (\tilde{\lambda}_1+\tilde{\lambda}_3)^2+240 (\tilde{\lambda}_1+\tilde{\lambda}_3)} & {-4 \pi ^2 \tilde{\lambda}_1 \tilde{\lambda}_3-2\pi
   ^2 \tilde{\lambda}_3^2+480 \tilde{\lambda}_3} \\
\end{array}
\right)+O\left(\frac{1}{N^2}\right).
\end{equation}

\begin{figure}
    \centering
    \includegraphics[width=0.8\textwidth]{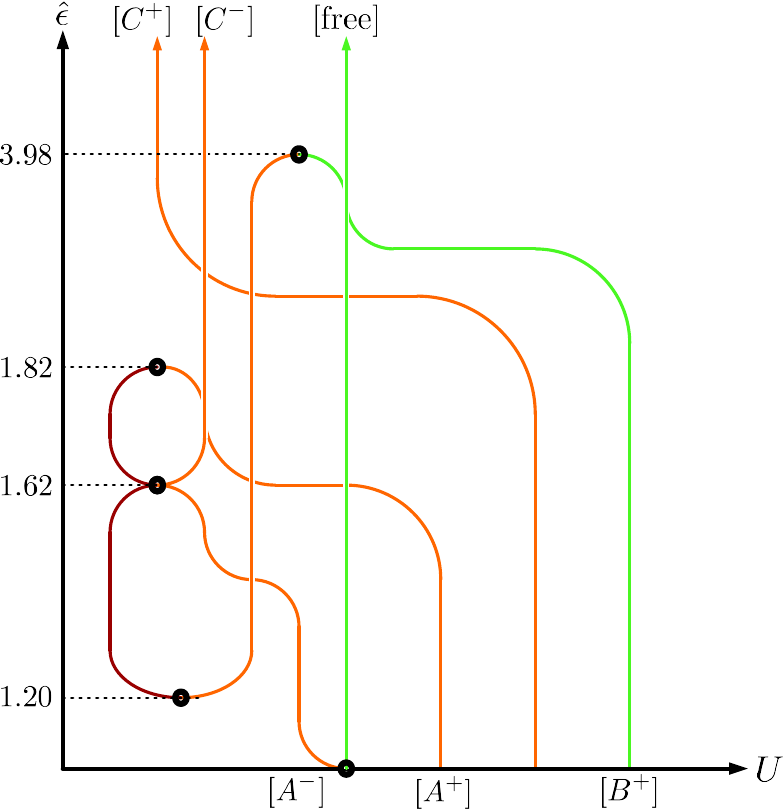}
    \caption{A schematic plot of the number of fixed-points as a function of $\hat{\epsilon}$, for $M=2$. The horizontal axis is schematic -- for any given value of $\hat{\epsilon}$ fixed-points are arranged in order of increasing $U$, so RG flow is only possible from right to left. The colour of each line specifies the stability of each fixed-point: red, orange, and green respectively denote fixed-points with zero, one, and two unstable directions. Black dots indicate mergers of fixed-points.}
    \label{fig:m2epsilon}
\end{figure}

Numerically, we find that the total number of distinct fixed-points $n_{\text{fixed}}$ of the beta function varies with $\hat{\epsilon}$ as follows: 
\begin{equation}
   n_{\text{fixed}} = \begin{cases}
    4& \hat{\epsilon}=0 \\
    5& 0<\hat{\epsilon}<1.20 \\
    7 & 1.20< \hat{\epsilon} < \frac{18}{ \pi ^2} \\
    5 & \frac{18}{\pi ^2} < \hat{\epsilon} < 3.98 \\
    3& 3.98 < \hat{\epsilon}  
    \end{cases}
\end{equation}
A schematic plot illustrating how the fixed-points merge and vary as a function of $\hat{\epsilon}$ is presented in Fig. \ref{fig:m2epsilon}.

In the next subsection we discuss the behaviour of these fixed-points near $d=3$. In subsection \ref{large-epsilon-m=2}, we discuss the fixed-points which persist for $\hat{\epsilon}$ arbitrarily large, as these could conceivably correspond to fixed-points near $d=2$ or $4$. (However, our large-$N$ limit is defined such that $\hat{\epsilon}=N\epsilon \sim O(1)$, so our analysis does not apply to $\epsilon \sim O(1)$.)

\subsubsection{Fixed points near $d=3$}
There are a total of 5 fixed-points $\hat{\epsilon}$ very small but nonzero. Apart from the free fixed-point, two of these are the IR and UV solutions $[A^-]$, and $[A^+]$ given in \eqref{all-m-solution-IR} and \eqref{all-m-solution-uv}. 

The remaining two solutions are UV fixed-points, given by the following expressions near $\hat{\epsilon}=0$:
\begin{eqnarray}
   ~[B_{(2)}]&:~~& \tilde{\lambda} _1 \to \frac{360}{\pi ^2}-\frac{10 \hat{\epsilon} }{3},
    ~\tilde{\lambda}_3\to5
   \hat{\epsilon}\\
   ~[C_{(2)}^+]&:~~& \tilde{\lambda} _1\to 38.8118-1.01761 \hat{\epsilon},~\tilde{\lambda} _3\to -12.511-7.6831\hat{\epsilon}
\end{eqnarray}

\begin{figure}
    \centering
    \includegraphics[scale=.6]{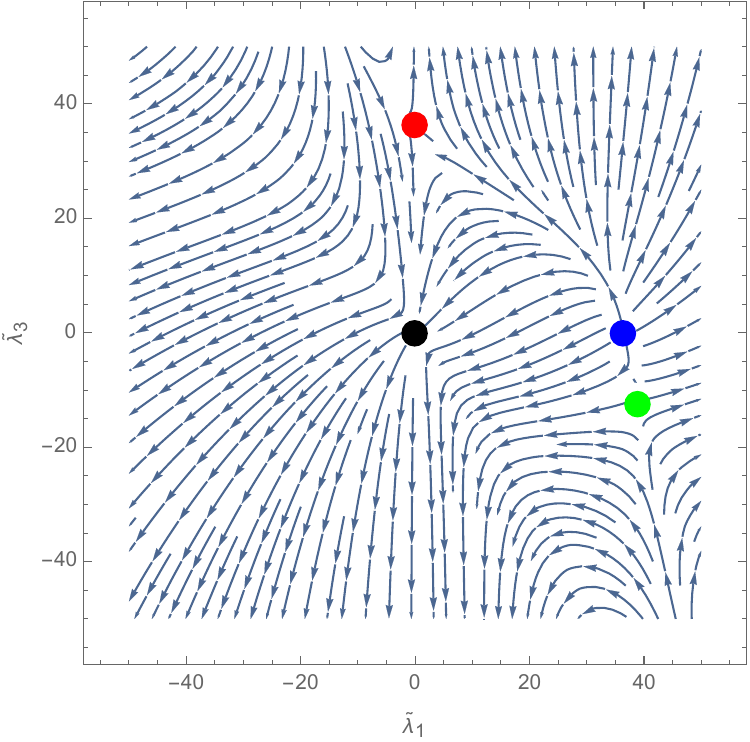}
    \caption{Flows in the $\tilde{\lambda}_1$-$\tilde{\lambda}_3$ plane in $d=3$ for the $M=2$ theory. The black dot is the free fixed-point, the red dot is  $[A^+]$, the blue dot is $[B^+_{(2)}]$ and the green dot is $[C^+_{(2)}]$. \label{flowsm2}}
\end{figure}

When $d=3$ exactly, there are only $4$ fixed-points, because the fixed-point $[A^-]$ becomes free when $\epsilon=0$. The eigenvalues of the stability matrix and anomalous dimensions of these fixed-points are given in Table \ref{m=2table}. We observe that the fixed-point $[B_{(2)}^+]$ is unstable in all directions, and acts as a stable UV fixed-point that may provide an asymptotically safe definition of the theory in this large-$N$ limit. Numerically determined flows in $d=3$ between these fixed-points in the $\tilde{\lambda_1}$-$\tilde{\lambda_3}$ plane are shown in Fig. \ref{flowsm2}. 

We were able to determine a formal analytic expression for the flows in $d=3$, given by
\begin{equation}
    C = \log {\lambda_1}+4 \sum_{x\\x^3-6x^2-x+1=0} \frac{2 x \log \left(\frac{\lambda _1}{\lambda _3}-x\right)+\log \left(\frac{\lambda _1}{\lambda _3}-x\right)}{3 x^2-12 x-1},
\end{equation}
where the sum is over the three roots of the polynomial $x^3-6x^2-x+1=0$, and $C$ is an integration constant.
\begin{table}
$$
\begin{array}{|c|rr|rr|r|r|}
 \hline 
 \text{Fixed Point}& (\partial_\lambda \beta_\lambda)^{(1 )}N &(\partial_\lambda \beta_\lambda)^{(2 )}N  & \gamma_{\phi^4}^{(1)} N & \gamma_{\phi^4}^{(2)}N & \gamma_\phi N^2&U N\\
 \hline
  ~[A^{+}] &-\frac{144}{\pi ^2}&\frac{96}{\pi ^2 } & -\frac{48}{\pi ^2} &0 &\frac{24}{\pi ^4 } & \frac{51840}{\pi^6} \approx 53.92\\
 ~[C_{(2)}^{+}] & -{31.26} & {6.01}  & -{12.53} & {5.79} &
   {1.80} & {804.02}\\
  ~[B^+_{(2)}] & \frac{48\left(-4-\sqrt{7}\right)}{\pi ^2} & \frac{48\left(-4+\sqrt{7}\right)}{\pi ^2}  & \frac{48\left(-1-\sqrt{2}\right)}{\pi ^2 } & \frac{48\left(-1+\sqrt{2}\right)}{\pi ^2} & \frac{168}{\pi ^4} &\frac{777600}{ \pi^6} \approx 808.83\\
    \hline 
\end{array}
$$
\caption{large-$N$ interacting $O(N)\times O(2)$ fixed-points in $d=3$. \label{m=2table}}
\end{table}

\subsubsection{Fixed points at large $\hat{\epsilon}$}
The fixed-point solutions $[B^+_{(2)}]$ and $[C^+_{(2)}]$ are presented as power series in $\hat{\epsilon}$, but we also studied their behaviour for arbitrary $\hat{\epsilon}$ numerically, which is depicted schematically in Fig. \ref{fig:m2epsilon}. The fixed-point, $[B_{(2)}]$ appears to only exist up to $\hat{\epsilon}=3.98$, while $[C_{(2)}^+]$ appears to survive for all $\hat{\epsilon}$. There is another solution $[C_{(2)}^-]$ that emerges at finite $\hat{\epsilon}$, and exists for arbitrarily large $\hat{\epsilon}$.

\label{large-epsilon-m=2}
Here we present the asymptotic form of the fixed-points that survive when $\hat{\epsilon}$ large. (But, recall that we still require $\hat{\epsilon} \ll N$ for our beta functions to be valid). We find two nontrivial fixed-points with the following asymptotic expressions when $\hat{\epsilon} \gg 1$:
\begin{equation}
\begin{array}{rll}
   ~[C_{(2)}^\mp]: &~~ \lambda _1\to \mp 29.79 \sqrt{\hat{\epsilon} } &  \lambda _3\to \pm 65.69
   \sqrt{\hat{\epsilon} } \\
\end{array}
\end{equation}
In the limit of large $\hat{\epsilon}$ both fixed-points have the same anomalous dimensions, which are:
\begin{equation}
    \gamma_{\phi}=\frac{1.22 \hat{\epsilon} }{N^2}, \quad  \gamma_{\phi^2}=\frac{39.19 \hat{\epsilon} }{N^2}
\end{equation}
The quartic anomalous dimension matrix is
\begin{equation}
    [\gamma_{\nu_1\nu_2}]=\left(
\begin{array}{cc}
 -\frac{19.46 \hat{\epsilon} }{N} & -\frac{19.46\hat{\epsilon}}{N} \\
 -\frac{7.07 \hat{\epsilon} }{N} & -\frac{4.41 \hat{\epsilon} }{N} \\
\end{array}
\right)
\end{equation}
The eigenvalues of the quartic anomalous dimension matrix are $-\frac{25.87  \hat{\epsilon} }{N}$ and $\frac{2  \hat{\epsilon} }{N}$.
\begin{equation}
    [M_{ab}]=\left(
\begin{array}{cc}
 -\frac{17.46  \hat{\epsilon}}{N} & -\frac{9.73  \hat{\epsilon}}{N} \\
 -\frac{85.34  \hat{\epsilon}}{N} & -\frac{34.69 \hat{\epsilon} }{N} \\
\end{array}
\right)
\end{equation}
The eigenvalues of the stability matrix are $-\frac{56.15  \hat{\epsilon} }{N}$ and $\frac{4  \hat{\epsilon} }{N}$, so in this limit, the fixed-points are stable in one direction and unstable in the other. 
\begin{figure}
    \centering
    \includegraphics[scale=.6]{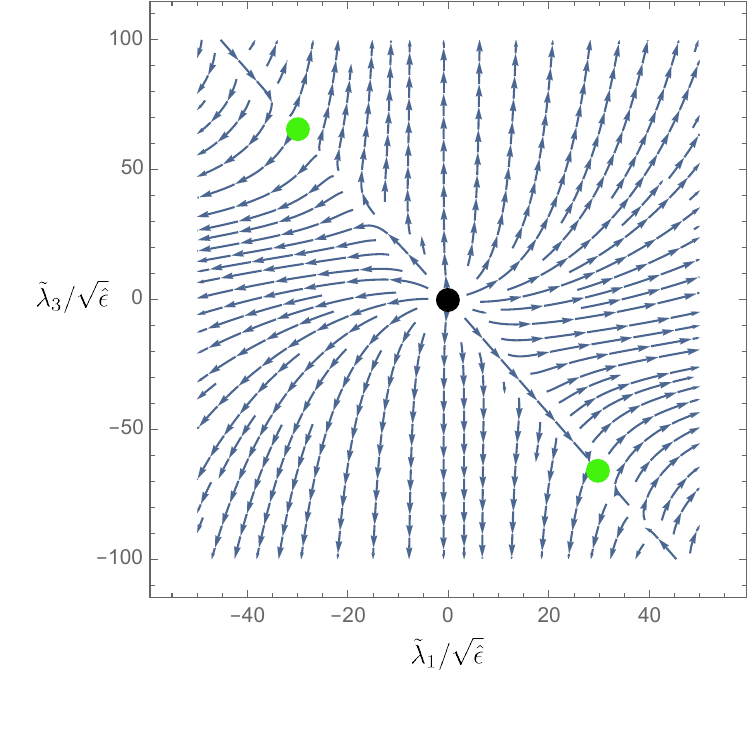}
    \caption{Flows in the $\tilde{\lambda}_1$-$\tilde{\lambda}_3$ plane for asymptotically large $\hat{\epsilon}$. The green dots denote $[C^\pm]$.  \label{largeepsilonflows}}
\end{figure}
The value of the potential at either fixed-points approaches:
\begin{equation}
    U=-\frac{110.21 \hat{\epsilon} ^2}{N}
\end{equation}
when $\hat{\epsilon} \to \infty$. Flows for large $\hat{\epsilon}$ are shown in Fig. \ref{largeepsilonflows}.

\subsection{$M = 3$}
\label{m=3}
We next study the theory when $M = 3$. Although the same beta functions apply for $M=3$ and $M> 3$, the structure of the solutions is slightly different for $M=3$, so we present this case separately. 

\begin{figure}
    \centering
    \includegraphics[width=0.8\textwidth]{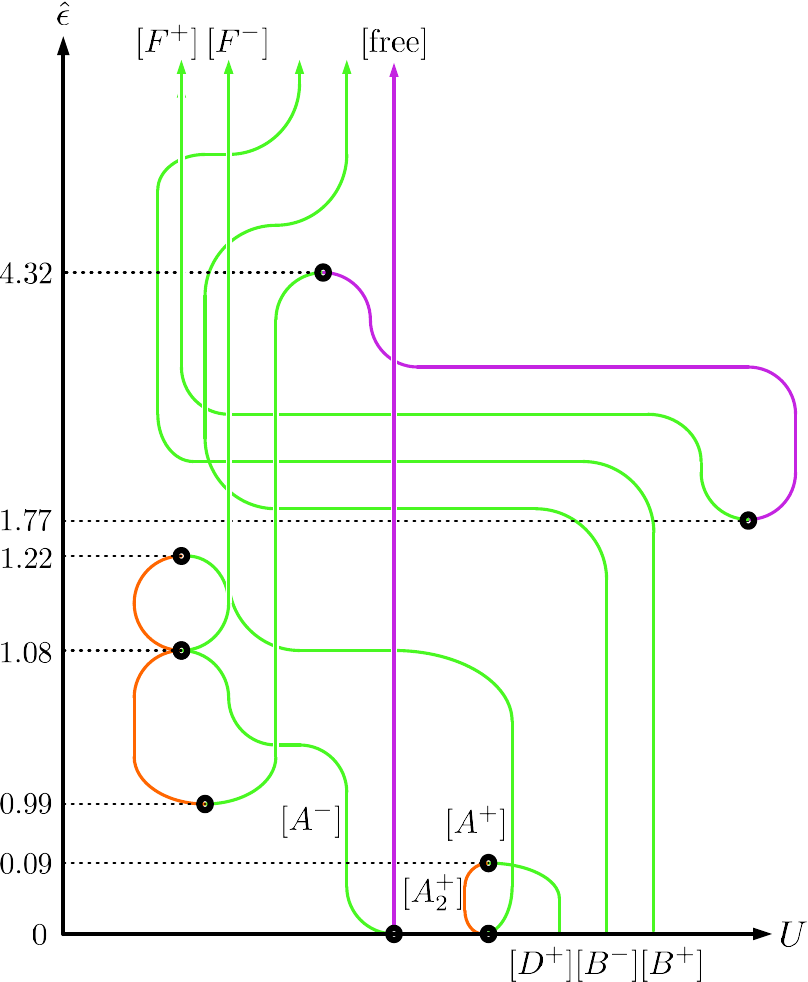}
    \caption{A schematic plot illustrating the number of fixed-point solutions as a function of $\hat{\epsilon}$ for $M=3$. The horizontal axis is schematic -- for any given value of $\hat{\epsilon}$ fixed-points are arranged in order of increasing $U$, so RG flow is only possible from right to left. The colour of each line specifies the stability of each fixed-point: red, orange, green and violet respectively denote fixed-points with zero, one, two and three unstable directions. Black dots indicate mergers of fixed-points. For $\hat{\epsilon}$ sufficiently large, all fixed-points become infrared fixed-points relative to the free theory.}
    \label{fig:m3epsilon}
\end{figure}

Numerically, for $M=3$, we find that the total number of distinct solutions $n_{\text{fixed}}$ to the beta function varies with $\hat{\epsilon}$ as follows: 
\begin{equation}
   n_{\text{fixed}} = \begin{cases}
    5& \hat{\epsilon}=0 \\
    7& 0<\hat{\epsilon}<0.09 \\
    5 & 0.09 < \hat{\epsilon} < 0.99 \\
    7 & 0.99 < \hat{\epsilon} < \frac{12}{\pi^2} \\
    5& \frac{12}{\pi^2} < \hat{\epsilon} <1.77\\
    7& 1.77< \hat{\epsilon} <4.32\\
    5& 4.32< \hat{\epsilon}
    \end{cases}
\end{equation}
A schematic plot of these fixed-points, according by potential, as a function of epsilon is given in Fig. \ref{fig:m3epsilon}.

\subsubsection{Fixed points near $d=3$}
There are a total of 5 fixed-points for $\hat{\epsilon}=0$, including the free fixed-point. Two of the fixed-points can be determined analytically. One of them is $[A^{+}]$ (which branches into two fixed-points, $[A^{+}]$ and  $[A_2^{+}]$, as $\hat{\epsilon}$ is increased.) The other is $[B^+]$ which has a closed-form series expression and exists for all $M$, as described in section \ref{all-m-solutions-2}. 

The remaining two UV solutions, which we determined numerically, are given by
\begin{eqnarray}
~[D^+]&:~~& \lambda _1= -5.34+13.22\hat{\epsilon},~\lambda _2= 12.98-22.70\hat{\epsilon},~\lambda _3= 28.97+0.46\hat{\epsilon}\label{m=3,uv1}\\
~[B^-]&:~~& \lambda _1= -16.78+1.85\hat{\epsilon},~\lambda _2= 103.57-16.33\hat{\epsilon},~\lambda _3= -37.95+9.01\hat{\epsilon}\label{m=3,uv2}
\end{eqnarray}
Fixed points $[D^+]$ and $[A_2^{+}]$ exist only up to $\hat{\epsilon}=0.085$. Fixed point $[D^+]$ is part of a larger family of fixed-points that exist for all $M>2$. $[B^-]$ and $[B^+]$ are real for all $\hat{\epsilon} \geq 0$ when $M=3$. 

The only IR fixed-point when $M=3$ is $[A^{-}]$. 

When $d=3$, we find the anomalous dimensions
\begin{equation}
\begin{array}{|c|rrr|rr|r|r|}
 \hline 
 \text{Fixed Point}& (\partial_\lambda \beta_\lambda)^{(1 )}N &(\partial_\lambda \beta_\lambda)^{(2 )}N & (\partial_\lambda \beta_\lambda)^{(3 )}N & \gamma_{\phi^4}^{(1)} N & \gamma_{\phi^4}^{(2)}N & \gamma_\phi N^2&U N\\
 \hline
   ~[A^{+}] & -9.73 & 6.48 & 0 & -3.24 & 0 & 0.11 & 23.97\\
   ~[D^{+}] & -9.92 & 5.82 & -0.76 & -3.44 & -2.07 &
   0.13 & 24.32\\
   ~[B^-] & -26.92 & 13.51 & -11.74 & -10.31 & -6.42 &
   0.79 & 230.59\\
 ~[B^{+}] & -35.50 & -8.32 & 6.13 & -13.86 & 1.71 & 2.71 & 2042.29\\
 
    \hline 
\end{array}
\end{equation}

\subsubsection{Fixed points at large $\hat{\epsilon}$}
Four fixed-points with $M=3$ survive when $\hat{\epsilon}$ large, as shown in Fig. \ref{fig:m3epsilon}. These are given by the following asymptotic forms:
\begin{eqnarray}
   ~[B^\mp]:~&\lambda _1\to& \mp 27.01 \sqrt{\hat{\epsilon} },\lambda _2\to \pm 81.03 \sqrt{\hat{\epsilon} },   \lambda _3\to \mp 54.02
   \sqrt{\hat{\epsilon} } \label{m=3,largeepsilon1}\\~[F^\pm]:~
    &\lambda _1\to& 10.15, \lambda _2\to \pm 64.36 \sqrt{\hat{\epsilon} },   \lambda _3\to \mp 98.57
   \sqrt{\hat{\epsilon} }
\label{m=3,largeepsilon2}
\end{eqnarray}
The anomalous dimensions and the value of the potential $U$ for these solutions are:
\begin{equation}
\begin{array}{|c|rrr|rr|r|r|}
 \hline 
 \text{Fixed Point}& (\partial_\lambda \beta_\lambda)^{(1 )}N &(\partial_\lambda \beta_\lambda)^{(2 )}N & (\partial_\lambda \beta_\lambda)^{(3 )}N & \gamma_{\phi^4}^{(1)} N & \gamma_{\phi^4}^{(2)}N & \gamma_\phi N^2 & U N\\
 \hline
 ~[F^\pm] & 4\hat{\epsilon} & -12.10\hat{\epsilon} & -33.03\hat{\epsilon} & 2\hat{\epsilon} &-13.92\hat{\epsilon} &
   0.66 \hat{\epsilon} & -89.72 \hat{\epsilon}^2\\
 ~[B^\pm]_{(M=3)} &  4 \hat{\epsilon} &  -11.33 \hat{\epsilon} & -2\hat{\epsilon} &  -9.33 \hat{\epsilon} & 0 &
   0.53 \hat{\epsilon} & -70.92 \hat{\epsilon}^2
 \\
    \hline 
\end{array}
\end{equation}
The fixed-point $[B^-]$ exists only for $M=3$. When $M>3$, the fixed-point $[B^+]$ does not survive for large $\hat{\epsilon}$, and instead merges with another fixed-point, $[E^-]$, that exists for $M>3$. The solution $[F^\pm]$ in \eqref{m=3,largeepsilon2} can be generalized to all $M \geq 3$.

\subsection{$M>3$}
\label{m>3}
We now discuss solutions for $M>3$.
\subsubsection{Fixed points in $d=3$}
\begin{figure}
    \centering
    \includegraphics[width=.8\textwidth]{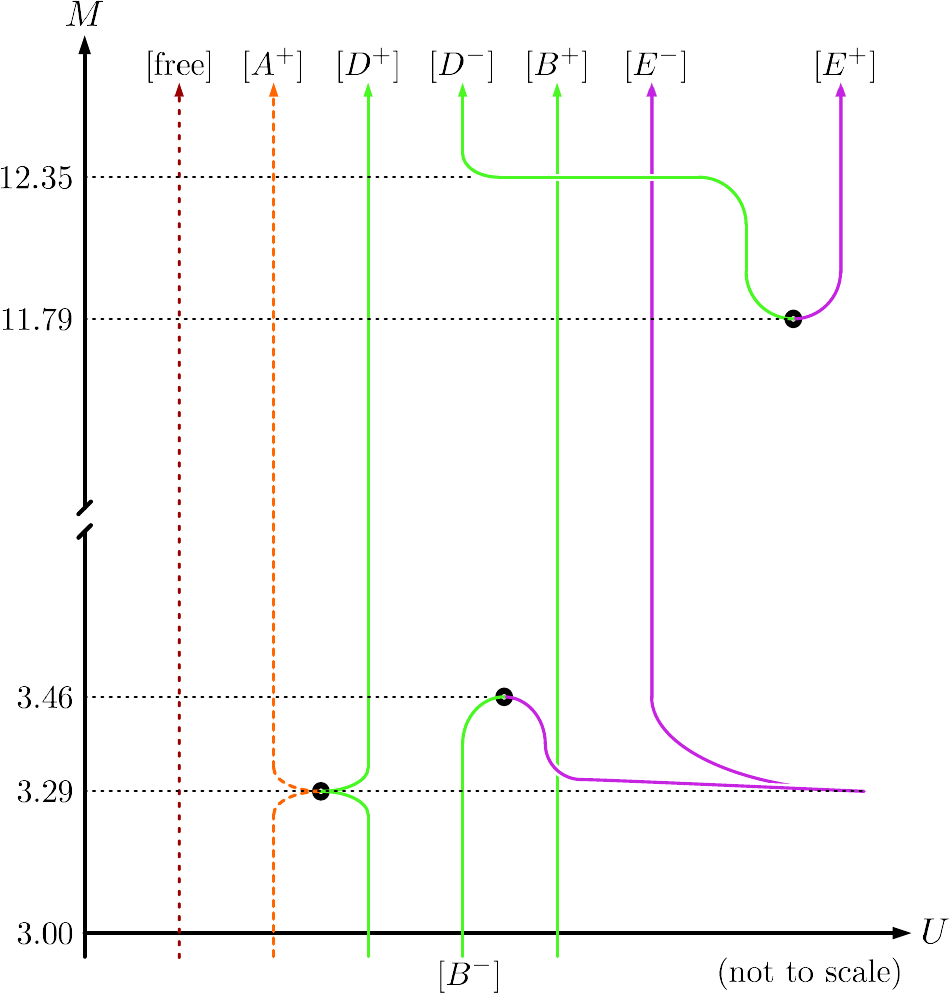}
    \caption{The above figure provides a schematic depiction of the large-$N$, $O(N)\times O(M)$ sextic fixed-points in $d=3$ as a function of $M$. The horizontal axis is schematic -- for any given value of $\hat{\epsilon}$ fixed-points are arranged in order of increasing $U$, so RG flow is only possible from right to left. The colour of each line specifies the stability of each fixed-point: red, orange, green and violet respectively denote fixed-points with zero, one, two and three unstable directions. Dashed lines indicate the presence of marginal directions. Black dots indicate mergers of fixed-points. For $\hat{\epsilon}$ sufficiently large, all fixed-points become infrared fixed-points relative to the free theory.}
    \label{fig:fixed-points-d3}
\end{figure}

In $d=3$ the number of UV fixed-points as a function of $M$, which we treat as a continuous parameter, is depicted schematically in Fig. \ref{fig:fixed-points-d3}. Fig. \ref{fig:fixed-points-d3} also depicts the stability of each fixed-point, and arranges the fixed-points in order of increasing value of the potential, which indicates which flows are possible.

For large values of $M$, there are 6 interacting fixed-points. We could determine two of these fixed-points, $[A^+]$ and $[B^+]$, analytically for arbitrary $M$ in $d=3$ -- these are presented in section \ref{all-m-solutions-2}. In addition there are four fixed-points, $[D^+]$, $[D^-]$, $[E^+]$ and $[E^-]$, which we could only determine numerically. $[E^-]$ emerges as a fixed-point solution for $M>3.29$. $[D^-]$ and $[E^+]$ emerge as solutions when $M>11.8$.  We present the numerical values of coupling constants and anomalous
dimensions of all the fixed-points that exist when $d=3$ and $M=4$ in Table \ref{tab:m4}. A similar table for $M=14$ is given in Table \ref{tab:m13}. We also determined asymptotic expressions for these fixed-points in the limit of $M \gg 1$, which are given in Table \ref{tab:largem}.
\begin{table}
    \centering
 $$ \begin{array}{|c|ccc|rrr|rr|r|r|}
 \hline 
 \text{Fixed Point}& \lambda_1 & \lambda_2 & \lambda_3 & (\partial_\lambda \beta_\lambda)^{(1 )}N &(\partial_\lambda \beta_\lambda)^{(2 )}N & (\partial_\lambda \beta_\lambda)^{(3 )}N & \gamma_{\phi^4}^{(1)} N & \gamma_{\phi^4}^{(2)}N & \gamma_\phi N^2 & U N\\
 \hline
  ~[E^-] & 41.41 & -30.20 & 41.55 & -39.14 & -23.49 & -3.41 & -18.07 & -6.90 &
   4.22 & 4281.38 \\
   ~[B^+] & 36.48 & 0 & 0 & -38.73 & -10.26 & 5.21 & -16.06 & 1.47 & 3.94 & 4259.84\\
 ~[D^+] & 11.01 & -44.76 & 73.70 & -10.87 & 10.76 & -1.66 & -3.93 & 3.11 &
    0.30 & 25.88 \\
   ~[A^+] & 0 & 0 & 36.48 & -7.30 & 4.86 & 0 & -2.43 & 0 & 0.06 & 13.48\\
    \hline 
\end{array} $$
    \caption{Fixed points, stability and anomalous dimensions when $M=4$ and $d=3$. Eigenvalues of sextic and quartic anomalous dimension matrices presented in an arbitrary order. Here we find a UV-stable fixed-point.}
    \label{tab:m4}
\end{table}
\begin{table}
    \centering
 $$ \begin{array}{|c|ccc|rrr|rr|r|r|}
 \hline 
 \text{Fixed Point} & \lambda_1 & \lambda_2 & \lambda_3 & (\partial_\lambda \beta_\lambda)^{(1 )}N &(\partial_\lambda \beta_\lambda)^{(2 )}N & (\partial_\lambda \beta_\lambda)^{(3 )}N & \gamma_{\phi^4}^{(1)} N & \gamma_{\phi^4}^{(2)}N & \gamma_\phi N^2 & U N\\
 \hline
  ~[A^+] & 0 & 0 & 36.48 & -2.08 & 1.39 & 0 & -0.69 & 0 & .005 & 1.10\\
   ~[D^+] & 27.48 & -222.39 & 473.59 & -355.43 & 66.36 & -44.65 & -134.07 & 10.47 & 18.89 & 88558 \\
   ~[D^-] & 34.54 & 62.16 & -266.68 & -170.33 & -68.38 & 15.30 & -71.10 & -13.19 & 29.03 & 212852 \\
 ~  [B^+] & 36.48 & 0 & 0 & -72.24 & -34.21 & 1.89 & -39.51 & 0.60 & 29.81 & 214016 \\
 ~[E^-] & 36.49 & -0.85 & 14.93 & -72.35 & -35.32 & -2.16 & -39.53 & -0.82 & 29.82 & 214017 \\
  ~[E^+] & 36.29 & 8.23 & -108.16 & -72.49 & -37.67 & -9.65 & -40.33 & -6.10 & 29.76 & 214092 \\
    \hline 
\end{array} $$
    \caption{Fixed points, stability and anomalous dimensions when $M=14$ and $d=3$. Eigenvalues of sextic and quartic anomalous dimension matrices presented in an arbitrary order.}
    \label{tab:m13}
\end{table}

\begin{table}
 $$ \hspace*{-2cm}\begin{array}{|c|ccc|rrr|}
 \hline 
  \text{\small{Fixed Point}} &\lambda_1 & \lambda_2 & \lambda_3 & (\partial_\lambda \beta_\lambda)^{(1 )}N &(\partial_\lambda \beta_\lambda)^{(2 )}N & (\partial_\lambda \beta_\lambda)^{(3 )}N 
  \\
 \hline
 ~[A^+] & 0 & 0 & \frac{360}{\pi ^2} & -\frac{288}{\pi^2 M} & \frac{192}{\pi^2 M} & 0  
 \\
   ~[D^+] &  \frac{360}{\pi ^2} & - \frac{180\ 2^{1/6} \sqrt{3}M^{2/3}}{\pi^2}  & \frac{30\ 2^{5/6} \sqrt{3}M^{4/3}}{\pi^2} & -\frac{36\ 2^{2/3} M^{5/3}}{\pi ^2}& \frac{48 M}{ \pi^2 } & -\frac{36 M}{\pi ^2} 
   \\
    ~[D^-] & \frac{360}{\pi ^2} & \frac{180\ 2^{1/6} \sqrt{3} M^{2/3}}{\pi^2}  & - \frac{30\ 2^{5/6} \sqrt{3}M^{4/3}}{\pi^2} & -\frac{36\ 2^{2/3} M^{5/3}}{\pi ^2} & -\frac{36 M}{\pi ^2} & \frac{48 M}{ \pi^2 } 
    \\
    ~[B^+] &
    \frac{360}{\pi ^2} & 0 & 0 & -\frac{36 M}{\pi ^2} & -\frac{24 M}{ \pi^2 } & \frac{288}{\pi ^2 M} 
    \\
    ~  [E^-] &
     \frac{360}{\pi ^2} &\frac{4320 \left(1-\sqrt{2}\right)}{\pi ^2 M^2} & \frac{360\left(-1+\sqrt{2}\right)}{\pi ^2} & -\frac{36 M}{\pi ^2} & -\frac{24 M}{ \pi^2 } & -\frac{576 \left(2-\sqrt{2}\right)}{\pi ^2 M}  \\
 ~[E^+] &
  \frac{360}{\pi ^2} & \frac{4320 \left(1+\sqrt{2}\right)}{\pi ^2 M^2} & \frac{360\left(-1-\sqrt{2}\right)}{\pi ^2} & -\frac{36 M}{\pi ^2} & -\frac{24 M}{ \pi^2 } & -\frac{576 \left(2+\sqrt{2}\right)}{\pi ^2 M}   \\
    \hline 
\end{array} $$
 $$ \hspace*{-2cm}\begin{array}{|c|rr|r|r|}
 \hline 
  \text{\small{Fixed Point}} & \gamma_{\phi^4}^{(1)} N & \gamma_{\phi^4}^{(2)}N & \gamma_\phi N^2 & UN
  \\
 \hline
 ~[A^+] & -\frac{96}{\pi^2 M}  & 0 & \frac{96}{\pi^2 M^2}  & \frac{207360}{\pi^6 M^2}
 \\
   ~[D^+] &   -\frac{12\ 2^{2/3} M^{5/3}}{\pi ^2} & \frac{16\ 2^{5/6} \sqrt{3} M^{1/3}}{\pi ^2} & \frac{12 M^2}{\pi^4} & \frac{3240 M^4}{\pi ^6}
   \\
    ~[D^-] & -\frac{12\ 2^{2/3} M^{5/3}}{\pi ^2} & -\frac{16\ 2^{5/6} \sqrt{3} M^{1/3}}{\pi ^2}  & \frac{12 M^2}{\pi^4} & \frac{3240 M^4}{\pi ^6}
   \\
    ~[B^+] &-\frac{24 M}{ \pi^2 } & \frac{96}{\pi ^2 M} & \frac{12 M^2}{\pi^4} & \frac{3240 M^4}{\pi ^6}
   \\
    ~  [E^-]  & -\frac{24 M}{ \pi^2 } &-\frac{192 \left(2-\sqrt{2}\right)}{\pi ^2 M } & \frac{12 M^2}{\pi^4} & \frac{3240 M^4}{\pi ^6}
   \\
 ~[E^+] &
   -\frac{24 M}{ \pi^2 } & -\frac{192 \left(2+\sqrt{2}\right)}{\pi ^2 M }  & \frac{12 M^2}{\pi^4} & \frac{3240 M^4}{\pi ^6}
   \\
    \hline 
\end{array} $$
    \caption{Fixed points, stability and anomalous dimensions for large $M$ in $d=3$. Eigenvalues of sextic and quartic anomalous dimension matrices presented in an arbitrary order.}
    \label{tab:largem}
\end{table}

\subsubsection{Fixed points in $d=3-\epsilon$}
\begin{figure}
    \centering
    \includegraphics[width=\textwidth]{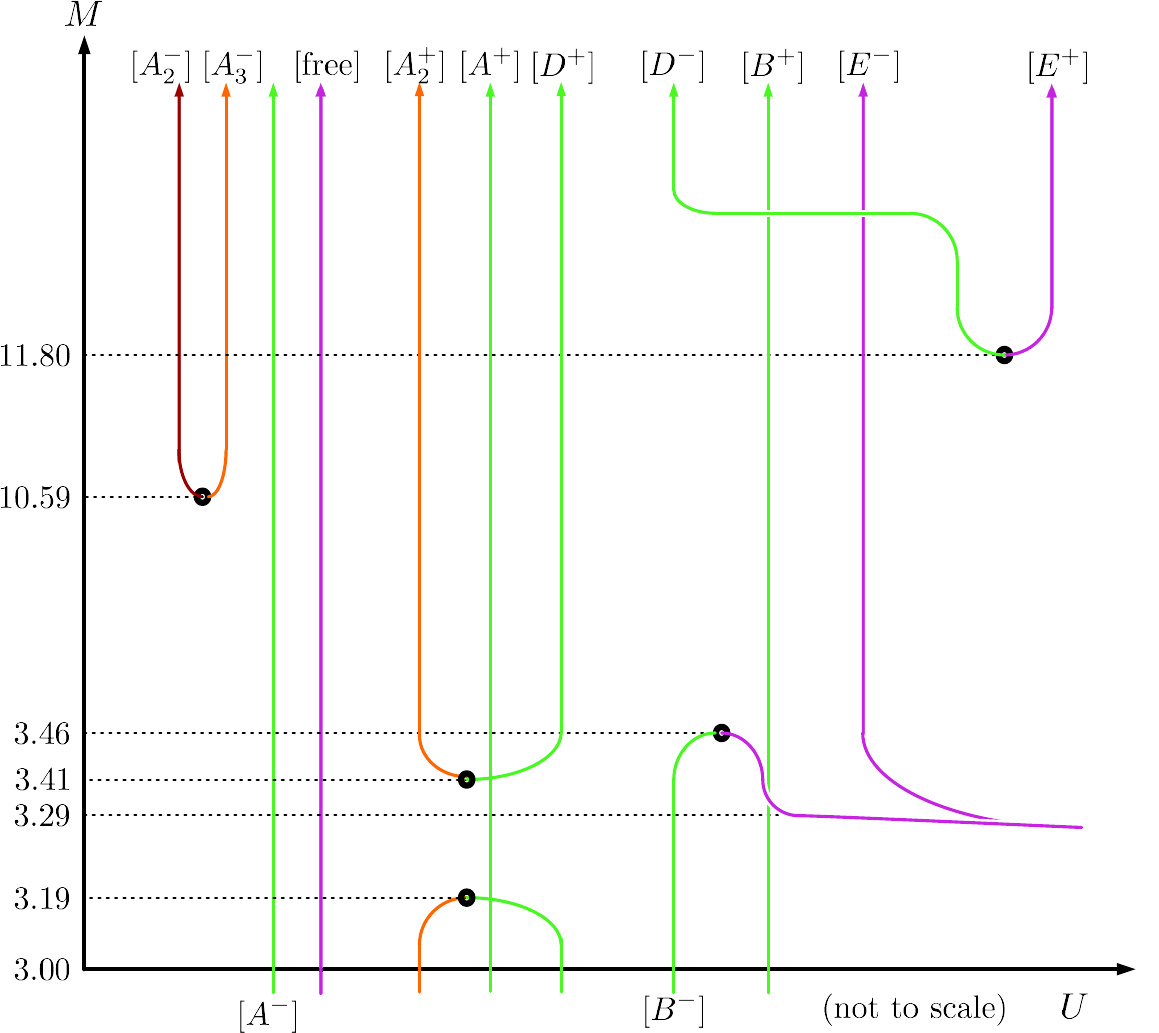}
    \caption{This plot illustrates the fixed-points as a function of $M$ for $0<\hat{\epsilon} \ll 1/M$. Some fixed-points which were marginal when $\hat{\epsilon}=0$ split into several fixed-points, while the other fixed-points remain unchanged. For any finite $\hat{\epsilon}$, several solutions merge for sufficiently large $M$, as indicated from the mergers for $M=14$ shown in Fig. \ref{fig:m14}. Numerical values of $M$ in this plot are determined using $\hat{\epsilon}=0.01$.}
    \label{fig:small-epsilonvsm}
\end{figure}

Let us discuss the solutions when $\hat{\epsilon}$ is small but nonzero.
When first order corrections in $\epsilon$ are included, we find the free fixed-point and the fixed-point $[A^-]$, both of which contained marginal directions in $d=3$, split into several IR fixed-points, depending on $M$, as shown in Fig. \ref{fig:small-epsilonvsm}. The other fixed-points remain essentially unchanged.

For $10 \geq M\geq 4$, there is only one nontrivial IR solution $[A^{-}]$, given by \eqref{all-m-solution-IR}.  
When $M \geq 11$, there are three real IR fixed-points. One of these solutions is $[A^-]$ given by Eq. \eqref{all-m-solution-IR}.
The other two solutions appear rather complicated for finite $M$, so we present them numerically for various values of $M$ in Tables \ref{other-table} and \ref{bifundamental-table}. 
\begin{table}
$$ \begin{array}{|c|rrr|r|rr|rrr|} \hline M &\lambda_1 & \lambda_2 & \lambda_3& \gamma_{\phi} N^2 & \gamma_{\phi^4}^{(1)} N & \gamma_{\phi^4}^{(2)}N & (\partial_\lambda \beta_\lambda)^{(1 )}N &(\partial_\lambda \beta_\lambda)^{(2 )}N & (\partial_\lambda \beta_\lambda)^{(3 )}N \\ \hline
 11 & 1.65 \hat{\epsilon}  & -9.04 \hat{\epsilon}  & 29.41 \hat{\epsilon}  & 0.0418 \hat{\epsilon} ^2 & 2.80 \hat{\epsilon}  & 0.90 \hat{\epsilon}  & 2.33 \hat{\epsilon}  & 2 \hat{\epsilon}  & 0.42 \hat{\epsilon} 
   \\
 12 & 1.50 \hat{\epsilon}  & -7.98 \hat{\epsilon}  & 32.48 \hat{\epsilon}  & 0.0406 \hat{\epsilon} ^2 & 2.74 \hat{\epsilon}  & 1.03 \hat{\epsilon}  & 2.37 \hat{\epsilon}  & 2 \hat{\epsilon}  & 0.74 \hat{\epsilon} 
   \\
 13 & 1.37 \hat{\epsilon}  & -7.22 \hat{\epsilon}  & 35.05 \hat{\epsilon}  & 0.0397 \hat{\epsilon} ^2 & 2.70 \hat{\epsilon}  & 1.09 \hat{\epsilon}  & 2.37 \hat{\epsilon}  & 2 \hat{\epsilon}  & 0.92 \hat{\epsilon} 
   \\
 14 & 1.27 \hat{\epsilon}  & -6.63 \hat{\epsilon}  & 37.50 \hat{\epsilon}  & 0.0392 \hat{\epsilon} ^2 & 2.67 \hat{\epsilon}  & 1.13 \hat{\epsilon}  & 2.36 \hat{\epsilon}  & 2 \hat{\epsilon}  & 1.04 \hat{\epsilon} 
   \\
 15 & 1.18 \hat{\epsilon}  & -6.14 \hat{\epsilon}  & 39.92 \hat{\epsilon}  & 0.0388 \hat{\epsilon} ^2 & 2.65 \hat{\epsilon}  & 1.16 \hat{\epsilon}  & 2.34 \hat{\epsilon}  & 2 \hat{\epsilon}  & 1.13 \hat{\epsilon} 
   \\
 16 & 1.11 \hat{\epsilon}  & -5.73 \hat{\epsilon}  & 42.32 \hat{\epsilon}  & 0.0385 \hat{\epsilon} ^2 & 2.63 \hat{\epsilon}  & 1.19 \hat{\epsilon}  & 2.33 \hat{\epsilon}  & 2 \hat{\epsilon}  & 1.20 \hat{\epsilon} 
   \\
 17 & 1.04 \hat{\epsilon}  & -5.38 \hat{\epsilon}  & 44.73 \hat{\epsilon}  & 0.0383 \hat{\epsilon} ^2 & 2.62 \hat{\epsilon}  & 1.21 \hat{\epsilon}  & 2.31 \hat{\epsilon}  & 2 \hat{\epsilon}  & 1.26 \hat{\epsilon} 
   \\
 18 & 0.987 \hat{\epsilon}  & -5.07 \hat{\epsilon}  & 47.13 \hat{\epsilon}  & 0.0381 \hat{\epsilon} ^2 & 2.61 \hat{\epsilon}  & 1.22 \hat{\epsilon}  & 2.30 \hat{\epsilon}  & 2 \hat{\epsilon}  & 1.31 \hat{\epsilon} 
   \\
 19 & 0.937 \hat{\epsilon}  & -4.80 \hat{\epsilon}  & 49.54 \hat{\epsilon}  & 0.0381 \hat{\epsilon} ^2 & 2.60 \hat{\epsilon}  & 1.23 \hat{\epsilon}  & 2.29 \hat{\epsilon}  & 2 \hat{\epsilon}  & 1.35 \hat{\epsilon} 
   \\
 20 & 0.891 \hat{\epsilon}  & -4.56 \hat{\epsilon}  & 51.96 \hat{\epsilon}  & 0.0380 \hat{\epsilon} ^2 & 2.60 \hat{\epsilon}  & 1.24 \hat{\epsilon}  & 2.28 \hat{\epsilon}  & 2 \hat{\epsilon}  & 1.39 \hat{\epsilon} 
   \\
 100 & 0.193 \hat{\epsilon}  & -0.967 \hat{\epsilon}  & 250.4 \hat{\epsilon}  & 0.0401 \hat{\epsilon} ^2 & 2.62 \hat{\epsilon}  & 1.33 \hat{\epsilon}  & 2.07 \hat{\epsilon}  & 2 \hat{\epsilon}  & 1.88  \hat{\epsilon}  \\
 1000 & 0.0199 \hat{\epsilon}  & -0.0996 \hat{\epsilon}  & 2500 \hat{\epsilon}  & 0.0415 \hat{\epsilon} ^2 & 2.66 \hat{\epsilon}  & 1.33 \hat{\epsilon}  & 2.01 \hat{\epsilon}  & 2 \hat{\epsilon}  & 1.99
   \hat{\epsilon}  \\\hline
\end{array}$$
\caption{This series of fixed-points, which we denote as $[A_2^-]$, approaches the fixed-point in \eqref{new-sol-2} when $M \to \infty$. \label{other-table}}
\end{table}

\begin{table}
$$ \begin{array}{|c|rrr|r|rr|rrr|} \hline M &\lambda_1 & \lambda_2 & \lambda_3 & \gamma_{\phi}N^2 & \gamma_{\phi^4}^{(1)} N & \gamma_{\phi^4}^{(2)}N & (\partial_\lambda \beta_\lambda)^{(1 )}N &(\partial_\lambda \beta_\lambda)^{(2 )}N & (\partial_\lambda \beta_\lambda)^{(3 )}N \\ \hline
 11 & 1.65 \hat{\epsilon}  & -9.99 \hat{\epsilon}  & 24.67 \hat{\epsilon}  & 0.0417 \hat{\epsilon} ^2 & 2.84 \hat{\epsilon}  & 0.582 \hat{\epsilon}  & 2 \hat{\epsilon}  & 1.97 \hat{\epsilon}  & -0.43
   \hat{\epsilon}  \\
 12 & 1.50 \hat{\epsilon}  & -9.62 \hat{\epsilon}  & 22.88 \hat{\epsilon}  & 0.0402 \hat{\epsilon} ^2 & 2.80 \hat{\epsilon}  & 0.448 \hat{\epsilon}  & 2 \hat{\epsilon}  & 1.73 \hat{\epsilon}  & -0.78 \hat{\epsilon} 
   \\
 13 & 1.37 \hat{\epsilon}  & -9.20 \hat{\epsilon}  & 21.78 \hat{\epsilon}  & 0.0390 \hat{\epsilon} ^2 & 2.76 \hat{\epsilon}  & 0.373 \hat{\epsilon}  & 2 \hat{\epsilon}  & 1.57 \hat{\epsilon}  & -0.97 \hat{\epsilon} 
   \\
 14 & 1.27 \hat{\epsilon}  & -8.81 \hat{\epsilon}  & 20.93 \hat{\epsilon}  & 0.0381 \hat{\epsilon} ^2 & 2.73 \hat{\epsilon}  & 0.321 \hat{\epsilon}  & 2 \hat{\epsilon}  & 1.45 \hat{\epsilon}  & -1.11 \hat{\epsilon}
    \\
 15 & 1.18 \hat{\epsilon}  & -8.44 \hat{\epsilon}  & 20.22 \hat{\epsilon}  & 0.0374 \hat{\epsilon} ^2 & 2.71 \hat{\epsilon}  & 0.282 \hat{\epsilon}  & 2 \hat{\epsilon}  & 1.35 \hat{\epsilon}  & -1.22 \hat{\epsilon}
    \\
 16 & 1.10 \hat{\epsilon}  & -8.11 \hat{\epsilon}  & 19.59 \hat{\epsilon}  & 0.0368 \hat{\epsilon} ^2 & 2.69 \hat{\epsilon}  & 0.251 \hat{\epsilon}  & 2 \hat{\epsilon}  & 1.27 \hat{\epsilon} & -1.30 \hat{\epsilon} 
   \\
 17 & 1.04 \hat{\epsilon}  & -7.80 \hat{\epsilon}  & 19.02 \hat{\epsilon}  & 0.0364 \hat{\epsilon} ^2 & 2.68 \hat{\epsilon}  & 0.225 \hat{\epsilon}  & 2 \hat{\epsilon}  & 1.20 \hat{\epsilon} & -1.37 \hat{\epsilon}  
   \\
 18 & 0.98 \hat{\epsilon}  & -7.52 \hat{\epsilon}  & 18.49 \hat{\epsilon}  & 0.0360 \hat{\epsilon} ^2 & 2.67 \hat{\epsilon}  & 0.204 \hat{\epsilon}  & 2 \hat{\epsilon}  & 1.15 \hat{\epsilon}  & -1.42 \hat{\epsilon} 
    \\
 19 & 0.93 \hat{\epsilon}  & -7.26 \hat{\epsilon}  & 18.00 \hat{\epsilon}  & 0.0358 \hat{\epsilon} ^2 & 2.66 \hat{\epsilon}  & 0.186 \hat{\epsilon}  & 2 \hat{\epsilon}  & 1.10 \hat{\epsilon} & -1.47 \hat{\epsilon} 
   \\
 20 & 0.89 \hat{\epsilon}  & -7.01 \hat{\epsilon}  & 17.53 \hat{\epsilon}  & 0.0355 \hat{\epsilon} ^2 & 2.65 \hat{\epsilon}  & 0.171 \hat{\epsilon}  & 2 \hat{\epsilon}  & 1.06
   \hat{\epsilon} & -1.52 \hat{\epsilon}   \\
 100 & 0.193 \hat{\epsilon}  & -1.78 \hat{\epsilon}  & 4.84 \hat{\epsilon}  & 0.0356 \hat{\epsilon} ^2 & 2.63 \hat{\epsilon}  & 0.00862 \hat{\epsilon}  & 2 \hat{\epsilon}   & 0.65
   \hat{\epsilon} & -1.97 \hat{\epsilon}   \\
 1000 & 0.0199 \hat{\epsilon}  & -0.18 \hat{\epsilon}  & 0.482 \hat{\epsilon}  & 0.0369 \hat{\epsilon} ^2 & 2.66 \hat{\epsilon}  & 0.0000856 \hat{\epsilon}  & 2 \hat{\epsilon}  & 0.66
   \hat{\epsilon} & -2.00 \hat{\epsilon}   \\
 \hline
\end{array}$$
\caption{This series of fixed-points, which we denote as $[A_3^-]$, approaches the bifundamental fixed-point when $M \to \infty$. \label{bifundamental-table}}
\end{table}

The fixed-point in Table \ref{other-table}, which we denote as $[A_2^{-}]$, is stable in all directions. For large $M$, the coupling constants obey the following asymptotic form:
\begin{equation}
    \lambda_1=\frac{20}{M}\hat\epsilon,\text{ }\lambda_2=-\frac{100}{M}\hat\epsilon \text{ and }\lambda_3=\frac{5M\hat\epsilon}{2},
\end{equation} 
with higher-order corrections in $\hat{\epsilon}$ and $1/M$.
The field anomalous dimension approaches
\begin{equation}
\gamma_{\phi}=\frac{\hat\epsilon^2}{24 N^2}
\end{equation}
as $M$ grows large. 
The quartic anomalous dimensions are:
\begin{equation}
    \gamma_{\nu_1} = \frac{4}{3N}\hat{\epsilon}, ~ \gamma_{\nu_2} = \frac{8}{3N}\hat{\epsilon},
\end{equation}
and all three  eigenvalues of the stability matrix approach $\frac{2\hat{\epsilon}}{N} + O(1/M)$.

The fixed-point in Table \ref{bifundamental-table}, which we denote as $[A_3^-]$ approaches the bifundamental fixed-point described in Eq. \eqref{bifundamental-fixed-point}. For large $M$, the coupling constants obey the following asymptotic form:
\begin{equation}
    \lambda_1=\frac{20}{M}\hat\epsilon,\text{ }\lambda_2=-\frac{180}{M}\hat\epsilon \text{ and }\lambda_3=\frac{480}{M}\hat\epsilon.
\end{equation}
 For $M\geq11$, this fixed-point is stable in two directions and unstable in one direction. The anomalous dimensions approach those of \eqref{bifundamental-fixed-point} as $M$ grows large, as expected.

The fixed-points $[A^-]$ and $[A_2^-]$ both have $\lambda_3 \sim \frac{5 M N \epsilon }{2}$, as $M \to \infty$. They correspond to new zeros of the beta function in the bifundamental large-$N$ limit that appear when $O(1/N^2)$ and $O(1/N^4)$ corrections are included, given in \eqref{new-sol-1} and \eqref{new-sol-2} respectively, in Appendix \ref{corrections}. 

\subsubsection{Fixed points at large $\hat{\epsilon}$}

\begin{figure}
    \centering
    \includegraphics[width=\textwidth]{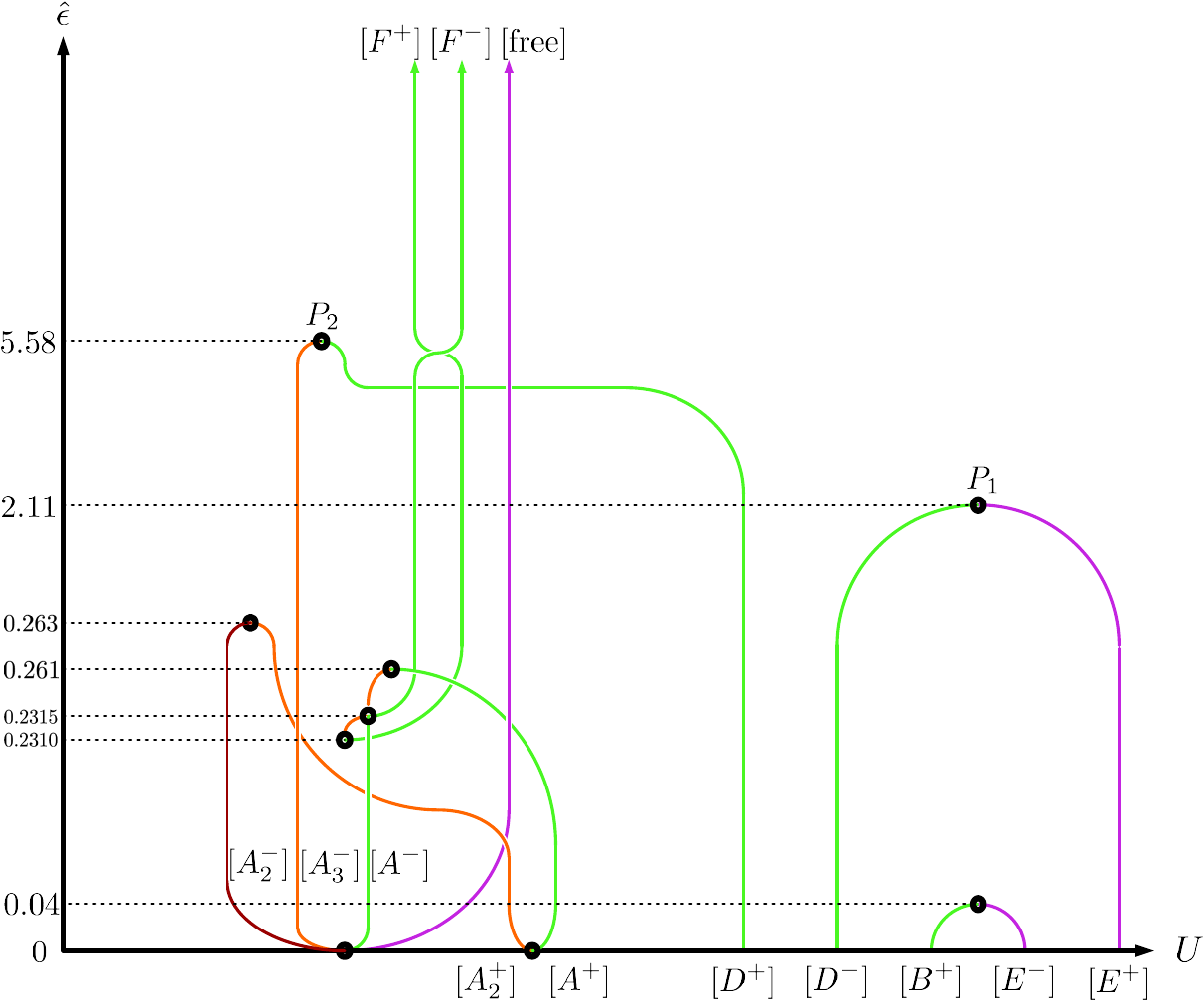}
    \caption{A plot of the fixed-points as a function of $\hat{\epsilon}$ for $M=14$. The horizontal axis is schematic -- for any given value of $\hat{\epsilon}$ fixed-points are arranged in order of increasing $U$, so RG flow is only possible from right to left. The colour of each line specifies the stability of each fixed-point: red, orange, green and violet respectively denote fixed-points with zero, one, two and three unstable directions. Black dots indicate mergers of fixed-points. For $\hat{\epsilon}$ sufficiently large, all fixed-points become infrared fixed-points relative to the free theory. When $M$ is increased, the mergers occur at $\hat{\epsilon}\sim O(1/M)$, except for the two mergers indicated by $P_1$ (between $[D^-]$ and $[E^+]$) and $P_2$ (between $[D^+]$ and $[A_3^-]$) which  occur at $\hat{\epsilon} \sim O(M)$. $P_1$ and $P_2$ become close to each other when $M$ is large. }
    \label{fig:m14}
\end{figure}

Fig. \ref{fig:m14} illustrates the number and stability of fixed-points as a function of $\hat{\epsilon}$ for $M=14$, and is somewhat representative of the behaviour of fixed-points for any generic $M>13$. 

Fig. \ref{fig:m14} shows that many of the fixed-points merge at finite values of $\hat{\epsilon}$. The structure of the mergers appears to be similar for generic large values of $M$. Numerically, we estimated the values of $\hat{\epsilon}$ at which the mergers occur vary asymptotically with $M$ when $M$ is large. $[A_3^-]$ merges with the fixed-point $[D^+]$ when $\hat{\epsilon} \sim \frac{9}{2 \pi ^2}M$ for large $M$. $[D^-]$ also merges with $[E^+]$ at a value of epsilon that also scales as $\hat{\epsilon} \sim \frac{9}{2 \pi ^2}M$ for large $M$. $[A_2^-]$ merges with the fixed-point $[A_2^+]$ at $\hat{\epsilon} \sim 3.6/M$, $[A^-]$ merges with $[A^+]$ at another value of $\hat{\epsilon} \sim 3.5/M$, and $[B^+]$ merges with $[E^-]$ at a value of $\hat{\epsilon} \sim 0.5/M$. Notice that, for this reason, the limits $M \to \infty$ and $\hat{\epsilon} \to 0$ do not commute. 

Numerically, we find the solution $[E^+]$ approaches the ultraviolet fixed-point of section \ref{sec:bifundamental}, for finite $\epsilon$. Near $\epsilon=0$, a discontinuity seems to emerge as $M \to \infty$ in the solution for $\lambda_3$ in $[E^+]$, and it rapidly changes from  $-\frac{1080}{\pi ^2}$ to $\frac{360 \left(-\sqrt{2}-1\right)}{\pi ^2}$, as $\epsilon$ approaches $0$ from above. Both $[A_3^-]$ and $[E^+]$ cease to exist near $\hat{\epsilon} \sim \frac{9}{2 \pi ^2}M$, in agreement with the results of \ref{sec:bifundamental}, although there appear to be some subtleties in the details of the $M\to \infty$ limits of the merger.

For any $M>3$ there are two nontrivial fixed-points that survive for large $\hat{\epsilon}$ which we denote by $[F^\pm]$. Both are unstable in two directions and stable in one direction. For $M=14$, when $\hat{\epsilon}$ is large, these two fixed-points have the following asymptotic behaviour:
\begin{equation}
    ~[F^\pm]:~~~\lambda_1 \sim 5.65, ~\lambda_2 \sim \pm 57.6 \sqrt{\hat\epsilon}, ~ \lambda_3 \sim \mp 143.3 \sqrt{\hat\epsilon}.
\end{equation}
It is possible to generalize this large $\hat{\epsilon}$ asymptotic solution to an analytic expression valid for arbitrary $M\geq 3$, but the expression is too unwieldy to reproduce here. For large $M$, the fixed-point solutions $[F^\pm]$ have the following asymptotic form:
\begin{eqnarray}
\lambda_1 & \sim & \frac{360 \sqrt[3]{2} \left(\frac{1}{M}\right)^{2/3}}{\pi ^2}+O\left(\frac{1}{M^{4/3}}\right) \\
\lambda_2/{\sqrt{\hat{\epsilon}}} & \sim & \pm\frac{90 \sqrt[6]{2} \sqrt[6]{M}}{\pi }\pm\frac{60 \sqrt{2} \sqrt{\frac{1}{M}}}{\pi }\mp\frac{15 \sqrt[6]{2} \left(\frac{1}{M}\right)^{5/6}}{\pi }+O\left(\frac{1}{M^{7/6}}\right) \\
\lambda_3/ {\sqrt{\hat{\epsilon}}} & \sim & \mp \frac{15\ 2^{5/6} M^{5/6}}{\pi } \mp \frac{100 \sqrt[6]{2} \sqrt[6]{M}}{\pi } \mp \frac{5 \sqrt[6]{\frac{1}{M}}}{\sqrt[6]{2} \pi } \mp \frac{80 \sqrt{2} \sqrt{\frac{1}{M}}}{\pi }+O\left(\frac{1}{M^{5/6}}\right)
\end{eqnarray}

The anomalous dimensions for $[F^\pm]$ when $M$ is large are both given by the following expressions. The anomalous dimension of $\phi$ is
\begin{equation}
    \gamma_\phi/\hat{\epsilon} \sim \frac{\sqrt[3]{2} \sqrt[3]{M}}{\pi ^2 N^2}+\frac{5\ 2^{2/3} \sqrt[3]{\frac{1}{M}}}{3 \pi ^2 N^2}+\frac{2 \sqrt[3]{2} \left(\frac{1}{M}\right)^{2/3}}{3 \pi ^2 N^2}+\frac{28}{9 \pi ^2 M N^2}+\frac{5\ 2^{2/3} \left(\frac{1}{M}\right)^{4/3}}{9 \pi ^2 N^2}+O\left(\frac{1}{M^{5/3}}\right).
\end{equation}
The anomalous dimensions of quartic operators are:
\begin{equation}
    \gamma_{\nu_1}/{\hat{\epsilon}} \sim -\frac{2^{2/3} M^{2/3}}{N}-\frac{10}{3 N}+O\left({\frac{1}{M^{1/3}}}\right), ~ \gamma_{\nu_2}/{\hat{\epsilon}} \sim \frac{2}{N},
\end{equation}
and the eigenvalues of the stability matrix are:
\begin{equation}
   N (\partial_\lambda \beta)_1/\hat{\epsilon} \sim -2-6 \sqrt[3]{2} \left(\frac{1}M\right)^{2/3}, ~ N(\partial_\lambda \beta)_2/\hat{\epsilon} \sim -\sqrt[3]{108}M^{2/3},~ N(\partial_\lambda \beta)_3/\hat{\epsilon} \sim 4.
\end{equation}

\section{$O(N)\times O(2)/\mathbb Z_2$ fixed-points for finite $N$}
\label{sec:finiteN}
In this section, we briefly study perturbative fixed-points of our theory for finite $M$ and $N$. Finite $N$ fixed-points of the form $O(N)\times O(N)$ were studied extensively in \cite{Jepsen:2020czw}. We restrict attention to $M=2$, since this is presumably the most physical example. 

When $M=2$, the three couplings are not independent, as discussed in section \ref{m=2}. We use the two independent couplings $\tilde{g}_1$ and $\tilde{g}_2$ given by
\begin{equation}
   (8\pi)^2\tilde{g}_1=g_1+\frac{2g_2}{3}, \quad  (8\pi)^2\tilde{g}_3=g_3+\frac{g_2}{3}.
\end{equation}
The two-loop beta functions for these couplings when $M=2$, for finite $N$ are:
\begin{eqnarray}
 \beta_{\tilde{g}_1}& =&-2 \tilde{g}_1 \epsilon+ \frac{4}{15} \tilde{g}_1 (3 \tilde{g}_1 (N+7)+4 \tilde{g}_3 (N+9)), \\
 \beta_{\tilde{g}_3} &=& -2 \tilde{g}_3 \epsilon+\frac{1}{30} \left(\tilde{g}_1^2 (3 N+29)+16 \tilde{g}_1 \tilde{g}_3 (N+7)+16 \tilde{g}_3^2 (3 N+11)\right).
\end{eqnarray}

It is possible to solve the zeros of this beta function analytically, for arbitrary $N$. There are in general four fixed-points -- the free fixed-point and three interacting fixed-points that we denote by $[a]$, $[b^+]$ and $[b^-]$.

Fixed point $[a]$ is given by
\begin{equation}
    [a]: ~\tilde{g}_1 = 0, ~ \tilde{g}_3= \frac{15 \epsilon }{12 N+44}.
\end{equation}

Fixed points $[b^\pm]$  are given by
\begin{equation}
\begin{split}
    [b^\pm]: ~\tilde{g}_1 &= \frac{15 \left(4 N^2\pm \sqrt{2} \sqrt{-N^2+3 N+126} N \pm 9 \sqrt{2} \sqrt{-N^2+3 N+126}+16 N-84\right) \epsilon }{4 (N+6) \left(9 N^2+88 N+159\right)} \\
    ~\tilde{g}_3&= \frac{15 \left(6 N^2\mp3 \sqrt{2} \sqrt{-N^2+3 N+126} N \mp 21 \sqrt{2} \sqrt{-N^2+3 N+126}+98 N+408\right) \epsilon }{16 (N+6) \left(9 N^2+88 N+159\right)}.
    \end{split}
\end{equation}

\begin{figure}
    \centering
    \includegraphics{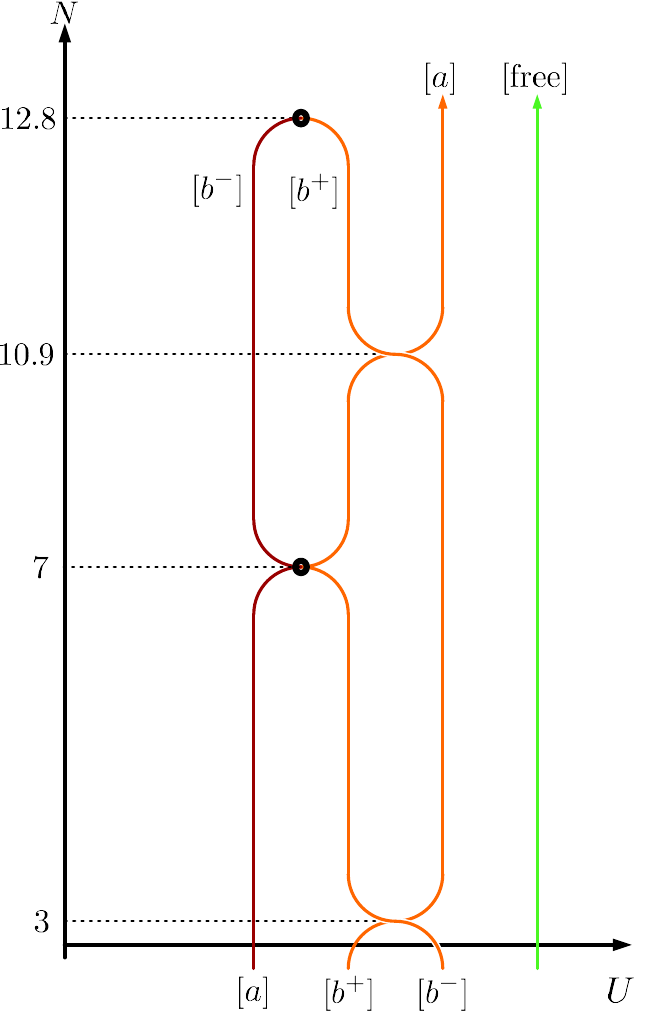}
    \caption{This figure schematically depicts the behaviour of fixed-point solutions of the $O(N)\times O(2)$ two-loop beta functions as a function of $N$. Conventions are the same as earlier figures.}
    \label{fig:finite-N}
\end{figure}

The schematic behaviour of these fixed-points as a function of $N$ is shown in Fig. \ref{fig:finite-N}.
When $N \to \infty$, the fixed-point $[a]$ approaches the fixed-point $[A_-]$ of section \ref{sec:vector-model}. Fixed points $[b^+]$ and $[b^-]$ merge and become complex when $N=N_* = \frac{3}{2} \left(1+\sqrt{57}\right) \approx 12.8$. 

The stability of these fixed-points is also depicted in Fig. \ref{fig:finite-N}. For all $N<N_*$, $[b^+]$ is stable in one direction and unstable in the other direction. For $N<7$, fixed-point $[a]$ is stable in two directions, while $[b^-]$ is stable in one direction and unstable in the other. When $N=7$, the fixed-points $[b^-]$ and $[a]$ merge into a single fixed-point with one marginal and one stable direction. For $N>7$, $[a]$ has one stable and one unstable direction, while $[b^-]$ has two stable directions (until it becomes complex at $N_*$.) 

\begin{figure}
    \centering
    \includegraphics[scale=0.7]{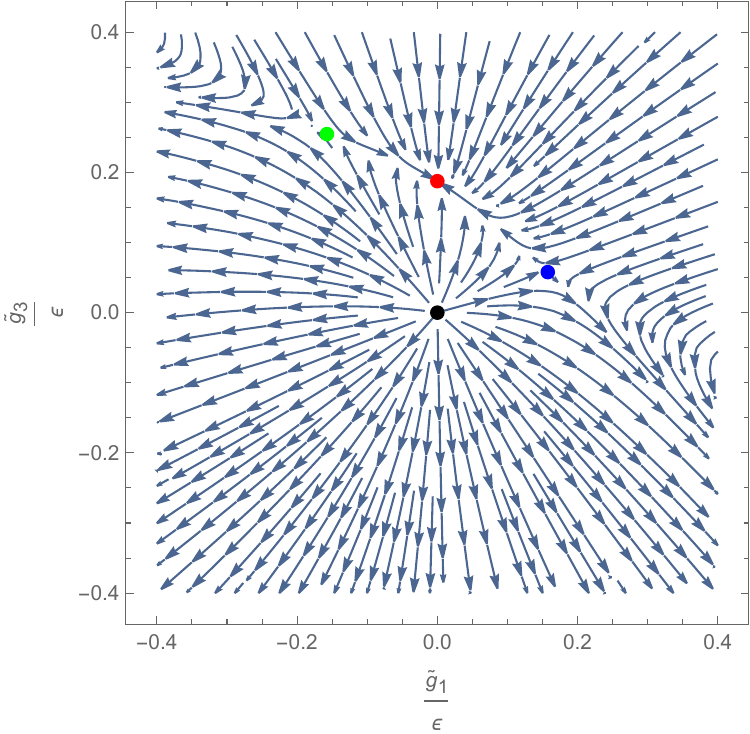}
    \caption{Flows in the $\tilde{g}_1$-$\tilde{g}_3$ plane in $d=3-\epsilon$ when $M=2$ and $N=3$. The black dot is the free fixed-point, the red dot is  $[a^+]$, the blue dot is $[b^+]$ and the green dot is $[b^-]$.}
    \label{flowsn3}
\end{figure}

We list anomalous dimensions for each of these fixed-points for $N=3$, $N=7$ and $N=12$ in Tables \ref{tab-3}, \ref{tab-7} and \ref{tab-12}. We also depict flows between the various fixed-points for $N=3$ in Fig. \ref{flowsn3}.

\begin{table}
$$
\begin{array}{|c|rr|rr|rr|r|r|}
 \hline 
 \text{Fixed Point}&\tilde{g}_1/\epsilon&\tilde{g}_3/\epsilon& (\partial_{\tilde{g}} \beta_{\tilde{g}})^{(1 )}/\epsilon &(\partial_{\tilde{g}} \beta_{\tilde{g}})^{(2 )}/\epsilon  & \gamma_{\phi^4}^{(1)}/\epsilon  & \gamma_{\phi^4}^{(2)}/\epsilon & \gamma_\phi/\epsilon^2 &U/\epsilon^3 \\
 \hline
  ~[a] & 0 & \frac{3}{16} &2  &\frac{2  }{5} & 1 & \frac{1 }{5} & \frac{1}{480} & -\frac{3}{8}\approx- 0.38 \\
 ~[b^{+}] & \frac{5}{12 \sqrt{7}}& \frac{5}{672} \left(21-5 \sqrt{7}\right) &2 & -\frac{2 }{3} & \frac{1}{18} \left(9+\sqrt{71}\right)   & \frac{1}{18}
   \left(9-\sqrt{71}\right)   & \frac{5 }{2592} & -\frac{25 }{72}\approx- 0.35 \\
  ~[b^-] &  -\frac{5}{12 \sqrt{7}}& \frac{5}{672} \left(21+5 \sqrt{7}\right)&2   & -\frac{2  }{3} & \frac{1}{18} \left(9+\sqrt{71}\right)  & \frac{1}{18}
   \left(9-\sqrt{71}\right)   & \frac{5 }{2592} & -\frac{25 }{72}\approx- 0.35 \\
    \hline 
\end{array}
$$
\caption{\label{tab-3}Fixed points, stability and anomalous dimensions when $M=2$ and $N=3$ in $d=3-\epsilon$. Eigenvalues of sextic and quartic anomalous dimension matrices presented in an arbitrary order.}
\end{table}
\begin{table}
$$
\begin{array}{|c|rr|rr|rr|r|r|}
 \hline 
 \text{Fixed Point}&\tilde{g}_1/\epsilon&\tilde{g}_3/\epsilon& (\partial_{\tilde{g}} \beta_{\tilde{g}})^{(1 )}/\epsilon &(\partial_{\tilde{g}} \beta_{\tilde{g}})^{(2 )}/\epsilon  & \gamma_{\phi^4}^{(1)}/\epsilon  & \gamma_{\phi^4}^{(2)}/\epsilon & \gamma_\phi/\epsilon^2 &U/\epsilon^3 \\
 \hline
 ~[a] & 0& \frac{15}{128}&2  & 0 & \frac{9  }{8} & \frac{1 }{8} & \frac{3 }{1024} & -\frac{315 }{256}\approx -1.23\\
   ~[b^{+}]& \frac{105}{988}&\frac{375}{7904}&2   & -\frac{98  }{247} & \frac{11}{494} \left(26+\sqrt{541}\right)  & \frac{11}{494}
   \left(26-\sqrt{541}\right)  & \frac{10989 }{3904576} & -\frac{1153845 }{976144}\approx -1.18 \\
   \hline
\end{array}
$$
\caption{\label{tab-7} Fixed points, stability and anomalous dimensions when $M=2$ and $N=7$ in $d=3-\epsilon$. Eigenvalues of sextic and quartic anomalous dimension matrices presented in an arbitrary order.}
\end{table}
\begin{table}
$$
\begin{array}{|c|rr|rr|rr|r|r|}
 \hline 
 \text{Fixed Point}&\tilde{g}_1/\epsilon&\tilde{g}_3/\epsilon& (\partial_{\tilde{g}} \beta_{\tilde{g}})^{(1 )}/\epsilon &(\partial_{\tilde{g}} \beta_{\tilde{g}})^{(2 )}/\epsilon  & \gamma_{\phi^4}^{(1)}/\epsilon  & \gamma_{\phi^4}^{(2)}/\epsilon & \gamma_\phi/\epsilon^2 &U/\epsilon^3 \\
 \hline
 ~[b^-] &\frac{5}{108} &\frac{25}{432}& 2   & \frac{2  }{27} & \frac{2}{81} \left(27+\sqrt{449}\right)  & \frac{2}{81}
   \left(27-\sqrt{449}\right)   & \frac{91 }{26244} & -\frac{1820 }{729}\approx -2.50 \\
 ~[b^{+}] &\frac{25}{372}& \frac{65}{1488}&2   & -\frac{10 }{93} & \frac{4}{279} \left(45+\sqrt{1409}\right) & \frac{4}{279}
   \left(45-\sqrt{1409}\right)   & \frac{539 }{155682} & -\frac{21560 }{8649}\approx -2.49\\
    ~[a] & 0 & \frac{15}{188} & 2  & -\frac{10 }{47} & \frac{56 }{47} & \frac{4 }{47} & \frac{91 }{26508} & -\frac{5460 }{2209}\approx -2.47 \\
    \hline 
\end{array}
$$
\caption{\label{tab-12}Fixed points, stability and anomalous dimensions when $M=2$ and $N=12$ in $d=3-\epsilon$. Eigenvalues of sextic and quartic anomalous dimension matrices presented in an arbitrary order.}
\end{table}

\section{Discussion}

It would be interesting to explore whether bootstrap techniques, as described in \cite{Henriksson:2020fqi} and references therein, could shed more light on the existence or non-existence of these CFT's for $\alpha=1$. It may also be interesting to study the small $\alpha$ limit, or the finite $M$, large $M$ fixed-points in greater detail using analytic bootstrap \cite{Alday:2016njk, Alday:2016jfr, Alday:2017zzv, Henriksson:2018myn, Alday:2019clp, Henriksson:2020fqi}.

For the infrared fixed-point, we found that the anomalous dimensions of the unprotected single-trace quartic and sextic operators attained their maximum at $\alpha=1$. This is consistent with the general expectation that scaling dimensions of unprotected operators become large when one approaches the matrix-like large-$N$ limit from the vector model. The scaling dimensions of $\phi$ and $\phi^2$ were independent of $\alpha$ at this order, but we suspect that higher-order corrections would display similar $\alpha$ dependence as the quartic and sextic single-trace operators. It should be possible to confirm this by computing 6-loop corrections to the large-$N$ anomalous dimensions following \cite{Hager:2002uq}. It would also be of interest to compute the spectrum of higher-spin operators, to four or six loops, of the theory. To determine the order $\alpha^2$ correction to Eq. \eqref{Critical-epsilon}, which would allow a Pad\`e approximation of the critical dimension, an 8-loop calculation is required.

It is worth mentioning the non-supersymmetric sextic tensor model of \cite{GKPPT} in $d=3-\epsilon$ also ceased to exist at a finite value of $\epsilon$. There, a real melonic fixed-point with was found to exist perturbatively  in $d=3-\epsilon$, but large-$N$ techniques showed that it ceased to be real for a finite, but small critical value of $\epsilon$. However, for the supersymmetric tensor model of \cite{Popov:2019nja}, smooth interpolation from  $d=3-\epsilon$ to $d=1+\epsilon$ was possible, though as noted there, the existence of the theory in $d=2$ exactly is still uncertain.\footnote{We thank Igor Klebanov for discussions on this point.} Perhaps it would be of interest to consider fixed-points $\mathcal N=1$ supersymmetric version of the $O(N)\times O(M)$ theory, particularly in the limit $M/N \ll 1$ and compare results to our non-supersymmetric model.

We also studied the theory in the limit where $N$ is large and $M$ is finite. In this limit, it is possible to compute the complete beta function of the theory to first order in $1/N$. Studying the zeros of this beta function, we found a rich collection of large-$N$ fixed-points in $d=3$ and in $d=3-\epsilon$, which are determined to all orders in the parameter $\hat{\epsilon}=N\epsilon$. For most values of $M$, the fixed-points include at least one ultraviolet stable fixed-point, that could potentially serve as an asymptotically safe definition of the theory in the large-$N$ limit. Unlike the case of $M=1$, \cite{Pisarski:1983gn}, we also found fixed-points which survive to arbitrarily large values of $\hat{\epsilon}$.

Quartic $O(N)\times O(M)$ fixed-points in $d=3$ have applications to frustrated antiferromagnetic systems (for small values of $M$ and $N$), as mentioned in the introduction. While the calculations in this paper suggest that it may also be possible to define sextic $O(N)\times O(M)$ fixed-points in $d=3$, these fixed-points are ultraviolet fixed-points, and require a reasonably high degree of fine-tuning, as all quartic interactions must be tuned to zero, in addition to any relevant sextic interaction terms. Therefore, it seems unlikely that these fixed-points would be relevant to frustrated anti-ferromagnetic systems, or other real world condensed matter/statistical systems,  unless those systems possess a high degree of fine tuning. Nevertheless, it would be interesting to construct such statistical models which would presumably be generalizations of the tricritical Ising model that possess several chemical potentials that need to be tuned to critical values. Some of our fixed-points in sections \ref{sec:vector-model} and \ref{sec:finiteN} do appear to extend to $d=2$ without any obstruction, and it would also be interesting identify if they correspond to any known $d=2$ conformal field theories. 

There are several additional computations one could carry out related to the theory we study in this paper. We mention a few possibilities for future work below.

It is possible to study flows between the large-$N$ fixed-points determined for finite $M$, or at small $\alpha$. We determined some flows analytically, in the limit of small $\alpha$ and for $M=2$, as shown in Fig.s \ref{exact-flows} and \ref{flowsm2}. It would be interesting to better study the flows between the large-$N$ fixed-points for various values of $M$. We have not discussed the phenomena of spontaneous breaking of conformal invariance, \cite{Bardeen:1983rv, Amit:1984ri, David:1984we, David:1985zz}, which could occur along these flows. We hope to investigate the Bardeen-Moshe-Bender phenomenon in the theories considered here in future work, both in the limit where $M$ and $N$ are both large, but $\alpha=M/N$ is small; or for the theories in which $N$ is large and $M$ is finite. Such an analysis is important for confirming the existence of the UV fixed-points identified in this paper.

One could determine the higher-spin spectrum of this and related multiscalar CFT's perturbatively, or using the technique of \cite{Giombi:2016hkj}. In the absence of a gauge field, the scaling dimensions of spin-$s$ operators, (represented schematically, as $\phi \partial^s \phi$), are not expected to grow as $\log s$ when $s \rightarrow \infty$, and instead approach a finite value as $s \rightarrow \infty$, suggesting that the dual is not string-like. (This objection does not apply for the non-supersymmetric bi-fundamental Chern-Simons theories, where the Chern-Simons gauge field, although non-dynamical, still gives rise to a $\log s$ dependence of anomalous dimensions of spin $s$ operators for large $s$ \cite{Giombi:2016zwa,GuruCharan:2017ftx, Jain:2019fja}.)

When $M$ is finite and $N$ is large, the theory is a large-$N$ vector model, so there are various computations possible in principle that may be worth exploring. One may also be able to study these fixed-points in the presence of a Chern-Simons gauge field as in \cite{Giombi:2011kc, Aharony:2011jz,Aharony:2012nh,GurAri:2012is, Bedhotiya:2015uga}. For $M=1$ the fixed-points of the sextic interaction were studied in \cite{Aharony:2018pjn} -- however, many of the terms in the beta function at present remain uncomputed, due to the difficulty of $1/N$ calculations in Chern-Simons vector models. Generalizing these discussions to higher values of $M$ is possible in principle.\footnote{See also \cite{Sakhi:2019rfj} for related computations.} Let us also point out that the sextic $O(N)$ theory was studied in the presence of a boundary in \cite{Herzog:2020lel}, and similar computations could be attempted for arbitrary $M$.

\cite{Yabunaka:2017uox} studies the existence and merger of the UV fixed-point with IR fixed-point in the sextic $O(N)$ theory as a function of $N$ and $\epsilon$ using the non-perturbative renormalization group technique -- they seem to find that a nonperturbtive UV fixed-point and IR fixed-point merge at $N(d)\approx 3.6/(3-d)$ in apparent agreement with \cite{Pisarski1982}.  The authors of \cite{Yabunaka:2017uox} also discuss quartic but not sextic $O(N)\times O(2) /\mathbb Z_2$ fixed-points.  The same group use similar techniques to also point out potential subtleties associated with the $N \to \infty$ limit \cite{Yabunaka:2018mju, Yabunaka:2021fow}. It may be interesting and worthwhile to repeat the non=perturbative renormalization group analysis of \cite{Yabunaka:2017uox, Yabunaka:2018mju, Yabunaka:2021fow} for the theories considered in this paper -- both for the large-$N$ bifundamental fixed-points of section \ref{small-alpha-section} in the $\alpha-d$ plane, as well as the finite $M$ fixed-points of section \ref{sec:vector-model} in the $N-d$ plane.

In addition, drawing inspiration from the discussion in \cite{Yabunaka:2017uox, Defenu:2020cji}, one could also study bifundamental fixed-points with $\phi^{2m}$ interactions for $m>3$. Such interactions become relevant for $d<\frac{2}{1-\frac{1}{m}}$, and could define multi-critical interacting conformal field theories in $d=2$. 

Another avenue for future research arises from the observation that, for non-integer $M$ in the $-3<M<2$, our theory can be used to define large-$N$ non-unitary fixed-points in $d=3$, that could exhibit unconventional behaviour, as in \cite{Jepsen:2020czw, Jepsen:2021rhs}. Unfortunately, as described in Appendix \ref{spooky-appendix}, we did not find a fixed-point whose stability matrix contains purely imaginary eigenvalues, so our particular model does not contain (perturbative) limit cycles. However, it may be possible to construct slightly more complicated theories, with sextic interactions that are effectively triple-trace, (e.g. $O(M_1)\times O(M_2)\times O(N)$ tensor models with $N$ large and $M_1$ and $M_2$ finite) to generate non-unitary large-$N$ fixed-points with unconventional behaviour in $d=3$. 

We briefly discussed the perturbative fixed-points of the $O(N)\times O(2)$ theory when $N$ is finite. As the focus of the paper was mainly on large-$N$ limits, we have not studied fixed-points with $O(N)\times O(M)/\mathbb Z_2$ symmetry when both $M$ and $N$ are finite and greater than $2$. Such fixed-points were studied when $M=N$ in \cite{Jepsen:2020czw}, but the case of $M\neq N$ appears not to have yet been considered in the literature, and could be investigated in order to better map out the space of multiscalar fixed-points. 

Finally, let us comment on the possibility of asking an analogous question about bifundamental multiscalar fixed-points in $6-\epsilon$ dimensions. A cubic $O(N)\times O(M)$ theory in $6-\epsilon$ dimensions was put forth in \cite{Herbut:2015zqa} generalizing the cubic vector model studied in \cite{Fei:2014yja, Fei:2014xta, Gracey:2015tta}. See also \cite{Gracey:2017oqf}, \cite{Gracey:2018khg}, and \cite{Osborn:2017ucf}.

\section*{Acknowledgements} We thank Igor Klebanov and Fedor Popov for discussions and comments on a draft of this manuscript. We also thank an anonymous referee for comments and suggestions. SP also thanks Simone Giombi, and Gregory Tarnopolsky for discussions on related topics in 2018 during the course of collaboration on \cite{GKPPT}. SP acknowledges the support of a DST-SERB MATRICS grant (MTR/2018/001077), and DST-SERB grant (CRG/2021/009137). 

\appendix

\section{Four-loop beta functions at finite $M$ and $N$}
\label{beta-function}
In this appendix we present the four-loop beta functions for $g_i$ at finite $N$ and $M$. These expressions are also available in the electronic supplemental information \cite{Supplemental}.  The expressions are rather long, and we organize them by powers of $\epsilon$ and $N$ as follows:
\begin{eqnarray}
\beta_{g_i} & = & -2 \epsilon g_i + \frac{1}{90 (8
   \pi)^2} \tilde{\beta}_{g_i}^{(2\text{ loop})}+\frac{1}{4(90)^2 (8\pi)^4}\tilde{\beta}_{g_i}^{(4\text{ loop})},
\end{eqnarray}
where 
\begin{equation}
    \tilde{\beta}_{g_i}^{(p\text{ loop})} = \sum_{j=0}^{p-1}  \tilde{\beta}_{g_i}^{(p\text{ loop},~k)}N^k. \label{beta-temp}
\end{equation}
The 2-loop contributions to Eq. \eqref{beta-temp} are given by 
\begin{eqnarray}
\tilde{\beta}^{(2\text{ loop},~1)}_{g_1} & = & \left(9 g_1^2+8 g_2^2\right) M+36 g_1^2+96 g_2 g_1, \\
\tilde{\beta}^{(2\text{ loop},~0)}_{g_1} & = & 12 g_1 \left(3 g_1+8 g_2\right) M + 4 \left(75 g_1^2+12 \left(9 g_2+10 g_3\right) g_1+64 g_2^2\right), \\
\tilde{\beta}^{(2\text{ loop},~1)}_{g_2} & = & \left(12 g_1 g_2+48 g_3 g_2\right) M+27 g_1^2+24 g_2 g_1+144 g_3 g_1+56 g_2^2, \\
\tilde{\beta}^{(2\text{ loop},~0)}_{g_2} & = & \left(27 g_1^2+24 g_2 g_1+144 g_3 g_1+56 g_2^2\right) M \nonumber \\ && +144 g_1^2+456 g_2 g_1+288 g_3 g_1+128 g_2^2+864 g_2 g_3, \\ 
\tilde{\beta}^{(2\text{ loop}, 1)}_{g_3} & = & \left(4 g_2^2+72 g_3^2\right) M+4 g_2^2+12 g_1 g_2+96 g_3 g_2, \\
\tilde{\beta}^{(2\text{ loop}, 0)}_{g_3} & = & \left(4 g_2^2+12 g_1 g_2+96 g_3 g_2\right) M \nonumber \\ && +21 g_1^2+36 g_2 g_1+144 g_3 g_1+84 g_2^2+528 g_3^2+96 g_2 g_3,
\end{eqnarray}
and the 4-loop contributions are given by the following for $g_1$,
\begin{eqnarray}
\tilde{\beta}^{(4\text{ loop}, 3)}_{g_1} & = & -3 \pi ^2 \left(3 g_1^2+8 g_2^2\right) g_1 M-36 \pi ^2 \left(g_1+4 g_2\right) g_1^2,  \\
\tilde{\beta}^{(4\text{ loop}, 2)}_{g_1} & = &
-6 \left(51 g_1^3+4 \left(\left(30+\pi ^2\right) g_2^2-96 g_3^2\right) g_1+384 g_2^2 g_3\right) M^2 \nonumber \\ && +\biggr(-27 \left(124+7 \pi ^2\right) g_1^3-36 \left(312+17 \pi ^2\right) g_2 g_1^2 -\frac{32}{3} \left(480+23 \pi ^2\right) g_2^3  \nonumber \\ && - 216 g_2 \left(\left(12+\pi ^2\right) g_2  +4 \left(20+\pi ^2\right) g_3\right) g_1\biggr) M \nonumber \\ && -1332 \pi ^2 g_1^3-15264 g_1^3-2880 \pi ^2 g_2 g_1^2-38016 g_2 g_1^2-2592 \pi ^2 g_3 g_1^2-40320 g_3 g_1^2 \nonumber \\ && -2880 \pi ^2 g_2^2 g_1-44928 g_2^2 g_1, \\
\end{eqnarray}
\begin{eqnarray}
\tilde{\beta}^{(4\text{ loop}, 1)}_{g_1} 
& = & 
\left(-9 \pi ^2 g_1^3-24 \pi ^2 g_2^2 g_1\right) M^3 
\nonumber 
\\ && +\biggr(-189 \pi ^2 g_1^3-3348 g_1^3-612 \pi ^2 g_2 g_1^2-11232 g_2 g_1^2-216 \pi ^2 g_2^2 g_1-2592 g_2^2 g_1 \nonumber \\ && -864 \pi ^2 g_2 g_3 g_1-17280 g_2 g_3 g_1-\frac{736}{3} \pi ^2 g_2^3-5120 g_2^3\biggr) M^2 \nonumber \\ && +\biggr(-4752 \pi ^2 g_1^3-63936 g_1^3-10980 \pi ^2 g_2 g_1^2-137376 g_2 g_1^2-11016 \pi ^2 g_3 g_1^2 \nonumber \\ && -122688 g_3 g_1^2  -12792 \pi ^2 g_2^2 g_1-170208 g_2^2 g_1-12960 \pi ^2 g_3^2 g_1-107136 g_3^2 g_1-6048 \pi ^2 g_2 g_3 g_1 \nonumber \\ && -79488 g_2 g_3 g_1-\frac{4384}{3} \pi ^2 g_2^3-18176 g_2^3-12096 \pi ^2 g_2^2 g_3-142848 g_2^2 g_3\biggr) M \nonumber \\ && -19944 \pi ^2 g_1^3-223344 g_1^3-\frac{51968}{3} \pi ^2 g_2^3-181248 g_2^3-47808 \pi ^2 g_1 g_2^2-509184 g_1 g_2^2 -59760 \pi ^2 g_1^2 g_2  \nonumber \\ && -680832 g_1^2 g_2-46656 \pi ^2 g_1^2 g_3-473472 g_1^2 g_3-107136 \pi ^2 g_1 g_2 g_3-1009152 g_1 g_2 g_3,\\
\tilde{\beta}^{(4\text{ loop}, 0)}_{g_1} 
& = & 
\left(-36 \pi ^2 g_1^3-144 \pi ^2 g_2 g_1^2\right) M^3 
\nonumber 
\\ 
&& +\big(-1332 \pi ^2 g_1^3-15264 g_1^3-2880 \pi ^2 g_2 g_1^2-38016 g_2 g_1^2  - 2592 \pi ^2 g_3 g_1^2 \nonumber \\ && -40320 g_3 g_1^2-2880 \pi ^2 g_2^2 g_1-44928 g_2^2 g_1\big) M^2 \nonumber \\ && +\biggr(-19944 \pi ^2 g_1^3-223344 g_1^3-59760 \pi ^2 g_2 g_1^2-680832 g_2 g_1^2-46656 \pi ^2 g_3 g_1^2-473472 g_3 g_1^2 \nonumber \\ &&  - 47808 \pi ^2 g_2^2 g_1-509184 g_2^2 g_1-107136 \pi ^2 g_2 g_3 g_1-1009152 g_2 g_3 g_1-\frac{51968}{3} \pi ^2 g_2^3-181248 g_2^3\biggr) M \nonumber \\ && -86040 \pi ^2 g_1^3-863136 g_1^3-\frac{198656}{3} \pi ^2 g_2^3-614400 g_2^3-252672 \pi ^2 g_1 g_2^2-2330496 g_1 g_2^2 \nonumber \\ && -207360 \pi ^2 g_1 g_3^2-1709568 g_1 g_3^2-246960 \pi ^2 g_1^2 g_2-2339712 g_1^2 g_2  -256608 \pi ^2 g_1^2 g_3-2200320 g_1^2 g_3 \nonumber \\ && -221184 \pi ^2 g_2^2 g_3-1990656 g_2^2 g_3-387072 \pi ^2 g_1 g_2 g_3-3442176 g_1 g_2 g_3,
\end{eqnarray}
and, for $g_2$, 
\begin{dmath}
\tilde{\beta}^{(4\text{ loop}, 3)}_{g_2} =
     -24 \pi ^2 g_2^2 g_3 M^2+\big(-32 \pi ^2 g_2^3-18 \pi ^2 g_1^2 g_2-144 \pi ^2 g_1 g_3 g_2\big) M-27 \pi ^2 g_1^3-36 \pi ^2 g_2 g_1^2-216 \pi ^2 g_3 g_1^2-168 \pi ^2 g_2^2 g_1,
\end{dmath}
\begin{dmath}
    \tilde{\beta}^{(4\text{ loop}, 2)}_{g_2} = -24 \pi ^2 g_2^2 g_3 M^3+\left(-48 \pi ^2 g_2^3-576 g_2^3-96 \pi ^2 g_3 g_2^2-18 \pi ^2 g_1^2 g_2-540 g_1^2 g_2-864 \pi ^2 g_3^2 g_2-15552 g_3^2 g_2-288 \pi ^2 g_1 g_3 g_2-3456 g_1 g_3 g_2\right) M^2+\left(-162 \pi ^2 g_1^3-2430 g_1^3-234 \pi ^2 g_2 g_1^2-3024 g_2 g_1^2-972 \pi ^2 g_3 g_1^2-13392 g_3 g_1^2-1008 \pi ^2 g_2^2 g_1-12816 g_2^2 g_1-2592 \pi ^2 g_3^2 g_1-53568 g_3^2 g_1-864 \pi ^2 g_2 g_3 g_1-6912 g_2 g_3 g_1-144 \pi ^2 g_2^3-864 g_2^3-2688 \pi ^2 g_2^2 g_3-39168 g_2^2 g_3\right) M-675 \pi ^2 g_1^3-8100 g_1^3-1440 \pi ^2 g_2^3-18560 g_2^3-1848 \pi ^2 g_1 g_2^2-19584 g_1 g_2^2-3168 \pi ^2 g_1^2 g_2-34560 g_1^2 g_2-2376 \pi ^2 g_1^2 g_3-26784 g_1^2 g_3-8928 \pi ^2 g_1 g_2 g_3-134784 g_1 g_2 g_3,
\end{dmath}
\begin{dmath}
\tilde{\beta}^{(4\text{ loop}, 1)}_{g_2} = 
    \left(-32 \pi ^2 g_2^3-18 \pi ^2 g_1^2 g_2-144 \pi ^2 g_1 g_3 g_2\right) M^3+\left(-162 \pi ^2 g_1^3-2430 g_1^3-234 \pi ^2 g_2 g_1^2-3024 g_2 g_1^2-972 \pi ^2 g_3 g_1^2-13392 g_3 g_1^2-1008 \pi ^2 g_2^2 g_1-12816 g_2^2 g_1-2592 \pi ^2 g_3^2 g_1-53568 g_3^2 g_1-864 \pi ^2 g_2 g_3 g_1-6912 g_2 g_3 g_1-144 \pi ^2 g_2^3-864 g_2^3-2688 \pi ^2 g_2^2 g_3-39168 g_2^2 g_3\right) M^2+\left(-2727 \pi ^2 g_1^3-30780 g_1^3-11574 \pi ^2 g_2 g_1^2-132192 g_2 g_1^2-7020 \pi ^2 g_3 g_1^2-95472 g_3 g_1^2-6936 \pi ^2 g_2^2 g_1-84384 g_2^2 g_1-5184 \pi ^2 g_3^2 g_1-107136 g_3^2 g_1-38880 \pi ^2 g_2 g_3 g_1-521856 g_2 g_3 g_1-6552 \pi ^2 g_2^3-81312 g_2^3-49248 \pi ^2 g_2 g_3^2-556416 g_2 g_3^2-6432 \pi ^2 g_2^2 g_3-77184 g_2^2 g_3\right) M-14742 \pi ^2 g_1^3-152280 g_1^3-10848 \pi ^2 g_2^3-121344 g_2^3-53328 \pi ^2 g_1 g_2^2-580608 g_1 g_2^2-88128 \pi ^2 g_1 g_3^2-919296 g_1 g_3^2-40644 \pi ^2 g_1^2 g_2-422064 g_1^2 g_2-54864 \pi ^2 g_1^2 g_3-652320 g_1^2 g_3-83904 \pi ^2 g_2^2 g_3-880128 g_2^2 g_3-66528 \pi ^2 g_1 g_2 g_3-819072 g_1 g_2 g_3,
\end{dmath}
\begin{dmath}
\tilde{\beta}^{(4\text{ loop}, 0)}_{g_2} = 
    \left(-27 \pi ^2 g_1^3-36 \pi ^2 g_2 g_1^2-216 \pi ^2 g_3 g_1^2-168 \pi ^2 g_2^2 g_1\right) M^3+\big(-675 \pi ^2 g_1^3-8100 g_1^3-3168 \pi ^2 g_2 g_1^2-34560 g_2 g_1^2-2376 \pi ^2 g_3 g_1^2-26784 g_3 g_1^2-1848 \pi ^2 g_2^2 g_1-19584 g_2^2 g_1-8928 \pi ^2 g_2 g_3 g_1-134784 g_2 g_3 g_1-1440 \pi ^2 g_2^3-18560 g_2^3\big) M^2+\big(-14742 \pi ^2 g_1^3-152280 g_1^3-40644 \pi ^2 g_2 g_1^2-422064 g_2 g_1^2-54864 \pi ^2 g_3 g_1^2-652320 g_3 g_1^2-53328 \pi ^2 g_2^2 g_1-580608 g_2^2 g_1-88128 \pi ^2 g_3^2 g_1-919296 g_3^2 g_1-66528 \pi ^2 g_2 g_3 g_1-819072 g_2 g_3 g_1-10848 \pi ^2 g_2^3-121344 g_2^3-83904 \pi ^2 g_2^2 g_3-880128 g_2^2 g_3\big) M-61020 \pi ^2 g_1^3-550800 g_1^3-80192 \pi ^2 g_2^3-761216 g_2^3-177216 \pi ^2 g_1 g_2^2-1713024 g_1 g_2^2-176256 \pi ^2 g_1 g_3^2-1838592 g_1 g_3^2-442368 \pi ^2 g_2 g_3^2-4161024 g_2 g_3^2-196704 \pi ^2 g_1^2 g_2-1845504 g_1^2 g_2-173232 \pi ^2 g_1^2 g_3-1785024 g_1^2 g_3-166272 \pi ^2 g_2^2 g_3-1654272 g_2^2 g_3-484416 \pi ^2 g_1 g_2 g_3-4672512 g_1 g_2 g_3,
\end{dmath}
and, for $g_3$,
\begin{dmath}
\tilde{\beta}^{(4\text{ loop}, 3)}_{g_3}  
 = -72 \pi ^2 g_3^3 M^3+\left(\frac{1}{3} (-4) \pi ^2 g_2^3-216 \pi ^2 g_3^2 g_2\right) M^2+\left(-\frac{1}{3} 4 \pi ^2 g_2^3-12 \pi ^2 g_1 g_2^2-192 \pi ^2 g_3 g_2^2-216 \pi ^2 g_1 g_3^2\right) M-\frac{1}{3} 112 \pi ^2 g_2^3-12 \pi ^2 g_1 g_2^2-18 \pi ^2 g_1^2 g_2-288 \pi ^2 g_1 g_3 g_2
\end{dmath}
\begin{dmath}
\tilde{\beta}^{(4\text{ loop}, 2)}_{g_3}   = \left(\frac{1}{3} (-4) \pi ^2 g_2^3-216 \pi ^2 g_3^2 g_2\right) M^3+\left(-8 \pi ^2 g_2^3-24 \pi ^2 g_1 g_2^2-288 g_1 g_2^2-528 \pi ^2 g_3 g_2^2-2592 g_3 g_2^2-216 \pi ^2 g_3^2 g_2-2448 \pi ^2 g_3^3-30528 g_3^3-648 \pi ^2 g_1 g_3^2+288 g_1^2 g_3\right) M^2+\left(-\frac{1}{3} 1012 \pi ^2 g_2^3-2848 g_2^3-108 \pi ^2 g_1 g_2^2-864 g_1 g_2^2-456 \pi ^2 g_3 g_2^2-2592 g_3 g_2^2-126 \pi ^2 g_1^2 g_2-1404 g_1^2 g_2-6048 \pi ^2 g_3^2 g_2-76032 g_3^2 g_2-2160 \pi ^2 g_1 g_3 g_2-12096 g_1 g_3 g_2-648 \pi ^2 g_1 g_3^2+864 g_1^2 g_3\right) M-90 \pi ^2 g_1^3-1080 g_1^3-224 \pi ^2 g_2^3-2560 g_2^3-1236 \pi ^2 g_1 g_2^2-9792 g_1 g_2^2-3456 \pi ^2 g_1 g_3^2-40320 g_1 g_3^2-306 \pi ^2 g_1^2 g_2-2808 g_1^2 g_2-1404 \pi ^2 g_1^2 g_3-12240 g_1^2 g_3-3552 \pi ^2 g_2^2 g_3-44928 g_2^2 g_3-1584 \pi ^2 g_1 g_2 g_3-8640 g_1 g_2 g_3,
\end{dmath}
\begin{dmath}
\tilde{\beta}^{(4\text{ loop}, 1)}_{g_3}  
 = \left(-\frac{1}{3} 4 \pi ^2 g_2^3-12 \pi ^2 g_1 g_2^2-192 \pi ^2 g_3 g_2^2-216 \pi ^2 g_1 g_3^2\right) M^3+\left(-\frac{1}{3} 1012 \pi ^2 g_2^3-2848 g_2^3-108 \pi ^2 g_1 g_2^2-864 g_1 g_2^2-456 \pi ^2 g_3 g_2^2-2592 g_3 g_2^2-126 \pi ^2 g_1^2 g_2-1404 g_1^2 g_2-6048 \pi ^2 g_3^2 g_2-76032 g_3^2 g_2-2160 \pi ^2 g_1 g_3 g_2-12096 g_1 g_3 g_2-648 \pi ^2 g_1 g_3^2+864 g_1^2 g_3\right) M^2+\left(-441 \pi ^2 g_1^3-3510 g_1^3-1008 \pi ^2 g_2 g_1^2-11772 g_2 g_1^2-4644 \pi ^2 g_3 g_1^2-35856 g_3 g_1^2-4248 \pi ^2 g_2^2 g_1-41328 g_2^2 g_1-13824 \pi ^2 g_3^2 g_1-174528 g_3^2 g_1-5328 \pi ^2 g_2 g_3 g_1-44928 g_2 g_3 g_1-896 \pi ^2 g_2^3-10464 g_2^3-44640 \pi ^2 g_3^3-494208 g_3^3-6048 \pi ^2 g_2 g_3^2-76032 g_2 g_3^2-16152 \pi ^2 g_2^2 g_3-193248 g_2^2 g_3\right) M-1674 \pi ^2 g_1^3-14580 g_1^3-6800 \pi ^2 g_2^3-72832 g_2^3-7440 \pi ^2 g_1 g_2^2-80640 g_1 g_2^2-10368 \pi ^2 g_1 g_3^2-120960 g_1 g_3^2-72576 \pi ^2 g_2 g_3^2-774144 g_2 g_3^2-8964 \pi ^2 g_1^2 g_2-82728 g_1^2 g_2-8748 \pi ^2 g_1^2 g_3-90288 g_1^2 g_3-12000 \pi ^2 g_2^2 g_3-147456 g_2^2 g_3-39888 \pi ^2 g_1 g_2 g_3-454464 g_1 g_2 g_3,
\end{dmath}
\begin{dmath}
\tilde{\beta}^{(4\text{ loop}, 0)}_{g_3}  
 = \left(-\frac{1}{3} 112 \pi ^2 g_2^3-12 \pi ^2 g_1 g_2^2-18 \pi ^2 g_1^2 g_2-288 \pi ^2 g_1 g_3 g_2\right) M^3+\left(-90 \pi ^2 g_1^3-1080 g_1^3-306 \pi ^2 g_2 g_1^2-2808 g_2 g_1^2-1404 \pi ^2 g_3 g_1^2-12240 g_3 g_1^2-1236 \pi ^2 g_2^2 g_1-9792 g_2^2 g_1-3456 \pi ^2 g_3^2 g_1-40320 g_3^2 g_1-1584 \pi ^2 g_2 g_3 g_1-8640 g_2 g_3 g_1-224 \pi ^2 g_2^3-2560 g_2^3-3552 \pi ^2 g_2^2 g_3-44928 g_2^2 g_3\right) M^2+\left(-1674 \pi ^2 g_1^3-14580 g_1^3-8964 \pi ^2 g_2 g_1^2-82728 g_2 g_1^2-8748 \pi ^2 g_3 g_1^2-90288 g_3 g_1^2-7440 \pi ^2 g_2^2 g_1-80640 g_2^2 g_1-10368 \pi ^2 g_3^2 g_1-120960 g_3^2 g_1-39888 \pi ^2 g_2 g_3 g_1-454464 g_2 g_3 g_1-6800 \pi ^2 g_2^3-72832 g_2^3-72576 \pi ^2 g_2 g_3^2-774144 g_2 g_3^2-12000 \pi ^2 g_2^2 g_3-147456 g_2^2 g_3\right) M-10260 \pi ^2 g_1^3-74520 g_1^3-\frac{38272}{3} \pi ^2 g_2^3-130944 g_2^3-195840 \pi ^2 g_3^3-1903104 g_3^3-42912 \pi ^2 g_1 g_2^2-421056 g_1 g_2^2-101952 \pi ^2 g_1 g_3^2-1080576 g_1 g_3^2-72576 \pi ^2 g_2 g_3^2-774144 g_2 g_3^2-27936 \pi ^2 g_1^2 g_2-263088 g_1^2 g_2-40824 \pi ^2 g_1^2 g_3-428544 g_1^2 g_3-100608 \pi ^2 g_2^2 g_3-991872 g_2^2 g_3-79200 \pi ^2 g_1 g_2 g_3-877824 g_1 g_2 g_3.
\end{dmath}

\section{Anomalous dimensions at finite $M$ and $N$}

These equations are included in electronic supplemental information. \cite{Supplemental} 
\label{anomalousApp}
The four-loop anomalous dimension of $\phi$ is:
\begin{equation}
    \begin{split}
        &\gamma_{\phi}^{(4)}=\frac{1}{32400 (8\pi)^4} \biggr(3 {g_1}^2 (36+24 N+ 24 M+21 N M +4 {N}^2+4 {M}^2+3{N}^2{M}+3{N}{M}^2+{N}^2 {M}^2) +\\
        &+ 24 {g_3}^2 (8 + 6 N M + {N}^2 {M}^2) + 24 g_1 g_2 (8 +7 N+ 7 M +4 N M+{N}^2+ {M}^2 + {N}^2 M + N {M}^2) +\\
        &+ 48 g_1 g_3 (4 +3 N+3 M +3 N M+{N}^2 + {M}^2) +
        48 g_2 g_3 (4+4 N+4 M+N M+N^2 M  + {N} {M}^2) +\\ &+4 {g_2}^2 (28+16N+16M+19N M+4 {N}^2+4{M}^2+ {N}^2 M +N{M}^2+ {N}^2{M}^2
        )\biggr)
    \end{split}
\end{equation}

The four-loop anomalous dimension of $\phi^2$ is 
\begin{equation}
    \gamma^{(4)}_{\phi^2}=32 \gamma_\phi^{(4)}.
\end{equation}

The four loop anomalous dimension matrix for quartic operators is given by
\begin{equation}
    \begin{pmatrix}
    \gamma_{\nu_1, \nu_1} &  \gamma_{\nu_1, \nu_2} \\
    \gamma_{\nu_2, \nu_1} &  \gamma_{\nu_2, \nu_2}
    \end{pmatrix}.
\end{equation}
At two-loops the entries of this matrix are:
\begin{eqnarray}
\gamma^{(2)}_{\nu_1 \nu_1} & = & \frac{3 g_1 M N+6 g_1 M+6 g_1 N+30 g_1+8 g_2 M+8 g_2 N+20 g_2+24 g_3}{1440 \pi ^2} \\
\gamma^{(2)}_{\nu_1 \nu_2} & = &  \frac{3 g_1 M+3 g_1 N+6 g_1+g_2 M N+8 g_2}{360 \pi ^2} \\
\gamma^{(2)}_{\nu_2 \nu_1} & = &  \frac{3 g_1 M+3 g_1 N+9 g_1+2 g_2 M N+2 g_2 M+2 g_2 N+18 g_2+12 g_3 M+12 g_3 N+12 g_3}{1440 \pi ^2} \\
\gamma^{(2)}_{\nu_2 \nu_2} & = &  \frac{3 g_1+2 g_2 M+2 g_2 N+2 g_2+3 g_3 M N+12 g_3}{360 \pi ^2}.
\end{eqnarray}
We do not present the four-loop expressions here, but they are available in \cite{Supplemental}.

The anomalous dimensions of sextic operators are determined from the eigenvalues of the stability matrix $\partial_{\lambda_b} \beta_a$.

\section{Four-loop and $1/N$ corrections to the  bifundamental fixed-point}
\label{corrections}
In this appendix, we discuss four-loop and $1/N$ corrections in the bifundamental large-$N$ limit, described in section \ref{sec:unique-infrared-fixed-point}.
\subsection{Four-loop large-$N$ fixed-point}
\label{four-loop}
In the large-$N$ bifundamental limit, the four-loop beta function is given by
\begin{eqnarray}
\beta_{\lambda_1} & = & -2 \lambda _1 \epsilon +\frac{\alpha  \lambda _1^2}{10} -\frac{\alpha  \left(\pi ^2 \left(\alpha ^2+1\right)+34 \alpha \right) \lambda _1^3}{3600} \\
\beta_{\lambda_2} & = & -2 \lambda _2 \epsilon  +\frac{1}{30} \alpha  \lambda _1 \left(9 (\alpha +1) \lambda _1+4 \lambda _2\right)  -\frac{\alpha \lambda _1^2}{3600}   \Big(3 (\alpha +1) \left(\pi ^2 \left(\alpha ^2+5 \alpha +1\right)+90 \alpha \right) \lambda _1 \nonumber \\ && +2 \left(\pi ^2 \left(\alpha ^2+\alpha +1\right)+30 \alpha \right) \lambda _2\Big)
\\
\beta_{\lambda_3} & = & -2 \lambda _3 \epsilon +\frac{1}{90} \alpha  \left(21 \alpha  \lambda _1^2+12 (\alpha +1) \lambda _2 \lambda _1+4 \lambda _2^2\right) \nonumber \\ && -\frac{\alpha}{97200}\Big(27 \alpha  \left(30 \left(4 \alpha ^2+13 \alpha +4\right)+\pi ^2 \left(10 \alpha ^2+49 \alpha +10\right)\right) \lambda _1^3 \nonumber \\ && +4 \pi ^2 (\alpha +1) \lambda _2^3   +54 \lambda _1^2 \left((\alpha +1) \left(\pi ^2 \left(\alpha ^2+6 \alpha +1\right)+78 \alpha \right) \lambda _2-16 \alpha  \lambda _3\right) \nonumber \\ && +36 \left(\pi ^2 (\alpha +1)^2+24 \alpha \right) \lambda _2^2 \lambda _1\Big)
\end{eqnarray}

The unique infrared fixed-point at four-loops is:
\begin{eqnarray}
\lambda_1^* & = & \frac{20 \epsilon }{\alpha } + \frac{10 \left(\pi ^2 \left(\alpha ^2+1\right)+34 \alpha \right) \epsilon ^2}{9 \alpha ^2}+O\left(\epsilon ^3\right) \\
\lambda_2^* & = & -\frac{180 (\alpha +1) \epsilon }{\alpha }-\frac{10 \left((\alpha +1) \left(\pi ^2 \left(3 \alpha ^2+\alpha +3\right)+22 \alpha \right)\right) \epsilon ^2}{\alpha ^2}+O\left(\epsilon ^3\right) \\
\lambda_3^* & = & \left(480 \alpha +\frac{480}{\alpha }+\frac{3020}{3}\right) \epsilon \nonumber \\ && +\frac{10\epsilon ^2}{27 \alpha ^2} \left(2 \alpha  \left(558 \alpha ^2+1241 \alpha +558\right)+\pi ^2 \left(558 \alpha ^4+1604 \alpha ^3+1977 \alpha ^2+1604 \alpha +558\right)\right) \nonumber \\ &&  +O\left(\epsilon ^3\right).
\end{eqnarray}

\subsection{$1/N$ corrections}
\label{1Nappendix}
The complete beta function allows us to compute $1/N$ and $1/N^2$ corrections to the fixed-point at two-loops and four-loops, without much difficulty. Here, however, we content ourselves with $1/N$ and $1/N^2$ corrections to the two-loop fixed-point.

We find that the $1/N^2$ corrections to the fixed-point are:
\begin{eqnarray}
\tlambda_1^* & = & \frac{20 \epsilon }{\alpha }-\frac{80 (\alpha +1) \epsilon }{\alpha ^2 N}+\frac{400 \left(6 \alpha ^2+7 \alpha +6\right)}{3 \alpha ^3} \frac{\epsilon}{N^2}\\
\tlambda_2^* & = & -\frac{180 (\alpha +1) \epsilon }{\alpha }-\frac{960 \epsilon }{\alpha  N} + \frac{160 \left(135 \alpha ^3+529 \alpha ^2+529 \alpha +135\right)}{\alpha ^3} \frac{\epsilon}{N^2}\\
\tlambda_3^* & = & \frac{20}{3} \left(72 \alpha +\frac{72}{\alpha }+151\right) \epsilon +\frac{80 \left(63 \alpha ^3+388 \alpha ^2+388 \alpha +63\right) \epsilon }{3 \alpha ^2 N} \nonumber \\ && -\frac{80 \left(4032 \alpha ^4+18673 \alpha ^3+28714 \alpha ^2+18673 \alpha +4032\right)}{3 \alpha ^3}\frac{\epsilon}{N^2}
\end{eqnarray}

The anomalous dimension of the scalar field in this limit is
\begin{equation}
    \gamma_{\phi} = \frac{\epsilon ^2}{27}-\frac{5 (\alpha +1) \epsilon ^2}{27 \alpha  N}.
\end{equation}

The anomalous dimensions of quartic operators mix, when $1/N$ corrections are taken into account. The corrections to the anomalous dimension matrix are:
\begin{equation}
    \gamma_{\nu_1 \nu_2}^{(1/N)}=\frac{1}{N}\left(\begin{array}{cc}-\frac{16 (\alpha+1) \epsilon }{3 \alpha } & -\frac{64 (\alpha +1) \epsilon }{3 \alpha  } \\-\frac{40 (\alpha +1) \epsilon }{3 \alpha  } & 0 \\
\end{array}
\right), \quad \gamma_{\nu_1 \nu_2}^{(1/N^2)}=\frac{1}{N^2}\left(
\begin{array}{cc}
 \frac{16 \left(12 \alpha ^2-11 \alpha +12\right) \epsilon }{9 \alpha ^2} & -\frac{64 \left(2 \alpha ^2+11 \alpha
   +2\right) \epsilon }{3 \alpha ^2} \\
 -\frac{8 \left(10 \alpha ^2+49 \alpha +10\right) \epsilon }{3 \alpha ^2} & \frac{64 \left(27 \alpha ^2+59 \alpha
   +27\right) \epsilon }{9 \alpha ^2} \\
\end{array}
\right)
\end{equation}
The eigenvalues of the above matrix are:
\begin{equation}
    \gamma_{\nu_1}=\frac{8
   \epsilon }{3}-\frac{2 \left(\pi ^2 \alpha ^2+3 \pi ^2 \alpha +38 \alpha +\pi ^2\right) \epsilon ^2}{27 \alpha }-\frac{16 (\alpha +1) \epsilon }{3 \alpha  N}+\frac{16 \left(72
   \alpha ^2+109 \alpha +72\right) \epsilon }{9 \alpha ^2 N^2} 
\end{equation}
\begin{equation}
    \gamma_{\nu_2}=\frac{64 \epsilon ^2}{27}+\frac{64 \left(12 \alpha ^2+29 \alpha +12\right) \epsilon }{9 \alpha ^2 N^2}
\end{equation}
with the eigenvectors:
\begin{equation}
    \left(
\begin{array}{c}
 1 \\
-\frac{5 (\alpha +1)}{\alpha  N}-\frac{(20 \alpha ^2+69 \alpha +20)}{\alpha ^2 N^2} \\
\end{array}
\right) \quad \text{ and, } \quad \left(
\begin{array}{c}
 \frac{8 (\alpha +1)}{\alpha  N}+\frac{8 \left(4 \alpha ^2+15 \alpha +4\right)}{\alpha ^2 N^2} \\
 1 \\
\end{array}
\right)
\end{equation}
The corrections to the stability matrix are:
\begin{equation}
   M_{ab}^{(1/N)}=\frac{\epsilon}{N}\left(
\begin{array}{ccc}
0  & 0 & 0 \\
 -\frac{32 \left(3 \alpha ^2+8 \alpha +3\right)  }{\alpha  } &-\frac{16 (\alpha +1)  }{3 \alpha  } & 0 \\
  -\frac{712 (\alpha +1)  }{3 } &  -\frac{8 \left(10 \alpha ^2+49 \alpha +10\right)  }{3 \alpha  } & 0  \\
\end{array}
\right)
\end{equation}
The eigenvalues and eigenvectors are:
\begin{equation}
   2 \epsilon,\quad  \frac{2 \epsilon }{3}-\frac{16(1+\alpha) \epsilon }{3 \alpha  N} \quad \text{and} \quad  -2 \epsilon
\end{equation}
\begin{equation}
    \left(
\begin{array}{c}
 3+\frac{3 \left(-540 \alpha ^3-2444 \alpha ^2-2444 \alpha -540\right)}{\alpha  \left(72 \alpha ^2+151 \alpha
   +72\right) N} \\
 -27 (\alpha +1)-\frac{9 \left(-756 \alpha ^4-4260 \alpha ^3-6896 \alpha ^2-4260 \alpha -756\right)}{\alpha  \left(72
   \alpha ^2+151 \alpha +72\right) N} \\
 72 \alpha ^2+151 \alpha +72 \\
\end{array}
\right),  \quad \left(
\begin{array}{c}
 0 \\
 -1+\frac{20 \alpha ^2+69 \alpha +20}{5 \alpha  (\alpha +1) N} \\
 5 (\alpha +1) \\
\end{array}
\right) \quad \text{and} \quad \left(
\begin{array}{c}
 0 \\
 0 \\
 1 \\
\end{array}
\right)
\end{equation}

\subsection{Additional fixed-points at $O(1/N^2)$ and $O(1/N^4)$}
\label{new-fixed-points}
When terms of order $1/N^2$ are included in the two-loop beta function, we find the beta function admits a new fixed-point, not present for the leading-order beta function, which is:
\begin{eqnarray}
\lambda_1 =  O(\epsilon^2),~\lambda_2 = O(\epsilon^2), \lambda_3 = \frac{5}{2}\alpha N^2\epsilon + O(\epsilon^2). \label{new-sol-1}
\end{eqnarray}

When terms of order $1/N^4$ are included in the two-loop beta function, we find that the beta function admits another new fixed-points, which is:
\begin{eqnarray}
\lambda_1 &=& \frac{20\epsilon}{\alpha }-\frac{80 (\alpha +1)\epsilon}{\alpha ^2 N} + O(\epsilon^2), \nonumber \\ 
\lambda_2 &=& -\frac{100 (\alpha +1)}{\alpha } \epsilon+\frac{80 \left(14 \alpha ^2+13 \alpha +14\right)}{3 \alpha ^2 N}\epsilon + O(\epsilon^2), \nonumber \\
\lambda_3 &=& \frac{5 \alpha N^2 \epsilon}{2}-\frac{80 \left(31 \alpha ^3-27 \alpha ^2-27 \alpha +31\right)}{27 \alpha ^2}\epsilon+\frac{5 \left(80 \alpha ^2-29 \alpha +80\right)}{9 \alpha N}\epsilon + O(\epsilon^2)\label{new-sol-2}
\end{eqnarray}

We mention these fixed-points only because, when $\alpha$ is small, these correspond to large $M$ limits of the two infrared fixed-points $[A^-]$ and $[A^-_2]$ determined in section \ref{vector-fixed-point}. Note, however, that, at finite $\alpha$, the fixed-points determined in this appendix are only perturbative fixed-points. Unlike the fixed-points of section \ref{vector-fixed-point}, we expect these fixed-points to receive corrections from all higher loop contributions to the beta function , which, in the bifundamental large-$N$ limit, are not suppressed by powers of $N$. 

\section{Beta-functions as the gradient of a potential}
\label{potential}
In a general scalar theory in $3-\epsilon$ dimensions, the beta-function up to four loops may be written as a gradient, $$\beta_a=T_{ab}\frac{\partial U}{\partial{\lambda_b}},$$  of the following potential, 
where $g_{ijklmn}$ is assumed to be a symmetric tensor,
\begin{equation}
    U=-\epsilon g_{ijklmn}g_{ijklmn}+\frac{1}{3}g_{ijklmn}\beta^{(2)}_{ijklmn}+\frac{1}{4}g_{ijklmn}\beta^{(4)}_{ijklmn}.
\end{equation}
This expression is similar to the ones used in \cite{ Benedetti:2020sye} and \cite{Jack:2015tka}. The inverse metric is trivial at this order:
\begin{equation}
    (T^{-1})_{ab}=\frac{\partial g_{ijklmn}}{\partial g_a}\frac{\partial g_{ijklmn}}{\partial g_b}.
\end{equation}

The expression for the potential when $M$ and $N$ are both finite is too long to present here. We will mainly make use of this potential in the limit where $N$ is large and $M$ is finite. In this limit, a potential which allows one to compute the beta functions in section \ref{sec:vector-model}, is 
\begin{equation}
    \begin{split}
        &U=-\frac{\epsilon}{360 M}  \biggr(16 \lambda _2^2+24 \lambda _3^2+48 \lambda _1 \lambda _3+48 \lambda _2 \lambda _3+3 \lambda_1^2 M^4+9 \lambda _1^2 M^3+12 \lambda _1^2 M^2+4 \lambda _2^2 M^2+\\
        &24 \lambda _1 \lambda _2 M^2+4 \lambda _2^2 M+24 \lambda _1 \lambda _2 M\biggr)+\frac{1}{97200 M^2 N}\biggr(896 \lambda _2^3+1728 \lambda _3^3+5184 \lambda _1 \lambda _3^2+5184 \lambda _2 \lambda _3^2+\\
        &4608 \lambda _2^2\lambda _3+6912 \lambda _1 \lambda _2 \lambda _3+27 \lambda _1^3 M^6+189 \lambda _1^3 M^5+756 \lambda _1^3 M^4+72\lambda _1 \lambda _2^2 M^4+648 \lambda _1^2 \lambda _2 M^4+\\
        &756 \lambda _1^3 M^3+216 \lambda _1 \lambda _2^2M^3+1944 \lambda _1^2 \lambda _2 M^3+416 \lambda _2^3 M^2+2592 \lambda _1 \lambda _2^2 M^2+2592 \lambda _1^2\lambda _2 M^2+\\
        &2592 \lambda _1^2 \lambda _3 M^2+288 \lambda _2^2 \lambda _3 M^2+1728 \lambda _1 \lambda _2 \lambda_3 M^2+416 \lambda _2^3 M+2304 \lambda _1 \lambda _2^2 M+2592 \lambda _1^2 \lambda _3 M+\\
        &288 \lambda _2^2 \lambda_3 M+1728 \lambda _1 \lambda _2 \lambda _3 M\biggr)-\frac{\pi ^2}{46656000 M^2 N} \biggr(1408 \lambda _2^4+1728 \lambda _3^4+3584 \lambda _1 \lambda _2^3+6912 \lambda _1 \lambda_3^3+\\
        &6912 \lambda _2 \lambda _3^3+10368 \lambda _1^2 \lambda _3^2+10368 \lambda _2^2 \lambda _3^2+20736 \lambda _1\lambda _2 \lambda _3^2+6656 \lambda _2^3 \lambda _3+18432 \lambda _1 \lambda _2^2 \lambda _3+\\
        &27 \lambda _1^4 M^6+189 \lambda _1^4 M^5+756 \lambda _1^4 M^4+144 \lambda _1^2 \lambda _2^2 M^4+864 \lambda _1^3 \lambda _2 M^4+756 \lambda _1^4 M^3+432 \lambda _1^2 \lambda _2^2 M^3+\\
        &2592 \lambda _1^3\lambda _2 M^3+160 \lambda _2^4 M^2+1664 \lambda _1 \lambda _2^3 M^2+5184 \lambda _1^2 \lambda _2^2 M^2+3456 \lambda _1^3 \lambda _2 M^2+3456 \lambda _1^3 \lambda _3 M^2+\\
        &128 \lambda _2^3 \lambda _3 M^2+1152 \lambda _1\lambda _2^2 \lambda _3 M^2+3456 \lambda _1^2 \lambda _2 \lambda _3 M^2+160 \lambda _2^4 M+1664 \lambda _1 \lambda_2^3 M+4608 \lambda _1^2 \lambda _2^2 M+\\
        &3456 \lambda _1^3 \lambda _3 M+128 \lambda _2^3 \lambda _3 M+1152 \lambda_1 \lambda _2^2 \lambda _3 M+3456 \lambda _1^2 \lambda _2 \lambda _3 M+13824 \lambda _1^2\lambda _2 \lambda _3\biggr).
    \end{split}
    \label{potentialVector}
\end{equation}
In the above expression, we rescaled $U$ by a factor of $N$, which was absorbed into the inverse metric in Eq. \eqref{inverse-metric} to make it independent of $N$.

\section{large-$N$ ``unconventional'' fixed-points in $d=3$}
\label{spooky-appendix}

In this Appendix, we extend our analysis in section \ref{sec:vector-model} to sextic theories with are $O(M)\times O(N)/\mathbb Z_2$ symmetry with non-integer values of $M$. $N$ is still assumed to be a (large) integer. We are motivated by the fact that $O(M)$ theories based on non-integer values of $M$ have recently been given a physical interpretation in \cite{Binder:2019zqc}, and have attracted some attention recently \cite{Jepsen:2020czw, Jepsen:2021rhs}. \cite{Binder:2019zqc} also explains how to formulate $O(M)$ symmetry for non-integer $M$ via category theory, and describes some of the (rather exotic) statistical models that may have such a symmetry. 

The essential idea of \cite{Jepsen:2020czw, Jepsen:2021rhs}  is that, when one considers fixed-points with non-integer $M$ one may obtain limit cycles or unusual behaviour. This can only occur if one works in a range of $M$ where the theory is non-unitary. For our theory, when $M = 2$, the three couplings are not independent -- this is reflected in the fact that the matrix $T$ in Eq. \eqref{inverse-metric} involved in rewriting the beta functions as the gradient of a potential has a zero eigenvalue when $M=2$. For non-integer in the range $M<2$, the matrix $T$ is not longer positive definite and possesses at least one negative eigenvalue. Therefore, although the beta functions are related to the gradient of a potential, the matrix $T$ relating the beta functions to the gradient is not positive definite, allowing, in principle for unconventional RG trajectories. We may also be able to find fixed-points for which the eigenvalues of $\frac{\partial \beta}{\partial g_i}$ include a pair of complex eigenvalues, or a pair of purely imaginary eigenvalues. Such fixed-points were dubbed as unconventional fixed-points in \cite{Jepsen:2020czw}. If a fixed-point were found in which the eigenvalues were purely imaginary, then we would expect small deformations around the fixed-point to result in limit cycles -- i.e., closed RG flows which start and end at the same point in coupling constant space. Such a fixed-point is known as a Hopf point. For fixed-points with complex eigenvalues, RG trajectories form spirals in coupling constant space, which are unusual, but, admittedly much less interesting than limit cycles. 

Our fixed-points differ from those in \cite{Jepsen:2020czw} in that our fixed-points are large-$N$ saddle points in integer dimension $d=3$, while the fixed-points there were perturbative finite $N$ fixed-points studied in $d=3-\epsilon$. So if we were to find limit cycles in our model, it would be quite interesting, and there could be a formal holographic dual to such RG flows. 

\begin{figure}
    \centering
    \includegraphics[width=.8\textwidth]{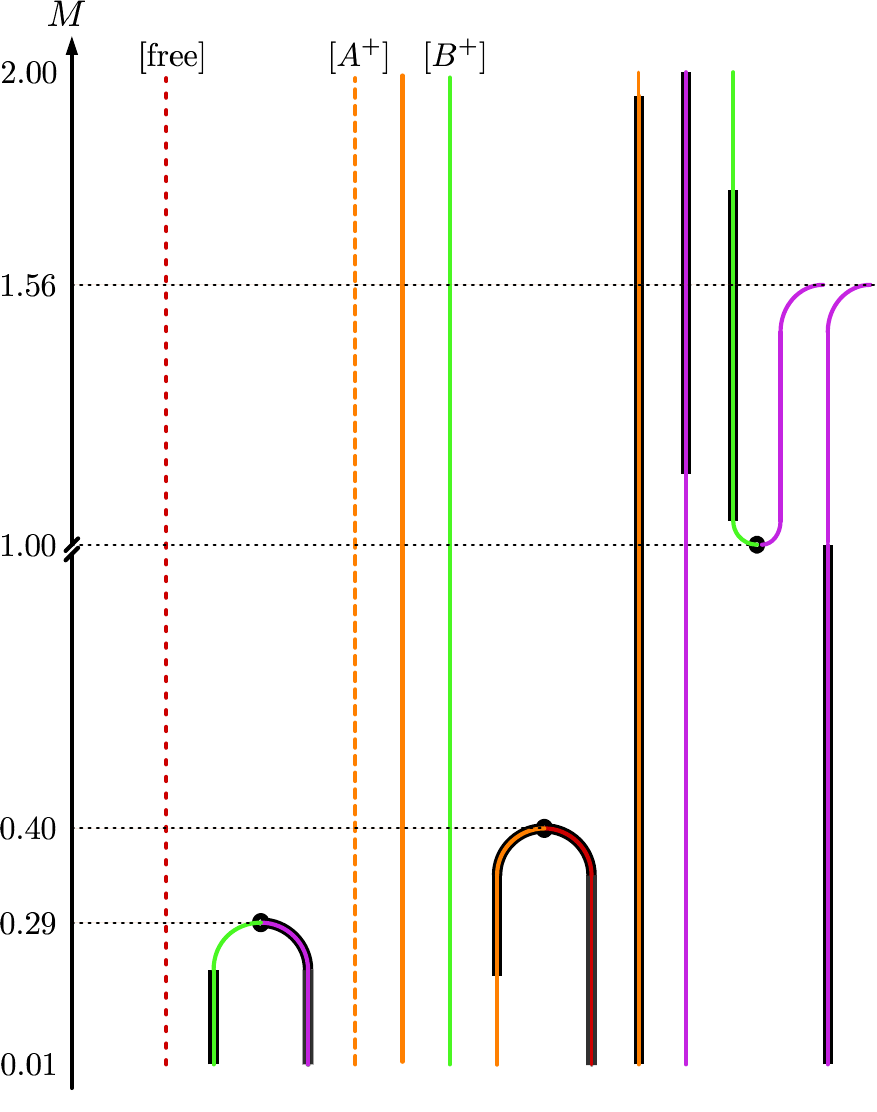}
    \caption{A plot of the fixed-points that exist in $d=3$, for non-integer $M$ between $0$ and $2$. We find that, for some ranges of $M$ fixed-points become ``spooky'' or unconventional \cite{Jepsen:2020czw} -- i.e., the stability matrix $\partial_{\lambda_i}\beta_j$ contains complex eigenvalues. These ranges are highlighted in black. The colour of each line denotes the number of unstable directions as in Fig. \ref{fig:fixed-points-d3}, where directions corresponding to complex eigenvalues of the stability matrix are considered unstable if their real part is negative. Unlike Fig. \ref{fig:fixed-points-d3}, fixed-points are not arranged according to the value of the potential.}
    \label{fig:spooky}
\end{figure}

The fixed-point solutions are depicted schematically in Fig. \ref{fig:spooky}. We find non-unitary fixed-points with whose stability matrix $\partial_{\lambda_i}\beta_j$ has a pair of complex eigenvalues, similar to some of the fixed-points in \cite{Jepsen:2020czw}. Unfortunately, however, we do not find any fixed-points with purely imaginary eigenvalues, hence we do not expect limit cycles in the theory.\footnote{We thank Fedor Popov and Igor Klebanov for discussions on this point.} We also studied the fixed-points in the range $-3<M<0$, and did not find any fixed-point with purely imaginary eigenvalues. For $M<-3$, the inverse metric $T$ in Eq. \eqref{inverse-metric} is negative definite, so no unconventional fixed-points are possible. 

 the fixed-points in the main text correspond to symmetry groups with integer values of $M$ and $N$, and therefore may describe phase transitions of some as-yet unknown, but not unconventional, statistical models that generalize the classic models such as the tricritical Ising model in a straightforward way -- they may be considered exotic only to the extent that presumably require the tuning of several chemical potentials, rather than just one.

\clearpage

\bibliography{tensor}
\bibliographystyle{ssg}

\end{document}